\pdfoutput=1
\documentclass[11pt,twoside,a4paper,cmspaper,final,collab]{cms-tdr}
\def\svnVersion{bc60a43}\def\svnDate{2026/07/10}\def\cmsCernNoTag{CERN-EP-2026-034}\def\cmsCernDate{\today}\def\cmsMessage{Published in the Journal of High Energy Physics as \href{http://dx.doi.org/10.1007/JHEP07(2026)054}{\doi{10.1007/JHEP07(2026)054}.}}
\begin{document}\cmsNoteHeader{SMP-24-012}

\newlength\cmsFigWidth
\ifthenelse{\boolean{cms@external}}{\setlength\cmsFigWidth{0.49\textwidth}}{\setlength\cmsFigWidth{0.65\textwidth}}
\ifthenelse{\boolean{cms@external}}{\providecommand{\cmsRight}{lower\xspace}}{\providecommand{\cmsRight}{right\xspace}}
\ifthenelse{\boolean{cms@external}}{\providecommand{\cmsLeft}{upper\xspace}}{\providecommand{\cmsLeft}{left\xspace}}
\newlength\cmsTabSkip\setlength{\cmsTabSkip}{0.2ex}

\DeclareRobustCommand{\PV}{{\HepParticle{V}{}{}}\Xspace}
\newcommand{\mw}{\ensuremath{m_{\PW}}\xspace}
\newcommand{\NTwo}{\ensuremath{N_{2}^{(1)}}\xspace}
\newcommand{\NTwoDDT}{\ensuremath{N_{2}^{(1),\mathrm{DDT}}}\xspace}
\newcommand{\pNetWMD}{\ensuremath{\pnscore{\mathrm{W vs. QCD}}}\xspace}
\newcommand{\pNetWMDDDT}{\ensuremath{P^{\mathrm{PN, DDT}}_{\mathrm{W vs. QCD}}}\xspace}
\newcommand{\softdropmass}{\ensuremath{m_\mathrm{SD}}\xspace}
\newcommand{\msd}{\softdropmass}
\newcommand{\msdgen}{\ensuremath{\softdropmass^\mathrm{ptcl}}\xspace}
\newcommand{\msdreco}{\ensuremath{\softdropmass^\mathrm{reco}}\xspace}
\newcommand{\ptgen}{\ensuremath{\pt^\mathrm{ptcl}}\xspace}
\newcommand{\rrho}{\ensuremath{\rho_{\mathrm{SD}}}\xspace}
\newcommand{\wqqjets}{\ensuremath{\PW(\PQq\PAQq')\text{+jets}}\xspace}
\newcommand{\zqqjets}{\ensuremath{\PZ(\PQq\PAQq)\text{+jets}}\xspace}
\newcommand{\wjets}{\ensuremath{\PW\text{+jets}}\xspace}
\newcommand{\zjets}{\ensuremath{\PZ\text{+jets}}\xspace}
\newcommand{\vjets}{\ensuremath{\PV\text{+jets}}\xspace}
\newcommand{\rrhoscaled}{\ensuremath{\hat{\rho}_\mathrm{SD}}\xspace}
\newcommand{\ptscaled}{\ensuremath{\hat{p}_{\mathrm{T}}}\xspace}
\newcommand{\POI}[2]{\ensuremath{\hat{\mu}_{p_{\mathrm{T}}^{#1},~m_{\mathrm{SD}}^{#2}}}\xspace}
\newcommand{\mybinom}[2]{\ensuremath{\begin{pmatrix}#1\\#2\end{pmatrix}}}
\newcommand{\nrho}{\ensuremath{n_{\rrho}}\xspace}
\newcommand{\npt}{\ensuremath{n_{\pt}}\xspace}
\newcommand{\brho}{\ensuremath{b_i^{\nrho}(\rrhoscaled)}\xspace}
\newcommand{\bpt}{\ensuremath{b_{j}^{\npt}(\ptscaled)}\xspace}
\newcommand{\pnscore}[1]{\ensuremath{P^{\mathrm{PN}}_{#1}}}
\newcommand{\pp}{\ensuremath{\Pp\Pp}\xspace}

\cmsNoteHeader{SMP-24-012} 
\title{Measurement of the jet mass in hadronic decays of boosted \texorpdfstring{\PW}{W} bosons at \texorpdfstring{13\TeV}{13 TeV} and extraction of the \texorpdfstring{\PW}{W} boson mass}

\date{\today}

\abstract{
The jet mass of \PW bosons decaying to a quark-antiquark pair is measured in \wjets events from proton-proton collisions at a center-of-mass energy of 13\TeV. The data used were collected by the CMS experiment at the CERN LHC and correspond to an integrated luminosity of 138\fbinv. Hadronic decays of \PW bosons with high momenta produce strongly collimated decay products due to the large Lorentz boost, and are reconstructed as single large-radius jets. These jets have a characteristic substructure that is exploited to distinguish them from the large background of quark- and gluon-initiated jets. The jet mass is computed using the soft-drop algorithm, which suppresses soft wide-angle radiation that leads to a broadening of the jet mass distribution. For the first time, unfolded measurements are presented of the double-differential \wjets cross section as a function of the jet transverse momentum and soft-drop mass. From these distributions, the \PW boson mass is obtained, with a value of $80.83 \pm 0.55\GeV$ in a scheme with mass-dependent width, achieving the smallest uncertainty available today from an all-jets final state at a hadron collider.
}

\hypersetup{
pdfauthor={CMS Collaboration},
pdftitle={Measurement of the jet mass in hadronic decays of boosted W bosons at 13 TeV and extraction of the W boson mass},
pdfsubject={CMS},
pdfkeywords={CMS, QCD, jet substructure, W mass}}

\maketitle

\section{Introduction\label{sec:Introduction}}

The \PW boson is produced at high rates in proton-proton \pp collisions at the CERN LHC.
The mass of the \PW boson, \mw, is among the most precisely known parameters of the standard model (SM) of particle physics, with measurements at LEP~\cite{ALEPH:2013dgf}, Tevatron~\cite{D0:2012kms,CDF:2022hxs}, and LHC~\cite{ATLAS:2017rzl, LHCb:2021bjt,ATLAS:2024erm}.
The most recent measurement by the CMS Collaboration yields $\mw=80.3602\pm0.0099$\GeV~\cite{CMS:2024lrd}, 
consistent with the SM expectation~\cite{ParticleDataGroup:2024cfk}.
The large center-of-mass energy at the LHC enables the study of \PW bosons with very high transverse momenta $\pt \gg \mw$.
The majority of \PW bosons decay into quark-antiquark pairs that appear in the detector as two separate jets. 
For \PW bosons with high \pt and therefore large Lorentz boosts, the quark-antiquark pair is highly collimated 
in the laboratory frame, such that the products of the parton showering and hadronization of the quark-antiquark pair 
are contained within a single large-radius jet. 
At the LHC, jets with high energy are produced abundantly and only a small fraction originates from \PW bosons. 
An analysis of the jet substructure can be used to identify the initiating particle of such jets.
A well-established quantity to distinguish such jets from quark- and gluon-initiated background jets 
is the jet mass computed after the removal of soft and wide-angle radiation from the parton shower 
with a jet grooming technique~\cite{Marzani:2019hun, Larkoski:2017jix}. 
Jets initiated by boosted \PW bosons are the subject of many direct searches for new physics and measurements 
of SM properties, which are often limited by imperfect modeling of the jet substructure~\cite{Kogler:2018hem, Kogler:2021kkw}.
Such modeling involves parton showering and hadronization, described in the simulation partly by perturbative quantum chromodynamics (QCD) and 
partly by phenomenological models.
A precise measurement and understanding of the \PW boson jet mass distribution is therefore important.

Previous measurements at SPS~\cite{UA2:1987vmn}, LEP~\cite{ALEPH:2013dgf} and Tevatron~\cite{CDF:1997iiz} had extracted \mw using the \PW boson decay to a final state with a pair of jets.
In this paper, we present the first measurement of the mass of large-radius jets 
originating from the hadronic decay of \PW bosons with large Lorentz boosts.
We aim at improving the precision compared to previous measurements in all-jets final states at hadron colliders.
The expected precision of the extracted \mw parameter is not yet competitive with previous measurements making use of leptonic final states. This measurement represents a first step towards a potential precision measurement of \mw in the all-jets final state at the high-luminosity LHC (HL-LHC) with prospects outlined in Ref.~\cite{Freytsis:2018cev}.
The measurement reported here is mainly aimed at testing and improving the modeling of jet substructure in \PW boson production.

\begin{figure}[htb]
    \centering
        \includegraphics[width=0.35\textwidth]{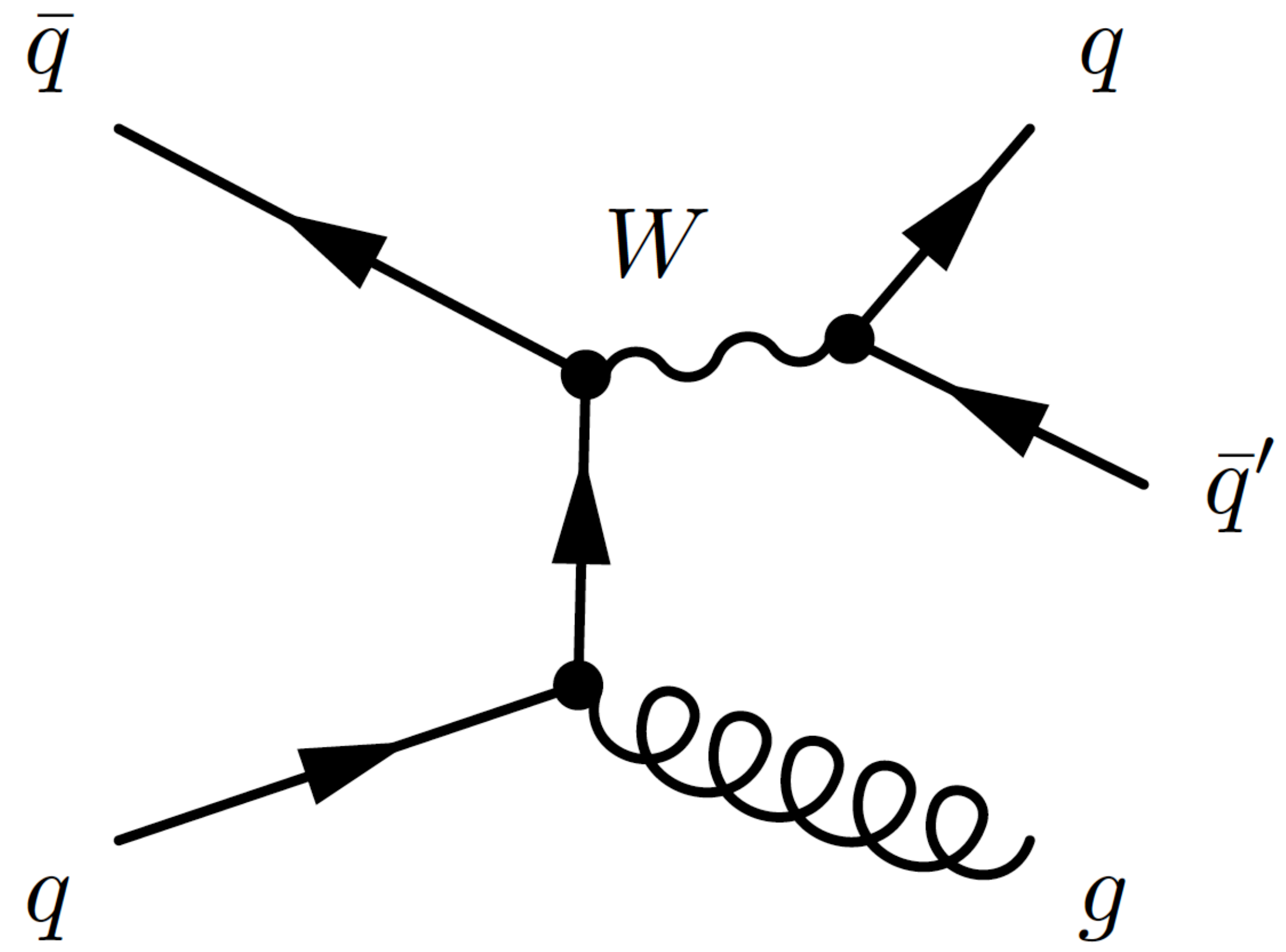}
    \caption{Example tree-level Feynman diagram contributing to \wqqjets production.
    \label{img:feynmanWqqjets}}
\end{figure}
An example Feynman diagram of a boosted \wqqjets event is depicted in Fig.~\ref{img:feynmanWqqjets}.
At leading order, the jet initiated by the boosted \PW boson and a recoiling jet from a quark or gluon carry the same transverse momentum making them difficult to distinguish.
The sample of boosted \wqqjets events is subject to a large background coming from multijet production through QCD processes.
This introduces one of the main challenges in the analysis and is addressed with 
dedicated \PW boson jet tagging approaches for background removal and an estimate of the remaining QCD multijet background using control samples in data.
Ordinary QCD jets acquire mass from splittings of quarks and gluons in the parton shower.
To mitigate the effects of soft and wide-angle radiation that obscure the jet mass, 
the modified mass drop algorithm~\cite{Dasgupta:2013ihk,Butterworth:2008iy}, 
which is a specific example of soft-drop (SD) grooming~\cite{Larkoski:2014wba}, is applied.
This technique removes soft and wide-angle radiation below a specified threshold.
Applying grooming significantly reduces the mass of quark- and gluon-initiated jets,
while bringing the mass of signal jets closer to the \PW boson mass, 
thereby enhancing the signal-to-background ratio for \PW boson identification.
In addition, events are categorized using a jet substructure tagger that distinguishes 
the characteristic `two-prong' substructure, corresponding to the fragmentation and hadronization of the two colored partons from the \PW boson decay,
from the `one-prong' background, 
where prongs are localized clusters of energy within the collection of final-state particles of the jet.

The analysis uses \pp collision data at $\sqrt{s}=13\TeV$ 
recorded by the CMS experiment~\cite{CMS:2008xjf} at the LHC during 2016--2018, 
corresponding to an integrated luminosity of 138\fbinv.
Measurements of the jet mass distribution have been previously performed at the LHC for light-quark and 
gluon jets~\cite{ATLAS:2012am,ATLAS:2017zda,CMS:2018vzn}, 
top quark jets~\cite{CMS:2017pcy, CMS:2019fak, CMS:2022kqg}, and \PZ boson jets~\cite{ATLAS:2019dty}.
The measurements of quark and gluon jets offer a testbed for perturbative and nonperturbative QCD predictions.
The study of the top quark jet mass enables the determination of the top quark mass using a sample of 
boosted \PW bosons for the jet mass calibration, which relies on a precise understanding of jet substructure 
in events with a color-connected top quark-antiquark pair (\ttbar).
The \wqqjets process offers a sample of boosted \PW bosons decaying to a quark-antiquark pair that is not color-connected to the remainder of the event.
In this paper, the groomed jet mass distribution of \wqqjets events is measured 
in intervals of the jet transverse momentum with $\pt>650 \GeV$.
A two-dimensional unfolding of the data in jet mass and \pt is performed using a maximum-likelihood template fit.
Finally, \mw is extracted using a template fit to the unfolded data.

The paper is organized as follows.
In Section~\ref{sec:cms} a brief description of the CMS detector and event reconstruction is given.
Section~\ref{sec:simulation} provides details on the simulation used for the analysis.
The jet mass observable and jet substructure algorithms are explained in Section~\ref{sec:substructure}.
This is followed by descriptions of the event selection in Section~\ref{sec:selection}, the background estimation procedure in Section~\ref{sec:background} and the uncertainties in Section~\ref{sec:uncertainties}.
The unfolding of the data for detector effects is described in Section~\ref{sec:unfolding} and the results of the cross section measurement and the estimation of the \mw parameter are presented in Section~\ref{sec:results}.
Section~\ref{sec:summary} summarizes the paper.

\section{The CMS detector and event reconstruction}\label{sec:cms}

The central feature of the CMS apparatus~\cite{CMS:2008xjf,CMS:2023gfb} is a superconducting solenoid of 6\unit{m} internal diameter, providing a magnetic field of 3.8\unit{T}. Within the solenoid volume are a silicon pixel and strip tracker, a lead tungstate crystal electromagnetic calorimeter (ECAL), and a brass and scintillator hadron calorimeter (HCAL), each composed of a barrel and two endcap sections. Forward calorimeters extend the pseudorapidity ($\eta$) coverage provided by the barrel and endcap detectors. Muons are detected in gas-ionization chambers embedded in the steel flux-return yoke outside the solenoid.
A more detailed description of the CMS detector, together with a definition of the coordinate system used and the relevant kinematic variables, can be found in Ref.~\cite{CMS:2008xjf,CMS:2023gfb}.

The silicon tracker used in 2016 measured charged particles within the range $\abs{\eta} < 2.5$. For nonisolated particles of $1 < \pt < 10\GeV$ and $\abs{\eta} < 1.4$, the track resolutions were typically 1.5\% in \pt and 25--90 (45--150)\mum in the transverse (longitudinal) impact parameter~\cite{CMS:2014pgm}. At the start of 2017, a new pixel detector was installed~\cite{Phase1Pixel}; the upgraded tracker measured particles up to $\abs{\eta} = 3.0$ with typical resolutions of 1.5\% in \pt and 20--75\mum in the transverse impact parameter~\cite{DP-2020-049} for nonisolated particles of $1 < \pt < 10\GeV$. According to simulation studies~\cite{DP-2017-015}, similar improvements are expected in the longitudinal direction. 

Events of interest are selected using a two-tiered trigger system. The first level, composed of custom hardware processors, uses information from the calorimeters and muon detectors to select events at a rate of around 100\unit{kHz} within a fixed latency of 4\mus~\cite{CMS:2020cmk}. The second level, known as the high-level trigger, consists of a farm of processors running a version of the full event reconstruction software optimized for fast processing, and reduces the event rate to a few kHz before data storage~\cite{CMS:2016ngn,CMS:2024aqx}.

The particle-flow (PF) algorithm~\cite{CMS:2017yfk} aims to reconstruct and identify each individual particle in an event, with an optimized combination of information from the various elements of the CMS detector. The energy of photons is obtained from the ECAL measurement. The energy of electrons is determined from a combination of the electron momentum at the primary interaction vertex as determined by the tracker, the energy of the corresponding ECAL cluster, and the energy sum of all bremsstrahlung photons spatially compatible with originating from the electron track. The energy of muons is obtained from the curvature of the corresponding track. The energy of charged hadrons is determined from a combination of their momentum measured in the tracker and the matching ECAL and HCAL energy deposits, corrected for the response function of the calorimeters to hadronic showers. Finally, the energy of neutral hadrons is obtained from the corrected ECAL and HCAL energies.

The primary vertex (PV) is taken to be the vertex corresponding to the hardest scattering in the event, evaluated using tracking information alone, as described in Section 9.4.1 of Ref.~\cite{CMS-TDR-15-02}.
Additional \Pp{}\Pp{} interactions within the same or nearby bunch crossings (pileup) can contribute additional tracks and calorimetric energy depositions to the jet momentum~\cite{CMS:2020uim}.
The pileup per particle identification algorithm (PUPPI)~\cite{Sirunyan:2020foa,Bertolini:2014bba} is used to mitigate the effect of pileup at the reconstructed particle level, making use of local shape information, event pileup properties, and tracking information. A local shape variable is defined, which distinguishes between collinear and soft diffuse distributions of other particles surrounding the particle under consideration. The former is attributed to particles originating from the PV and the latter to particles originating from pileup interactions. Charged particles identified to be originating from pileup vertices are discarded. For each neutral particle, a local shape variable is computed using the surrounding charged particles compatible with the PV within the tracker acceptance ($\abs{\eta} < 2.5$), and using both charged and neutral particles in the region outside of the tracker coverage. The momenta of the neutral particles are then rescaled according to their probability to originate from the PV, deduced from the local shape variable, superseding the need for jet-based pileup corrections~\cite{Sirunyan:2020foa}. 

For each event, hadronic jets are clustered from the reconstructed PF candidates with the PUPPI weights applied, using the infrared and collinear safe anti-\kt algorithm~\cite{Cacciari:2008gp, Cacciari:2011ma} with a distance parameter $R=0.8$.
The jet momentum is determined as the vectorial sum of all particle momenta in the jet. It is found from simulation to be, on average, within 5 to 10\% of the true momentum over the entire \pt spectrum and detector acceptance.
Jet energy corrections are derived from simulation to adjust the measured response of jets to that of particle-level jets on average. In-situ measurements of the momentum balance in dijet, \Pgg{}+jets, \zjets, and multijet events are used to account for any residual differences in jet energy scale in experimental data and simulation~\cite{CMS:2016lmd}.
The jet energy resolution amounts typically to 15\% at 10\GeV, 8\% at 100\GeV, and 4\% at 1\TeV.
Additional selection criteria are applied to each jet to remove jets potentially dominated by instrumental effects or reconstruction failures~\cite{Sirunyan:2020foa}.

In simulation, we also cluster particle-level jets using the same jet algorithm from final-state particles with proper lifetime $c\tau > 10\unit{mm}$, after hadronization, where $c$ is the speed of light. The distance parameter is $R=0.8$ unless quoted otherwise.

The techniques used for the reconstruction of the jet mass and jet substructure are discussed in Section~\ref{sec:substructure}.

\section{Event simulation}\label{sec:simulation}

The simulation of \pp collision events is performed using various software packages in multiple steps.
Processes of \wqqjets, and \zqqjets are generated at leading order (LO) using \MGvATNLO v2.6.5~\cite{MadGraph_aMCatNLO} with additional partons added to the matrix element computation. The \wqqjets sample is generated with up to three additional partons, while the \zqqjets one is generated with up to four partons.
The \MGvATNLO generator is interfaced to \PYTHIA~8.309~\cite{Bierlich:2022pfr} with the CP5 tune~\cite{Sirunyan:2019dfx} for the parton showering and hadronization.
\PYTHIA implements a dipole shower ordered in \pt and the hadronization of quarks and gluons 
into final hadrons is described by the Lund string model~\cite{Andersson:1983ia,Sjostrand:1984iu}. 
To avoid double-counting of jets from the matrix element calculations and the parton shower, jets are matched using the MLM scheme~\cite{MadGraph5_MLM}.
An alternative sample of \wqqjets is obtained interfacing \MGvATNLO with \HERWIG~7.2.2~\cite{Bellm:2015jjp} 
at LO with angular-ordered showering and the CH3 tune~\cite{Sirunyan:2020pqv}.
The sample is generated with up to three jets in the final state using FxFx matching~\cite{Frederix:2012ps}.
The \MGvATNLO generator is used to generate samples of QCD multijet events binned in the scalar sum of all 
jet transverse momenta (\HT) at LO with up to four $R=0.4$ jets in the final state that are showered 
with \PYTHIA using MLM matching.
The \POWHEG v2 generator~\cite{POWHEG_main1,POWHEG_main2,POWHEG_main3} is used for the simulation of the 
\ttbar process at next-to-LO (NLO) accuracy in perturbative QCD~\cite{POWHGE_BOX:hvq} and is
interfaced with \PYTHIA. 
All samples are generated using the NNPDF3.1~\cite{NNPDF:2017mvq} parton distribution function (PDF) sets at next-to-NLO (NNLO) accuracy.
The generated events are then passed through 
the \GEANTfour~\cite{GEANT4:2002zbu} package to simulate the CMS detector response before 
they are treated with the same reconstruction algorithms as the real data.

The simulated events are corrected for various effects, which are not well modeled in the simulation. 
These correction factors are applied to the simulated events as event weights. 
The \wqqjets and \zqqjets samples are corrected to higher-order precision using the same dedicated 
correction factors as derived in Ref.~\cite{CMS:2022fxs}. These corrections are multiplicative 
factors for the differential cross section of the \vjets (where \PV stands for \PW or \PZ) processes 
as a function of the \pt of the vector boson. They include higher-order QCD and electroweak (EW) effects. 
The QCD corrections were derived from NLO \vjets samples generated using \MGvATNLO, whereas the EW 
corrections are based on higher-order calculations at NLO EW precision~\cite{Lindert:2017olm}.
The NLO \POWHEG{}+\PYTHIA\ \ttbar samples used in the analysis show deviations in the \pt spectrum of 
top quarks when compared to data~\cite{TopPTdilepton,TopPTlepjets}. We apply a reweighting of \ttbar events 
based on the measurements of the top quark \pt.

As the detector conditions have varied throughout the data taking, four separate sets of simulations with corresponding correction factors matching four data-taking periods (early 2016, late 2016, 2017, and 2018) are used. 
The events are generated with a distribution in the number of pileup interactions per proton bunch crossing, 
which is not the same as in observed events. All simulated events are reweighted to match the profile of that in the respective 
data-taking period.

\section{Jet mass reconstruction and jet substructure tagging}\label{sec:substructure}

The jet with the largest \pt in the event, referred to as the leading jet, is considered the boosted \PW boson candidate.
While at leading order, the chance of selecting a jet from the recoiling quark or gluon in a \wqqjets event rather than the boosted \PW boson is 50\%, this fraction is strongly reduced by further selections on jet substructure as explained in the following.

The constituents of the leading jet are reclustered with the Cambridge--Aachen (CA) 
algorithm~\cite{Dokshitzer:1997in}, and then groomed using the SD algorithm. 
The SD algorithm splits the jet into two subjets when undoing the last step of the CA jet clustering.
It is then checked if the two subjets satisfy the condition
\begin{linenomath}
\begin{equation}
 \frac{ \min ( \pt^{(1)}, \pt^{(2)} ) }{\pt^{(1)} + \pt^{(2)}  } > z_{\text{cut}} \left(  \frac{ \Delta R_{12} }{R} \right) ^ {\beta_{\text{sd}}},
\end{equation}
\end{linenomath}
where $\pt^{(1)}$ and $\pt^{(2)}$ are the transverse momenta of the two subjets, $R$ is the jet radius parameter, $\Delta R_{12} = \sqrt{(\Delta y_{12})^2 + (\Delta \phi_{12})^2}$ is the distance between the two subjets with the rapidity and azimuthal differences $\Delta y_{12}$ and $\Delta \phi_{12}$, and $z_{\text{cut}}$ and $\beta_{\text{sd}}$ are tunable parameters of the SD algorithm, set to default parameters $z_{\text{cut}} = 0.1$ and $\beta_{\text{sd}} = 0$ used in CMS studies~\cite{CMS:2020poo}.
In Ref.~\cite{Freytsis:2018cev}, it was shown that the SD algorithm with this set of parameters is well suited for this measurement in combination with a jet substructure tagger to reduce sensitivity to nonperturbative effects.
If the condition is met, the vectorial sum of the two subjets is returned as the SD groomed jet.
If the condition is not met, the subjet with the lower \pt is rejected.
The declustering procedure is then repeated by splitting the higher-\pt subjet into two subjets, undoing another CA clustering step, and iterating until the condition is met.
This grooming procedure removes soft and wide-angle radiation from the jet, thereby making the jet mass more resilient to effects from pileup, underlying event, and initial-state radiation.
The jet \pt used for the measurement is taken from the original ungroomed anti-\kt jet, because anti-\kt jet \pt is resistant to soft and wide-angle radiation. 
The SD grooming is only applied to compute the jet mass.

The reconstructed SD jet mass, \msd, is corrected to match that of particle-level jets on average in two steps.
First, jet energy corrections~\cite{CMS:2016lmd} are applied to the momenta of the SD subjets of the uncalibrated CA jets, matching the response in data to that of particle-level jets.
As the \msd is strongly correlated with the momenta of the subjets, this procedure yields a reconstructed \msd in data close to the particle-level SD jet mass \msdgen.
Second, we apply \pt-dependent corrections of 2--6\%, derived from the ratio of the reconstructed SD jet mass to the particle-level SD jet mass in simulated \wqqjets events.
While this correction does not modify the data-to-simulation agreement, it simplifies the unfolding by yielding a response closer to unity.
Because of the PUPPI algorithm, the jet \msd is almost insensitive to pileup~\cite{Sirunyan:2020foa}.
It should be noted that this procedure does not rely on data samples containing boosted \PW bosons, but rather independent dijet, \Pgg{}+jets, \zjets, and multijet data samples for calibration, in order to avoid any bias to the mass measurement.

We use jet substructure taggers that distinguish the characteristic two-prong substructure of 
jets from boosted \PW bosons from light-flavor quark and gluon background jets.
We carry out the analysis with two different tagging strategies, first with a well-established 
jet substructure observable $\NTwo$~\cite{Moult:2016cvt,Larkoski:2013eya} and second with a 
neural-network classifier called \textsc{ParticleNet}~\cite{Qu:2019gqs}. 
The $\NTwo$-based tagger is closer to a theoretically well-defined particle-level observable, 
thus it can yield smaller extrapolation uncertainties between detector and particle levels.
The \textsc{ParticleNet} tagger provides a cleaner sample with less background, thus it can yield 
smaller experimental uncertainties. Whereas for a future measurement with a larger data set, e.g. at the HL-LHC, 
the $\NTwo$-based approach may be more suitable, for this measurement, the \textsc{ParticleNet} 
approach is presented as the main result, and the  $\NTwo$-based approach as a cross-check.

The $\NTwo$ observable is a combination of energy correlation functions (ECFs) \cite{Larkoski:2013eya, Moult:2016cvt} 
\begin{equation}
    \label{eq:ecf_base}
    _\nu e_n^{(\beta)} = \sum_{1\leq i_1 \leq i_2 < \cdots < i_n \leq n_J} \left(\prod_{a=1}^{n} z_{i_a}\right) \prod_{m=1}^{\nu} \min^{(m)}_{s<t\in{i_1,i_2,...,i_n}} \left(\Delta R_{s,t}\right)^\beta,
\end{equation}
with the energy fractions $z_i = \frac{p_{\mathrm{T}, i}}{p_{\mathrm{T}, J}}$, the particle and jet transverse momenta 
$p_{\mathrm{T}, i}$ and $p_{\mathrm{T}, J}$, the number of constituent particles $n_J$ in the jet, the number of particles to correlate $n$, the chosen number $\nu$ of angles $\Delta R_{s,t}$ between particles with indices $s$ and $t$ entering the product, and the angular exponent $\beta$.
The $N_i$ series of ECF ratios is designed to distinguish jets with $N$-prong substructure,  
\begin{equation}
    N_n^{(\beta)} = \frac{_{2}e_{n+1}^{(\beta)}}{\left(_{1}e_{n}^{(\beta)}\right)^2}. 
\end{equation}
In the context of this paper, the variable $\NTwo$ is used for the identification of \PW boson jets.
Using the definition of generalized ECFs in Eq.~\eqref{eq:ecf_base} with $\beta=1$, the variable 
\NTwo is defined using the following two-point and three-point correlation functions:
\begin{equation}
\begin{aligned}
    _{2}e^{(1)}_3 & = \sum_{1\leq i \leq j \leq k \leq n_J} z_i z_j z_k \min\left\{\Delta R_{ij}\Delta R_{ik},\Delta R_{ij}\Delta R_{jk},\Delta R_{ik}\Delta R_{jk}\right\}, \\
    _{1}e^{(1)}_2 & = \sum_{1\leq i \leq j \leq n_J} z_i z_j\Delta R_{ij}.
\end{aligned}
\end{equation}

The \textsc{ParticleNet}~\cite{Qu:2019gqs} algorithm is a graph neural network (GNN) based on 
a dynamic graph convolutional neural network~\cite{10.1145/3326362}, designed to discriminate hadronic decays of highly boosted top quarks, \PW, \PZ, and Higgs bosons from QCD multijet background. 
Whereas many other machine-learning approaches treat the jet as an ordered structure~\cite{CMS:2020poo}, 
the \textsc{ParticleNet} algorithm treats jets as particle clouds, i.e., an unordered, permutation-invariant set of particles.
The GNN yields probability scores (in the following called \pnscore{}) for each class of process used in the training. Each class corresponds to a decay mode of the considered particle-category (e.g., $\PW\to \Pq \Paq'$, where $\Pq \in [\PQu,\PQd, \PQs]$). The \textsc{ParticleNet} algorithm is trained with the jet constituents (PUPPI  weighted PF candidates) and secondary vertices associated with the jet.
Input for each particle is information about its kinematic properties (\pt, $\eta$, $\phi$, energy), its charge, and the difference of each kinematic property and the corresponding property of the jet.
The secondary vertices include information about their kinematics, displacement, and quality criteria.
The \pNetWMD discriminator is defined as the ratio of the sum of probability scores of the GNN for $\PW\to \Pq \Paq'$, where $\Pq \in [\PQu,\PQd,\PQs,\PQc]$, and the sum over the $\PW\to \Pq \Paq'$ and QCD multijet background modes includes all decay modes considered by \textsc{ParticleNet}.

The $\NTwo$ observable and the \textsc{ParticleNet} discriminator are correlated with \msd.
In this measurement, a technique to estimate the QCD multijet background described in Section~\ref{sec:background}  
relies on a mass-decorrelated tagging variable for the construction of the control region (CR) and signal region (SR).
We therefore construct mass-decorrelated versions of the $\NTwo$ and \textsc{ParticleNet} discriminants.
For \textsc{ParticleNet}, the GNN architecture is trained using a signal sample that consists of a generic 
spin-0 particle $\PX$ with a uniform mass distribution that decays into two highly Lorentz-boosted particles.
This dedicated mass-decorrelated (MD) training is referred to as \textsc{ParticleNet}-MD.
Additionally, the jets used in the training are reweighted both in the signal and in the QCD background sample 
to have a flat jet \pt and \msd spectrum~\cite{CMS-DP-2020-002}.
Finally, for both $\NTwo$ and \textsc{ParticleNet}-MD, we follow the 
design decorrelated tagger (DDT) \cite{Dolen:2016kst,CMS:2019emo} approach to modify the tagging variables.
This approach decorrelates the tagging variable by enforcing a constant selection efficiency for QCD multijet 
background jets across the probed phase space. 
The DDT approach introduces the variable $\rrho = 2 \ln(\frac{\msd}{\pt})$~\cite{Dasgupta:2013ihk}, whose distribution is approximately independent of \pt for QCD jets.
Whereas \msd and \pt could be used interchangeably for $\rrho$ and \pt, the latter allow to describe the phase space more conveniently for QCD jets.
Here, the approach pioneered by CMS~\cite{CMS:2019emo} is adopted, 
thus the decorrelation is flattening the $(\rrho,\pt)$-dependence of the tagging variable.
For this, a constant selection efficiency for QCD jets across the $(\rrho,\pt)$ plane is enforced 
when selecting on \NTwo or \pNetWMD.
The decorrelated variables $\NTwoDDT$ and $\pNetWMDDDT$, where the latter is also referred to as 
\textsc{ParticleNet}$^\mathrm{DDT}$, are defined as
\begin{equation}
\begin{aligned}
    \NTwoDDT = & \NTwo - P^X_{\NTwo}(\rrho, \pt),  \\
    \pNetWMDDDT = & P^X_{\pNetWMD}(\rrho, \pt) - \pNetWMD,
\end{aligned}
\end{equation}
where $P^X_i$ denotes the $X$-th QCD selection efficiency percentile of the tagger $i$ as a function of \rrho and \pt. 
The two taggers are then used by requiring the new variables to be negative, i.e., $\NTwoDDT<0$ and $\pNetWMDDDT<0$ to identify \PW jets.

\section{Event selection}\label{sec:selection}

The data were collected by the CMS experiment in the years 2016--2018 and correspond to an integrated luminosity of 138\fbinv.
Triggers requiring an anti-\kt jet with $R=0.8$ and a minimum \pt of 450 and 500\GeV are used to collect \wqqjets events.
The trigger with $\pt>450\GeV$ was prescaled during part of the data taking, resulting in a lower effective luminosity of 56.6\fbinv.
The trigger with $\pt>500\GeV$ was unprescaled, but inactive during a small fraction of data taking. The data  collected with this trigger correspond to an integrated luminosity of 135\fbinv.
The two triggers are used to select event samples with offline jets passing $575<\pt<650\GeV$ and 
$\pt>650\GeV$, respectively.
Both triggers are more than 99.9\% efficient.
To account for residual differences in trigger efficiency in data and the modeled efficiency in the simulation, 
data-to-simulation scale factors are measured and applied as event weights to the simulation.

After reconstruction and calibration, events are required to have at least one jet with $\pt>575\GeV$ and $\abs{\eta}<2.4$.
The measurement is carried out in \pt intervals (bins) with the low bin edges 
at 575, 650, 725, 800, 1000, 1200\GeV, and larger (denoted by $\infty$).
This choice of bin edges ensures that the bin width is well above the experimental resolution of the order 
5--10\% and the number of bins is small enough to have reasonable statistical uncertainties in all bins. 
Additional selection criteria are applied to remove jets potentially dominated by anomalous contributions 
from various subdetector components or reconstruction failures~\cite{Sirunyan:2020foa}.
In 2018, two sectors of the HCAL endcap in the region $-1.57< \phi< -0.87$ and $-3.2< \eta< -1.3$ were not functional 
for the last ${\approx}65\%$ of the data taking. In data recorded during this time, events are rejected if 
the leading jet is reconstructed within this region. Events in the simulation are weighted accordingly if the 
leading jet is reconstructed in this region to reflect the resulting inefficiency.
Events with a reconstructed electron or muon with $\pt>10\GeV$ and $\abs{\eta}<2.4$ and passing loose 
identification criteria~\cite{CMS:2020uim, CMS:2018rym} are rejected to suppress backgrounds with lepton final states.

The main backgrounds of the measurement are processes with high-\pt quark and gluon jets, 
which arise from QCD multijet processes or \wqqjets events where the leading jet is not initiated by the \PW boson decay, 
but rather by a recoiling quark or gluon.
At LO, the jet initiated by the boosted \PW boson and the recoiling jet from a quark or gluon 
carry opposite transverse momentum, while at higher orders, the \PW boson jet is more likely to be reconstructed 
as the leading jet. Two handles to suppress those backgrounds are first the \softdropmass and second 
jet substructure taggers that distinguish the characteristic two-prong substructure of boosted \PW boson jets from background. 

The leading jet is required to have $\msd>50\GeV$ to reject background from quark and gluon jets.
Further, the jet is required to have $\rrho < -2.1$ to reject events with high \msd but low jet \pt and avoid instabilities and edge effects from the SD algorithm and jet clustering~\cite{ATLAS:2019mgf}.
For jets from boosted \PW bosons this selection ensures that the decay products of the \PW boson are fully merged within a jet with $R=0.8$.

{\tolerance=800
Events are finally categorized into signal (pass) and control (fail) regions using a mass-decorrelated jet 
substructure tagger with a QCD background efficiency of 5\%. 
The corresponding signal and background efficiencies for both tagging approaches are summarized in 
Table~\ref{tab:efficiencies_reco_wjets}. 
The \pNetWMDDDT approach yields a significantly larger signal efficiency at a similar background efficiency.
\par}

\begin{table}[tb]
    \centering    
    \topcaption{Summary of signal ($\varepsilon_{\PW\!+\!\text{jets}}$) 
    and background ($\varepsilon_{\text{bkg}}$) efficiencies in the different signal (pass) and control (fail) regions for 
    the $\NTwoDDT$ and $\pNetWMDDDT$ taggers in the fully-hadronic \wqqjets selection. 
    The efficiencies are estimated from simulation and averaged over the years of data-taking.\label{tab:efficiencies_reco_wjets}}
    \begin{tabular}{lcc{c}@{\hspace*{5pt}}cc}
        & \multicolumn{2}{c}{$\NTwoDDT$} && \multicolumn{2}{c}{$\pNetWMDDDT$}\\
        & $\varepsilon_{\PW\!+\!\text{jets}}$ & $\varepsilon_{\text{bkg}}$&&  
        $\varepsilon_{\PW\!+\!\text{jets}}$ & $\varepsilon_{\text{bkg}}$\\
        \cline{2-3} \cline{5-6}
        Pass & 18.5\% & 4.1\% &&  38.2\% & 4.2\% \\
        Fail & 81.5\% & 95.9\% && 61.8\% & 95.8\% \\
    \end{tabular}
\end{table}

In this measurement, we carry out an unfolding of the data to the particle level (ptcl).
At the particle level, we require at least one anti-\kt jet with $R=0.8$ and with $\ptgen>500\GeV$ and $\msdgen>30\GeV$.
The measurement is carried out in $\ptgen$ and $\msdgen$ bins with lower bin edges at 500, 650, 800 and 1200\GeV,  
and 30, 70, 80 and 90\GeV, respectively.
While the $500<\ptgen<650\GeV$ bin is taken into account in the migration matrix, the final result 
is presented for $\ptgen>650\GeV$.
The binning of \ptgen is chosen to have twice the bin width of the binning at the detector level, 
which is a typical choice when using a maximum likelihood fit based unfolding.
The approach requires having more bins at the detector level than at the particle level to have an 
over-constrained fit.
The binning of \msdgen is chosen such that the stability and purity are above approximately 50\% for all bins.
The stability is the fraction of events in a given bin at the particle level that are reconstructed 
in the same bin at the detector level.
Similarly, purity is the fraction of events that are reconstructed in a given bin at the detector level 
and originate from the same bin at the particle level.
At detector level, a finer binning in steps of 5\GeV in \msd is chosen to offer more constraining power to the background estimation procedure.
In addition, we require $\NTwo<0.2$ at the particle level.
The corresponding fiducial cross sections at each stage of the particle-level selection 
are reported in Table~\ref{tab:crosssection}, computed with \MGvATNLO{}+\PYTHIA at NLO QCD and 
with EW corrections. 

\begin{table}[tb]
    \topcaption{Fiducial \wqqjets cross sections predicted by \MGvATNLO{}+\PYTHIA at NLO QCD with EW corrections 
    after different stages of the particle-level selection. The given uncertainty is the sum in quadrature 
    of parton shower variations and the hadronization model uncertainty, as well as the 
    uncertainties in the QCD and EW corrections.
\label{tab:crosssection}}
    \centering
    \begin{tabular}{lc}
        Selection & Cross section [fb]\\
        \hline
        $\ptgen>650\GeV$ & $1396^{+116}_{-146}$  \\
        $\ptgen>650\GeV$ \& $\msdgen>30\GeV$ & $888^{+74}_{-93}$ \\
        $\ptgen>650\GeV$ \& $\msdgen>30\GeV$ \& $\NTwo<0.2$ & $241^{+24}_{-21}$ \\
    \end{tabular}
\end{table}

\begin{figure}[tb]
    \centering
        \includegraphics[width=0.49\textwidth]{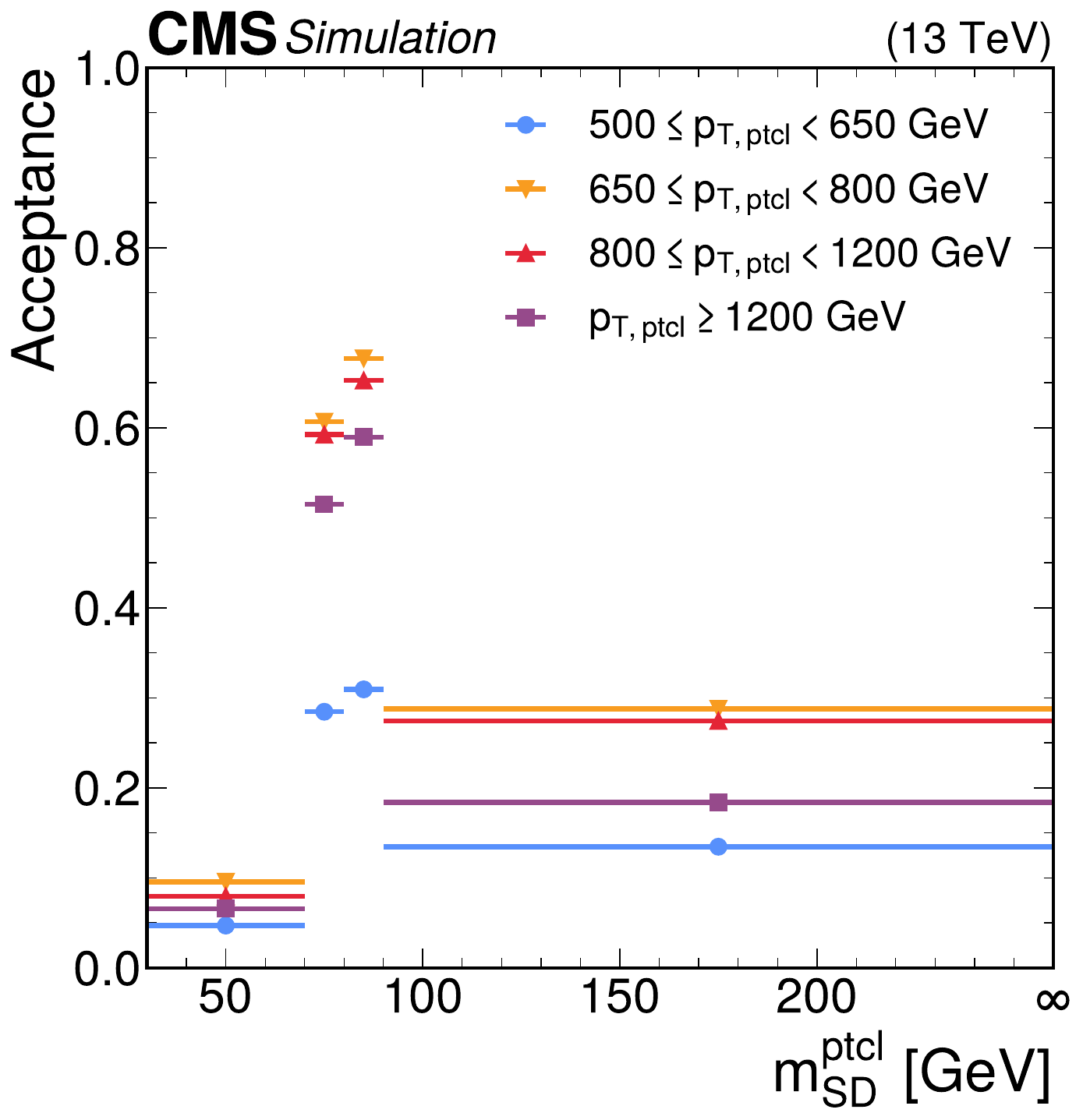}
        \includegraphics[width=0.49\textwidth]{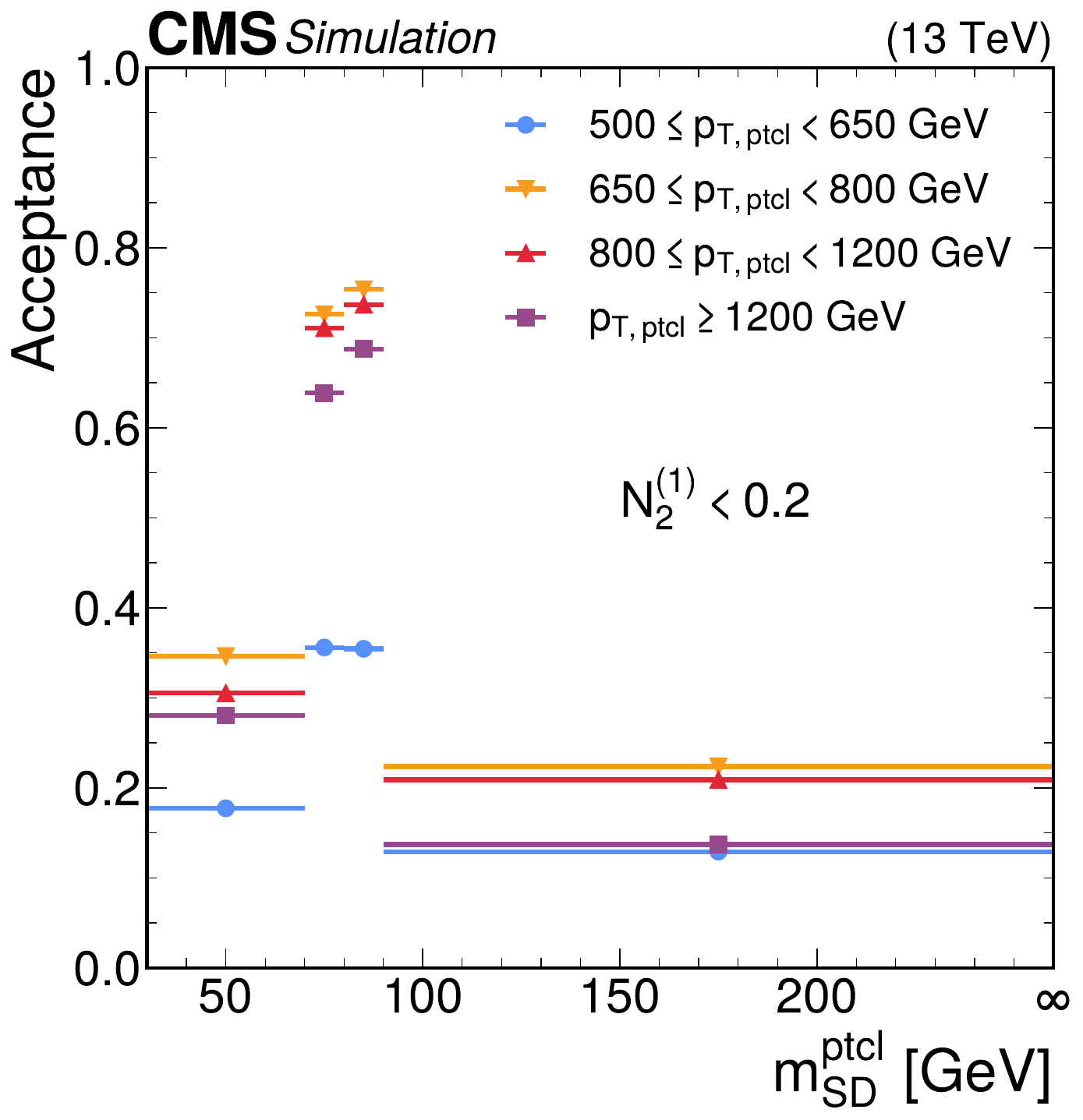}
    \caption{Acceptance as a function of \msdgen without (\cmsLeft) and with (\cmsRight) the requirement 
    $\NTwo<0.2$ at the particle level. The acceptance is calculated using the \wqqjets signal simulation with 2018 detector conditions. \label{img:acceptance_efficiency}}
\end{figure}
Figure~\ref{img:acceptance_efficiency} shows the acceptance of the detector-level selection, 
which includes the jet \pt, $\abs{\eta}$, and $\softdropmass$ requirements and a matching within $\Delta R<0.4$ between the detector-level jet and the particle-level jet corresponding to the simulated \PW boson.
The first and last bins in \msdgen have a low acceptance of 5--20\%, especially in the first and last \ptgen bins,
mainly because the selected detector-level jet often does not match to the simulated W boson, when outside the \PW boson peak region.
The central \msdgen bins in the \PW boson mass peak region have a much larger acceptance of 30--70\%.
When including the $\NTwo<0.2$ selection in the particle-level definition, the acceptance is significantly larger.
The efficiencies of the trigger and jet reconstruction are larger than 99\% and accounted for as well.

\begin{figure}[tbp]
    \centering
        \includegraphics[width=0.49\textwidth]{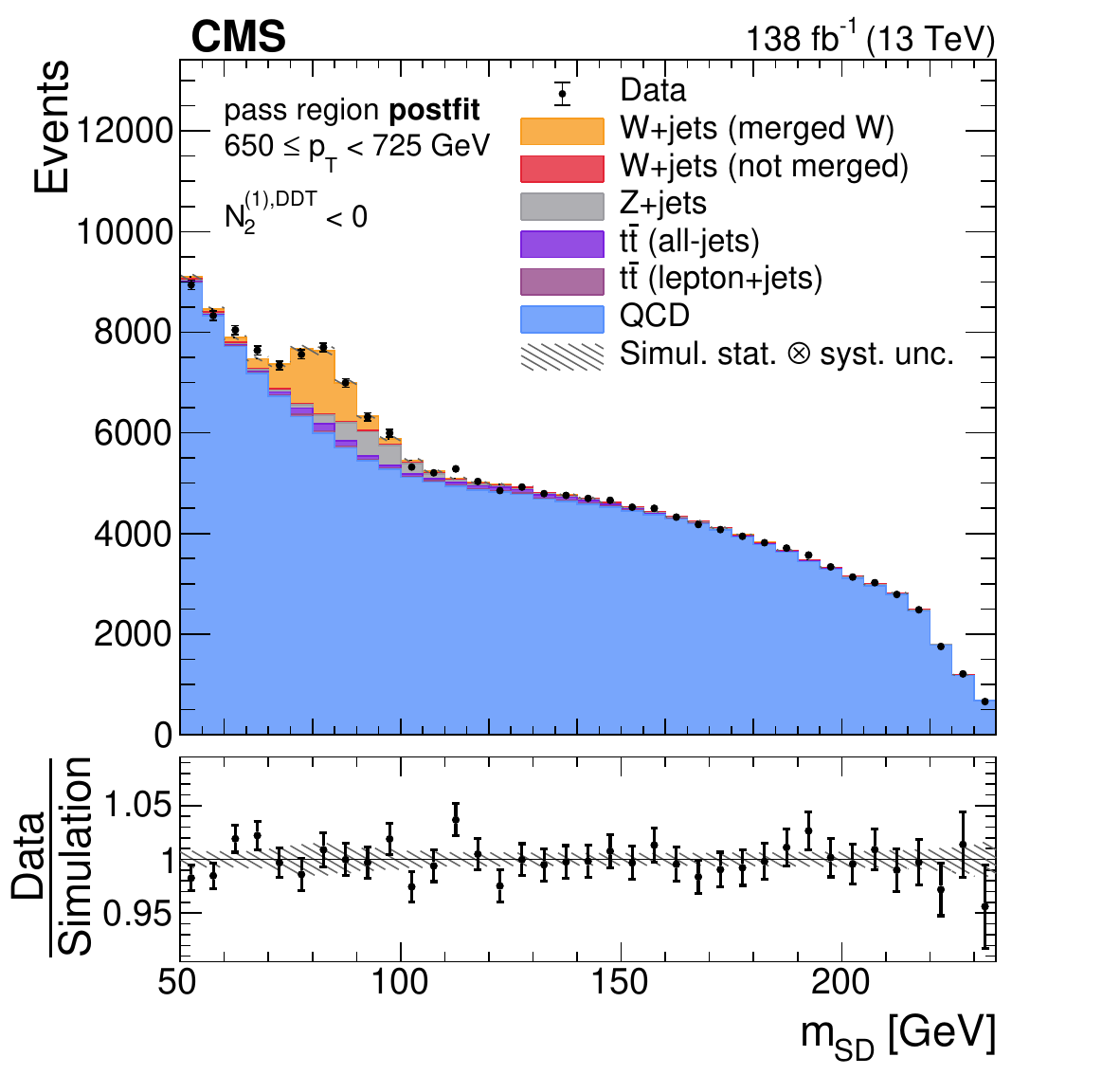}
        \includegraphics[width=0.49\textwidth]{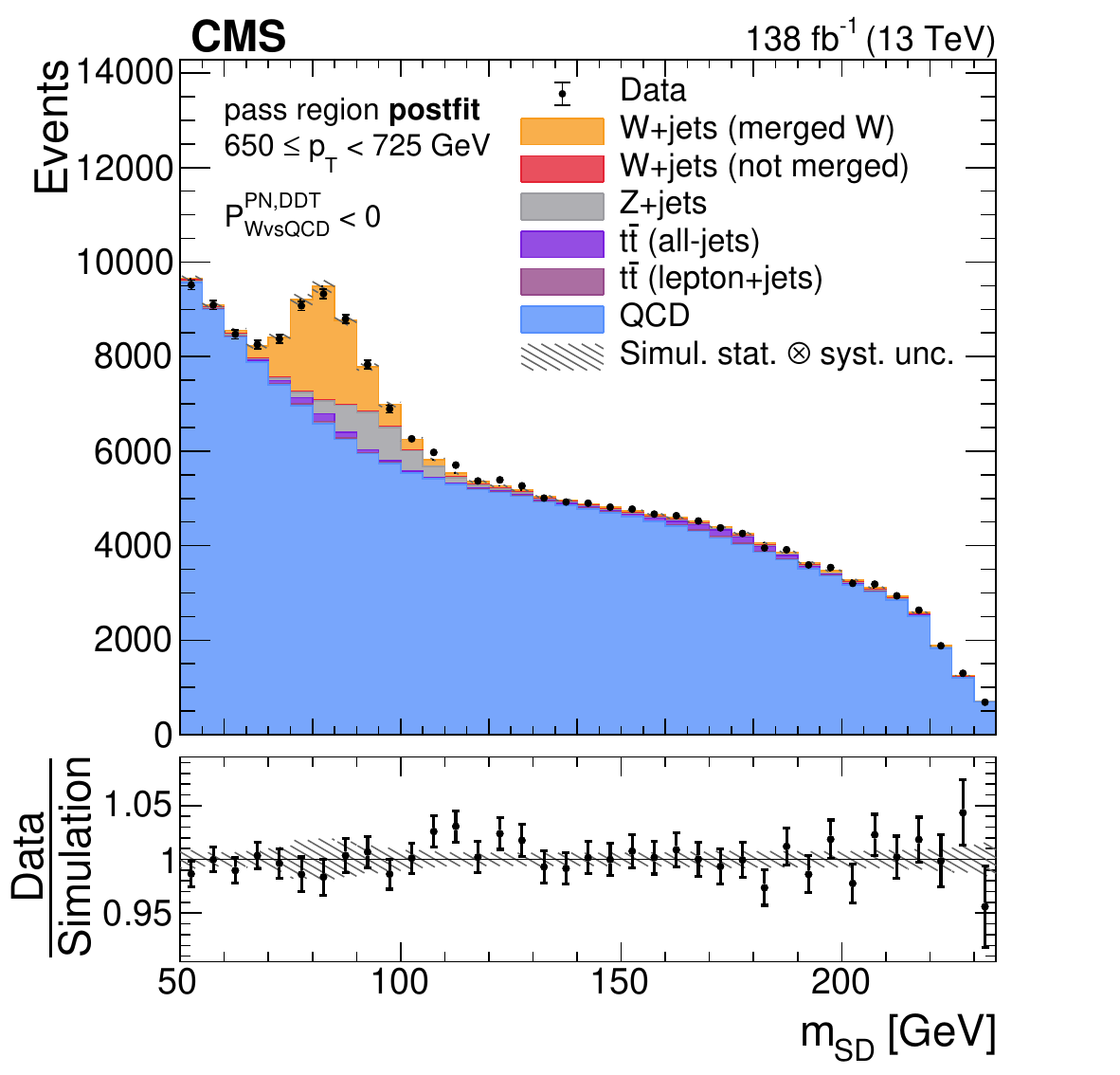}
        \includegraphics[width=0.49\textwidth]{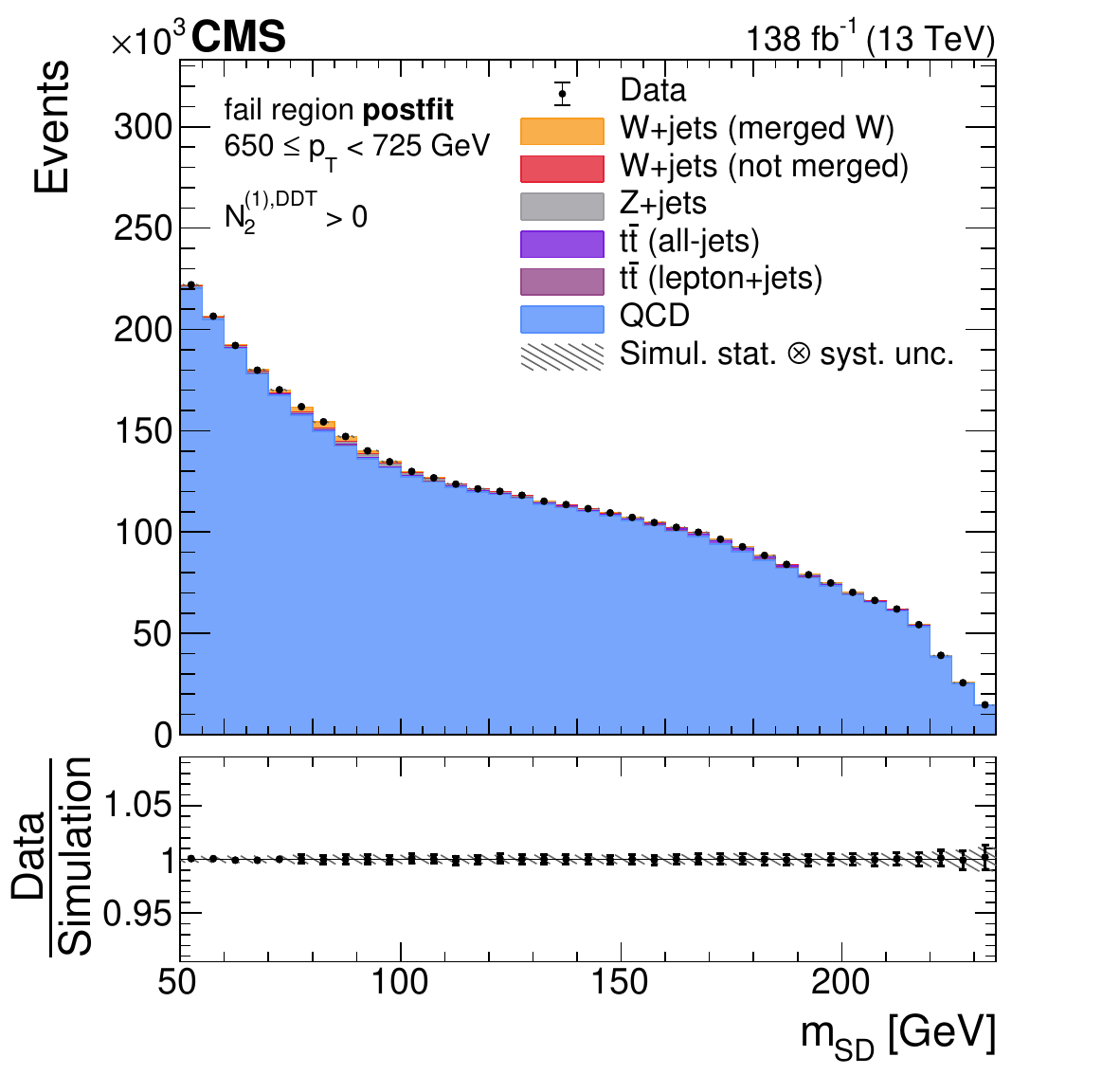}
        \includegraphics[width=0.49\textwidth]{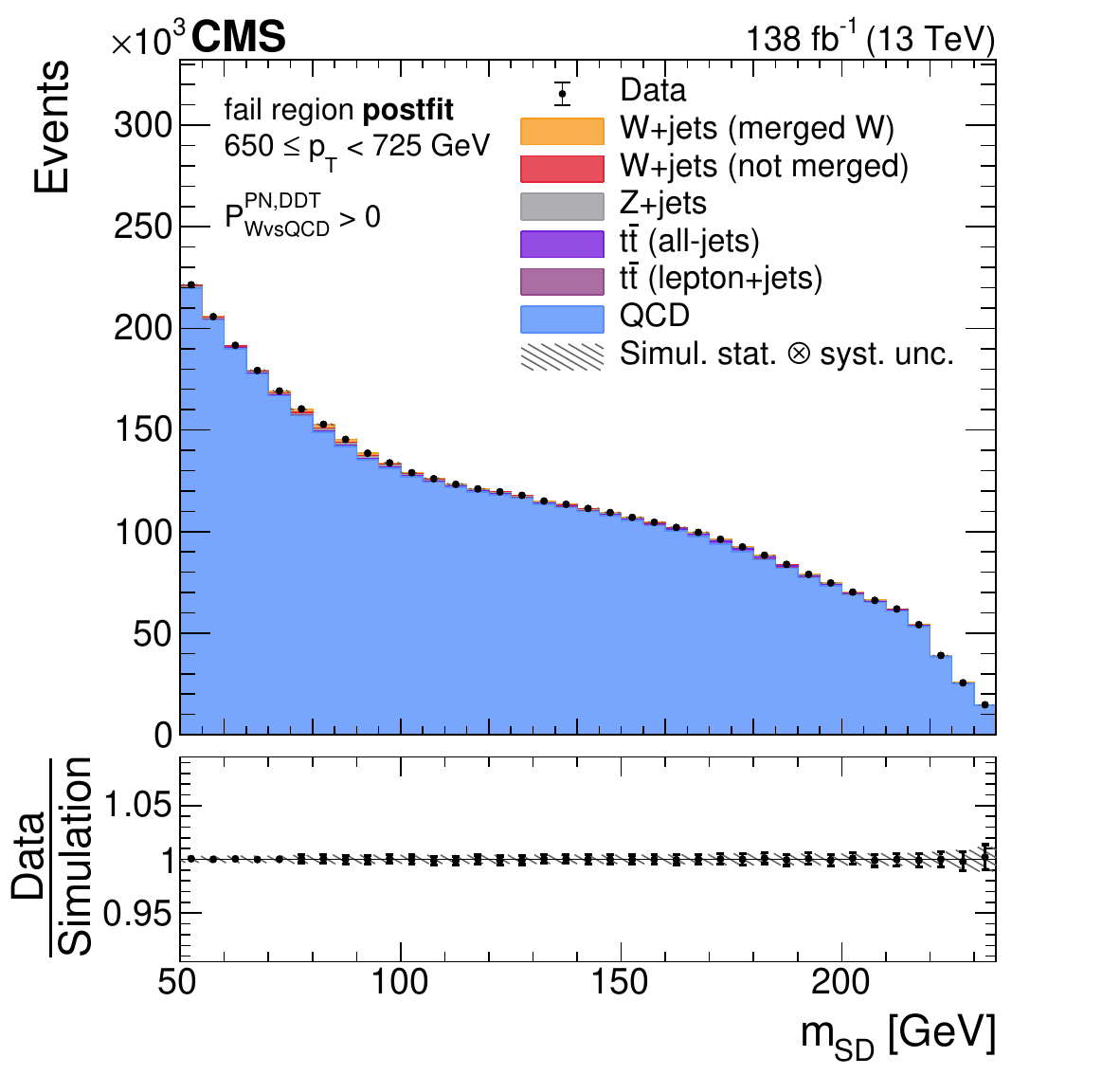}
    \caption{Reconstructed \msd distributions in the second \pt bin with $650 < \pt < 725\GeV$ 
    in the signal (upper row) and control (lower row) regions defined using the \NTwoDDT (\cmsLeft) and 
    \pNetWMDDDT (\cmsRight) taggers after the background estimation and a fit to the data, explained in Section~\ref{sec:background}. All four data-taking periods are 
    combined, resulting in a total integrated luminosity of 138\fbinv.
    The lower panels show the data-to-simulation ratio. The error bars correspond to the statistical uncertainty in the data. The dashed band is the total uncertainty on the background after a fit to the data distribution. \label{img:postfit_demo}}
\end{figure}
In Fig.~\ref{img:postfit_demo}, examples of the reconstructed \msd distributions in signal and control 
regions are shown for the second \pt bin after the background estimation and a fit to the data distribution, discussed in Section~\ref{sec:background}.
The full set of reconstructed \msd distributions can be found in Appendix~\ref{sec:appendix}, Fig~\ref{img:reco_msd_1}--\ref{img:reco_msd_3}.
The dominant background in both SR and CR is from QCD multijet production 
that follows a smoothly falling distribution.
The signal \wqqjets events and backgrounds from \zqqjets and \ttbar exhibit peaks at the respective 
\PW, \PZ, and \PQt quark masses in the SR, while they contribute negligibly to the CR.
The peaks are more pronounced in the SR obtained with the \pNetWMDDDT tagger 
compared to the \NTwoDDT tagger, due to the higher efficiency.
The detector-level jets from the \wqqjets simulation are matched to particle-level jets. 
The corresponding distributions are labeled ``merged \PW'', while 
jets not matched to the generated \PW boson are labeled ``not merged''. 

For the statistical analysis of the data and estimation of the background, 
a maximum likelihood template fit is performed using histograms of \msd in the SR and CR.
The construction of the likelihood functions and the minimization of the negative log-likelihood 
is performed using the \textsc{Combine} toolkit~\cite{CAT-23-001}.

\section{Background estimation}\label{sec:background}

After the event selection, the data sample is dominated by QCD multijet events, which are difficult to model in simulation.
Therefore, an approach based on control samples of observed events is used to estimate the QCD multijet background, while the remaining processes are estimated from simulation. 
This approach is based on an established
method as used in previous 
CMS analyses~\cite{CMS:2019emo,CMS:2022fxs}. The method is based on the assumption that the shape of 
the \msd distribution in QCD multijet events is similar in the pass and fail regions of a given substructure tagger that has been decorrelated with \msd.
Consequently, one can estimate the shape of the QCD background from the signal-depleted fail region from data, 
and use a transfer factor to extrapolate the shape to the signal-enriched pass region. 
Ideally, this transfer factor is a constant, equal to the selection efficiency $\varepsilon_{\text{QCD}}$ 
measured in a QCD multijet sample. If the data were perfectly described by simulation, 
the efficiency $\varepsilon_{\text{QCD}}$ could be obtained from simulated QCD multijet events since 
a pure background sample is impossible to obtain from data. 
However, the jet tagging efficiency is not perfectly modeled in simulation and is not a constant, 
but depends on the jet kinematics, primarily on the jet \pt. 
Additionally, the jet tagging variables are not entirely decorrelated with \msd, despite efforts to 
explicitly decorrelate them as described above.
We address this by using a two-dimensional transfer function that depends on \pt and \rrho of the jet, 
\begin{equation}
    \label{eq:transfer_factor}
        \mathrm{TF}(\ptscaled, \rrhoscaled) = \varepsilon_{QCD}(\ptscaled) r(\ptscaled, \rrhoscaled), \\
\end{equation}
where \ptscaled and \rrhoscaled are functions of \msd and \pt and correspond to the normalized observables \pt and \rrho, scaled to lie in the interval $[0,1]$.
The efficiency $\varepsilon_{QCD}(\ptscaled)$ is obtained from simulated QCD 
multijet events, separately for each data-taking period. The residual function $r(\ptscaled, \rrhoscaled)$ 
is determined in the fit to data and used to model differences between data and simulation in $\varepsilon_{\text{QCD}}(\ptscaled)$. 

As a basis for the two-dimensional residual function, Bernstein polynomials
are used.
The $i$-th Bernstein basis polynomial of $n$-th degree with $i \in [0,\cdots,n]$ is defined for $x \in [0,1]$ as
\begin{equation}
    b_{i}^{n} (x) = \mybinom{n}{i} x^{i} (1-x)^{n-i}.
\end{equation}
The Bernstein basis is chosen over a standard polynomial because with the variable $x$ bounded between 0 and 1 it is more stable numerically and the function is nonnegative~\cite{CMS:2022fxs}.
The residual function is a linear combination of Bernstein polynomials
\begin{equation}
    r(\ptscaled, \rrhoscaled) = \sum_{i}^{\npt} \sum_{j}^{\nrho} a_{ij} \bpt \brho,
\end{equation}
where \npt and \nrho are the orders of the Bernstein polynomials in \ptscaled and \rrhoscaled, respectively.
The coefficients $a_{ij}$ are free parameters in the fit to data. 

\begin{figure}[tb]
    \centering
        \includegraphics[width= 0.65\textwidth]{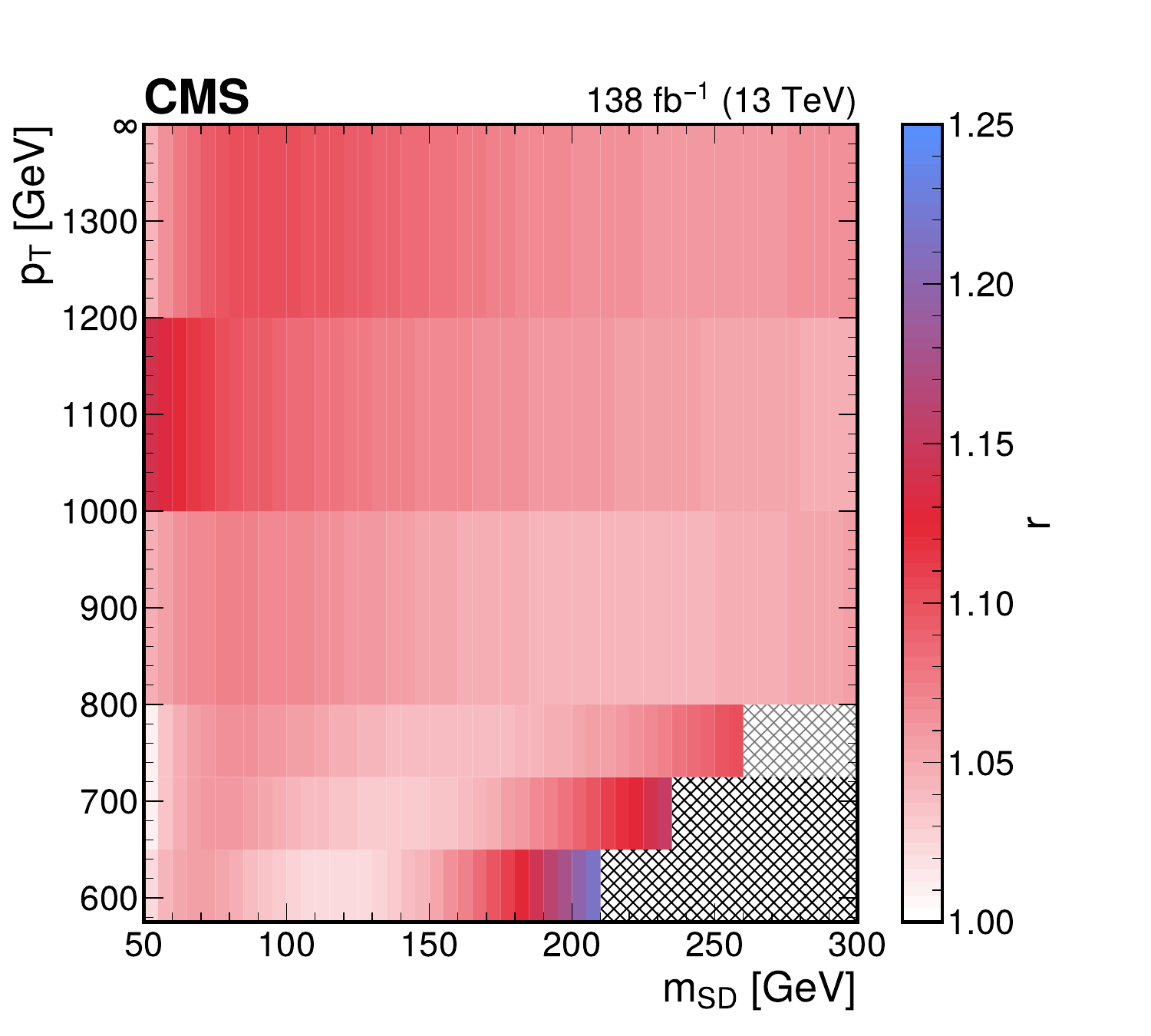}
    \caption{Residual function $r(\ptscaled, \rrhoscaled)$ obtained from a fit to data, when using the \pNetWMDDDT as jet tagger. The arguments of the function $r$, \ptscaled and \rrhoscaled are functions of \msd and \pt and correspond to the normalized observables \pt and \rrho, scaled to lie in the interval $[0,1]$.
    The hatched area is excluded from the analyses by selecting $\rrho < -2.1$.\label{img:postfit_bernstein_demo}}
\end{figure}

An example of the residual function after a fit to data is shown in Fig.~\ref{img:postfit_bernstein_demo}. 
To determine the optimal order of the Bernstein polynomials \nrho and \npt, the goodness of fit using the 
saturated model approach is evaluated for a series of orders. The resulting test statistics are then 
compared in F-tests and the orders of the polynomials are chosen to range from 1--2 for \npt and 
2--6 for \nrho depending on the data-taking period.

\section{Systematic uncertainties}\label{sec:uncertainties}

The measurement of the differential \wqqjets cross section is subject to multiple sources of systematic uncertainties.
These affect the \msd in different ways, causing differences in shape, normalization, or both.
Several such sources of systematic uncertainties are accounted for in the fit as nuisance parameters in the likelihood function.

Uncertainties related to the QCD background estimate  are accounted for through the unconstrained parameters $a_{ij}$ of the transfer function.
Alternative functional forms for the transfer function (Chebyshev polynomials, exponentials of Bernstein polynomials) have been found to have a significantly smaller impact on the result than the accounted uncertainties on the parameters $a_{ij}$~\cite{CMS:2022fxs}.

Uncertainties related to the jet energy scale calibration affect both the normalization through migrations of events 
between \pt bins, and the \msd shape through variations in the subjet energy.
The uncertainty in the measurement of the total luminosity of 138\fbinv is estimated to be 1.6\%~\cite{CMS-LUM-17-003,CMS-PAS-LUM-17-004,CMS-PAS-LUM-18-002}. This uncertainty is accounted for in the fits as normalization nuisance parameters for signal and background processes.
The uncertainty in the trigger scale factors originates from the statistical uncertainties in the trigger 
efficiency measurement. Variations in the trigger scale factors are propagated to the \pt and \msd and 
used as a shape uncertainty in the fit.

The number of interactions per bunch crossing is estimated from the measurements of the luminosity and the inclusive 
inelastic cross section of \pp collisions. The simulation is reweighted to match the distribution of the number 
of interactions per bunch crossing in data. The inclusive inelastic cross section is varied up and down by 4.6\%~\cite{CMS:2018mlc} 
and the reweighting is repeated. The uncertainty corresponds to the effect of these variations and is propagated to 
the \pt and \msd. It is treated as a shape uncertainty for all simulated events.

Differences in jet tagging efficiencies between data and simulation in \wqqjets signal events are accounted for by 
an unconstrained nuisance parameter $\theta$ in the fit. This parameter allows for anticorrelated changes of the 
normalization in the pass and fail regions while conserving the total number of events,
\begin{equation}
\begin{aligned}
    N_\text{total} &= N_\text{pass} + N_\text{fail} = \hat{N}_\text{pass} + \hat{N}_\text{fail} \\ 
    &= \hat{\theta} N_\text{pass} + \left[ (1-\hat{\theta}) \frac{N_\text{pass}}{N_\text{fail}} +1 \right] N_\text{fail}, 
\end{aligned}
\end{equation}
where $N_\text{total}$, $\hat{N}_\text{pass}$, and $\hat{N}_\text{fail}$ are the total number of events 
and the best fit values of the number of events in the pass and fail regions, respectively. 
The parameter $\hat{\theta}$ denotes the best fit value of $\theta$, while $N_\text{pass}$ and $N_\text{fail}$ 
are the initial number of events in the pass and fail regions, respectively. 

The uncertainty in the modeling of initial-state radiation (ISR) and final-state radiation (FSR) is estimated by varying the respective parton shower renormalization scale up and down by a factor of two,
and storing corresponding weights during the event generation.
The effect is treated as a shape uncertainty for signal and background events by propagating the variations to the \pt and \msd through a reweighting of simulated events.
To account for uncertainties in the hadronization model of the \wqqjets signal process, we assign a shape 
uncertainty equal to the difference between the central simulation sample (\MGvATNLO{}+\PYTHIA) and an 
alternative generator (\MGvATNLO+\HERWIG), which implements another hadronization model.
This uncertainty mainly affects the tails of the \PW boson mass distribution, where nonperturbative effects dominate.

The higher-order QCD and EW correction systematic uncertainties for the \wqqjets and \zqqjets simulation is estimated from 
the envelope of the scale variations for the NLO QCD and NLO EW predictions derived in Ref.~\cite{Lindert:2017olm}. 
The QCD and EW correction uncertainties are treated as separate shape uncertainties in the fit.
Uncertainties related to parton distribution functions have been estimated to be a factor ${\geq}3$ smaller than those in the considered \pt range~\cite{Lindert:2017olm} and neglected.

To account for uncertainties in the prediction of the \ttbar cross section and kinematics, 
a shape uncertainty is introduced. The up and down variations are constructed by comparing the \pt and \msd distributions 
with and without applying the top quark \pt reweighting.
In addition, normalization uncertainties are included as nuisance parameters with 
log-normal constraints amounting to 20\% for \zqqjets~\cite{Rubin:2010xp}.

\begin{figure}[tb]
    \centering
        \includegraphics[width=0.49\textwidth]{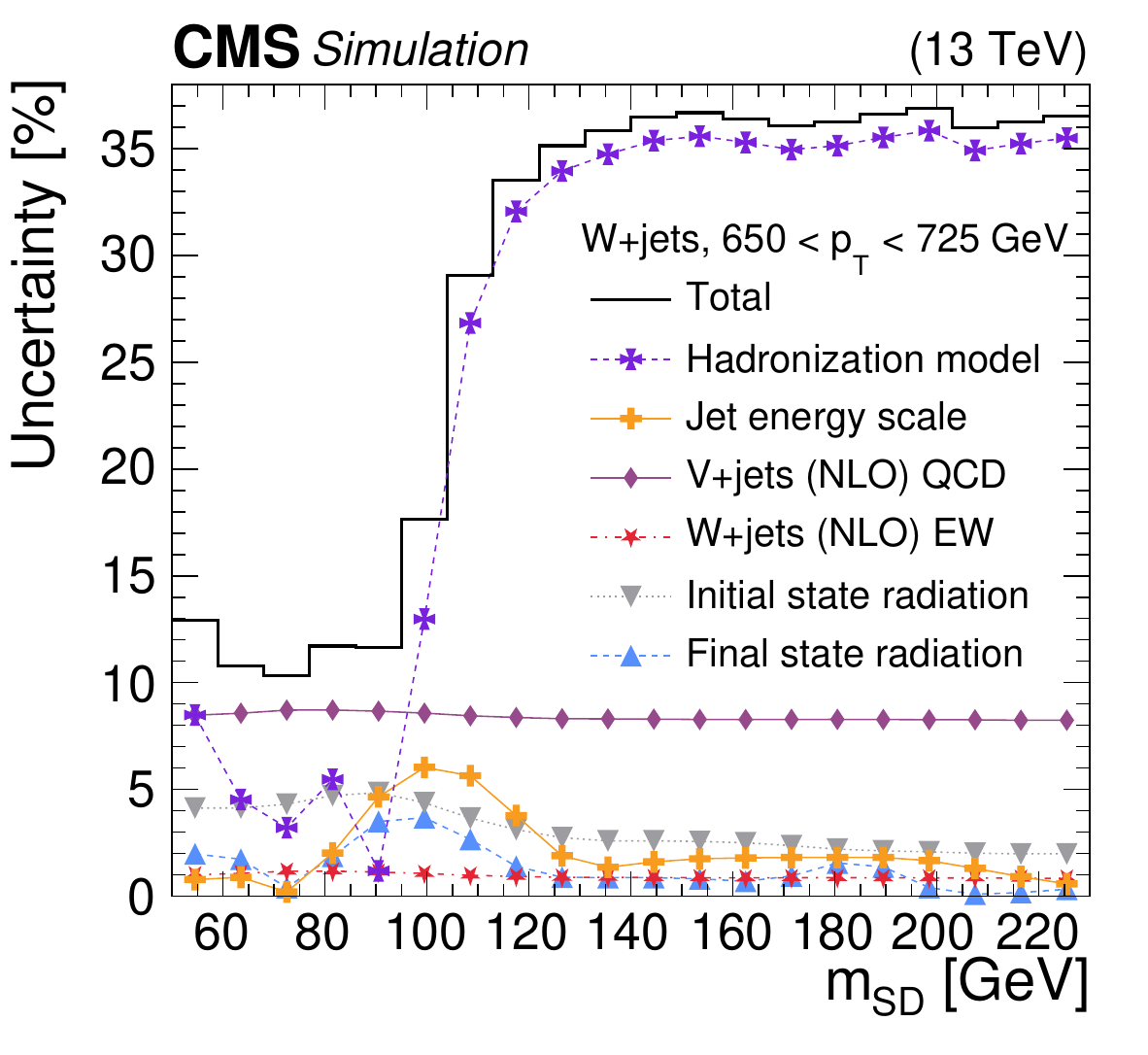}
    \caption{Summary of the effect of the systematic uncertainties in the reconstructed SD jet mass in the \wqqjets signal sample in a representative \pt bin.
    The dominant shape effects can be attributed to uncertainties in the hadronization model, jet energy scale, and final-state shower.
    The jet energy scale and final state shower mainly affect the region of the \PW boson mass peak ($70<\msd<110\GeV$). 
     \label{img:systematics_summary}}
\end{figure}

In Fig.~\ref{img:systematics_summary}, we summarize the effect of the systematic uncertainties 
in the reconstructed \msd.
The dominant shape effects can be attributed to uncertainties in the hadronization model, jet energy scale, and final-state shower.
The jet energy scale and final-state shower mainly affect the region of the \PW boson mass peak ($70<\msdreco<110\GeV$). 
These lead to a shift of events from the peak region on top of a falling background, thus affecting most strongly 
the number of events at the right-hand side of the \PW boson mass peak ($90<\msdreco<110\GeV$).
The size of the hadronization model uncertainty is smallest at the \PW boson mass peak and mainly affects the tails 
of the \PW boson mass distribution ($\msdreco<70\GeV$ and $\msdreco>90\GeV$) where light-flavor quark and gluon 
jets dominate~\cite{CMS:2018vzn}.

\section{Unfolding}\label{sec:unfolding}

To obtain a measurement at the particle-level, detector effects are corrected for by unfolding the measured detector-level distribution.
In the maximum-likelihood template fit with constraints for regularization, background subtraction and inversion of the detector response matrix $R_{ij}$ are performed simultaneously in a global fit to the recorded collision data.
This approach as detailed in the following is similar to the one outlined in Ref.~\cite{S_Schmitt_2012}, though a maximum likelihood fit instead of a least-squares fit is used.

\begin{figure}[tb]
    \centering
    \includegraphics[width=0.65\textwidth]{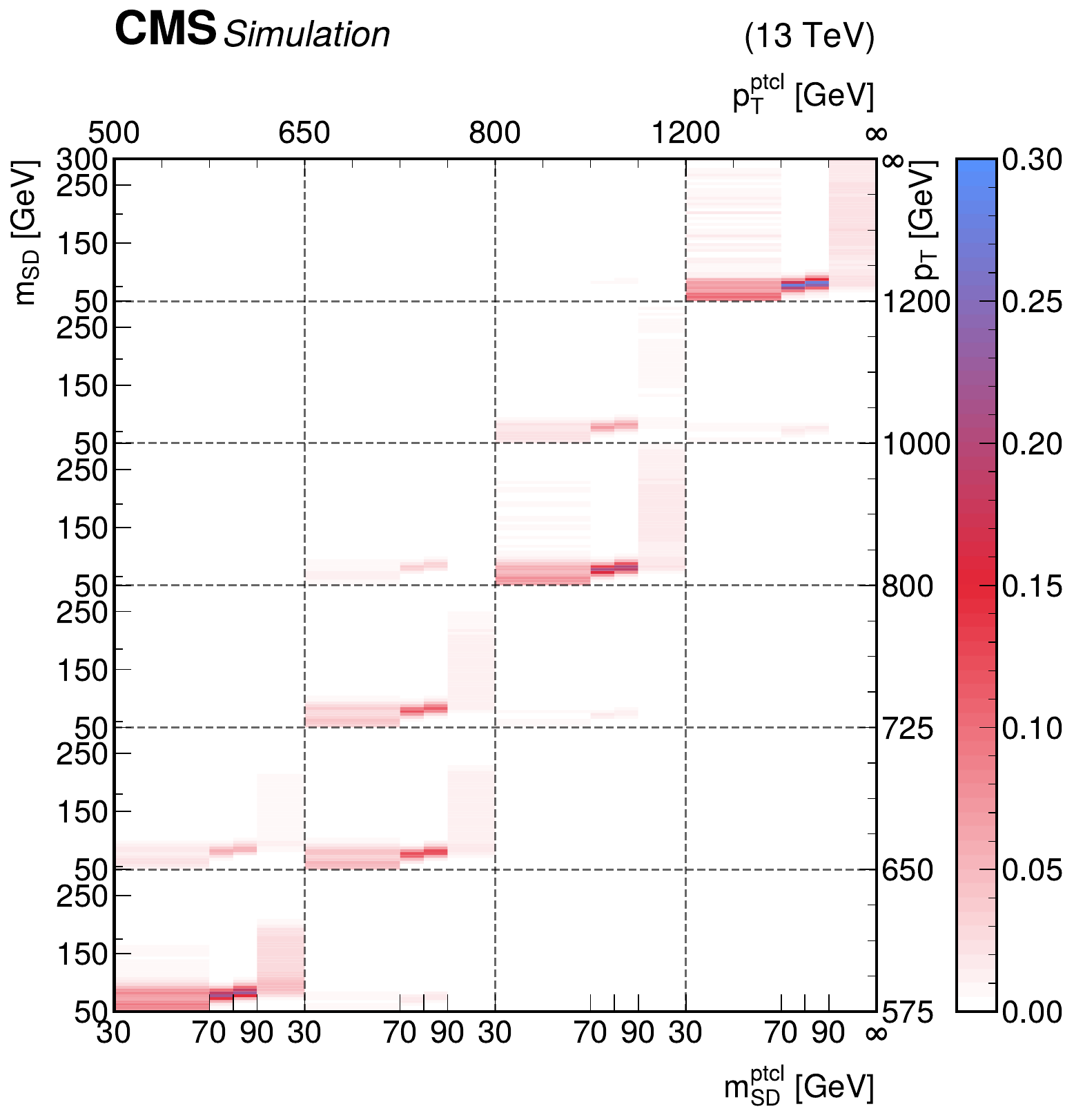}
    \caption{Response matrix obtained for selected events in simulation. The matrix is obtained from a sum 
    of all data-taking eras.
    The grey dashed lines separate the individual \pt bins.
    The binning within each \pt bin corresponds to the \msd binning.
\label{img:response_matrix_final}}
\end{figure}

The signal strength modifiers $\mu_j$ are the ratios of the number of events predicted by the 
fit to data to the expected number of events at the particle level in simulation. The total number of 
expected events in bin $i$ at the detector level is given by 
\begin{equation}
    y_i = \sum_{j=1}^M R_{ij} \mu_j x_j + b_i.
\end{equation}
Here, the expectation value of the true number of events in bin $j$ at the particle level is given by $x_j$ 
and $b_i$ denotes the total number of expected background events in bin $i$ at the detector level. 
The response matrix $R_{ij}$ is obtained from the simulation. It is constructed by filling events in bin $i$ 
at the detector level, given that they were generated in bin $j$ at the particle level. 
Events are only considered in the response matrix if the selected jets at the detector and particle level 
can be matched to the generated \PW boson. 
Acceptance and efficiency effects originating from events falling out of the fiducial phase space 
by failing either the detector- or particle-level selections, or the matching criterion, 
are not included in the response matrix. Instead, these are accounted for in a factorized approach. 
Events that pass the detector-level selection, but do not have a matching jet at the particle level or are not matched to a generated \PW boson are treated as background in the fit.
Events that fail the detector-level selection or the matching criterion but pass the particle-level selection are accounted 
for with an acceptance correction. The resulting response matrix is shown in Fig.~\ref{img:response_matrix_final}.

The best fit values of the parameters of interest $\vec{\mu}$ are obtained by minimizing the negative log-likelihood function
\begin{equation}
    - 2 \ln \mathcal{L}(\vec{\mu}) = - 2 \sum_i^N \ln P(n_i, y_i),\label{eq:likelihood_part1}
\end{equation}
where $n_i$ is the total number of events in data in bin $i$ at the detector level and $P(n_i, y_i)$ 
refers to the Poisson distribution.

We include all bins in \msd and \pt in the fit, where each bin is split into pass and fail regions 
from the jet substructure tagger requirements.  
The fit is performed in all four eras of data-taking simultaneously by using the product of the individual 
likelihoods with one common set of parameters of interest. 
We measure one signal strength modifier $\mu_j$ per particle-level bin $j$, which scales the prediction in 
each particle-level bin to obtain the measured cross sections.

The regularization of the multidimensional distribution is performed using Tikhonov regularization~\cite{Tikhonov1963}, adding a penalty term with a strength parameter $\tau$ to the negative log-likelihood function.
We choose a regularization strength parameter $\tau=0.59$ by minimizing an average global correlation coefficient which was found to vary only mildly (by 0.5\%) over a wide range ($0.1<\tau<2$) of the regularization strength parameter.
Regularization is necessary to obtain accurate measurements in the lowest and highest particle-level \msd bins, where acceptance is low and backgrounds are more dominant.

To estimate the bias and over- or under-coverage introduced by the regularization, 
a series of tests is performed using pseudo-data in the fit instead of collision data. 
Replacing data with signal simulation yields perfect closure with the expectation. 
To test the model against statistical fluctuations in the pseudo-data, 
the pseudo-data are perturbed by generating 500 toy data sets. We then compute  
the pull distribution of the maximum likelihood estimators $\hat{\mu}_j$ of the signal strength modifiers.
The mean of the distribution of pulls is consistent with zero within one standard deviation.
To further test for biases introduced by the regularization, the same tests are performed 
using statistically independent pseudo-data, yielding the same conclusions.

To better understand the dominant sources of uncertainty in representative \ptgen and \msdgen bins, 
the impacts of key groups of systematic uncertainties in the signal strength modifiers $\mu_j$ are summarized in 
Table~\ref{tab:impact_summary_particlenet} for the \pNetWMDDDT tagger. 
The largest contributions to the total uncertainty near the \mw peak (second column) are the \PW tagging efficiency, the \vjets QCD NLO corrections, the \zjets normalization, FSR and hadronization model, reaching up to 14.4\%. 
Outside the \mw peak region (third column), the jet energy scale and the QCD multijet estimate also contribute up to 23.9\%.
{\def\arraystretch{1.3}%
\begin{table}[tb]
    \centering
    \topcaption{Summary of systematic uncertainties and their impact on the measurement in the $650<\ptgen<800\GeV$ bin and two representative \msdgen bins, 
    using the \pNetWMDDDT tagger.
    \label{tab:impact_summary_particlenet}}
    \begin{tabular}{lcc}
        \multirow{2}{*}{{Uncertainty}} & \multicolumn{2}{c}{{Impact in $\msdgen$ bin}}\\
        & $80<\msdgen<90\GeV$  & $\msdgen>90\GeV$ \\
        \hline
        Jet energy scale &  4.8 \% & 23.9 \% \\
        \PW tagging efficiency & 11.9 \% & 14.3 \% \\
        Pileup reweighting & 1.1 \% & 3.1 \% \\
        Trigger scale factor & 0.7 \% & 1.1 \% \\
        Integrated luminosity & 1.6 \% & 1.3 \% \\
        QCD multijet estimate  & 3.6 \% & 13.9 \% \\
        \zjets normalization &  6.6 \% & 3.0 \% \\
        Top quark \pt reweight &  0.3 \% & 0.9 \% \\
        \vjets QCD NLO & 14.4 \% & 17.7 \% \\
        \wjets EW NLO & 1.3 \% & 0.7 \% \\
        \zjets EW NLO & ${<}0.1$ \% & 0.2 \% \\
        FSR &  12.2 \% & 17.4 \% \\
        ISR &  3.7 \% & 4.9 \% \\
        Hadronization model & 12.4 \% & 15.5 \% \\
    \end{tabular}
\end{table}%
}

\section{Results}\label{sec:results}

The measurement of the fiducial cross section is double differential in \msdgen and \ptgen and uses the \pNetWMDDDT approach.
Tabulated results are provided in the HEPData record for this analysis~\cite{hepdata}.
 \begin{figure}[tb]
    \centering
        \includegraphics[width= 0.49\textwidth]{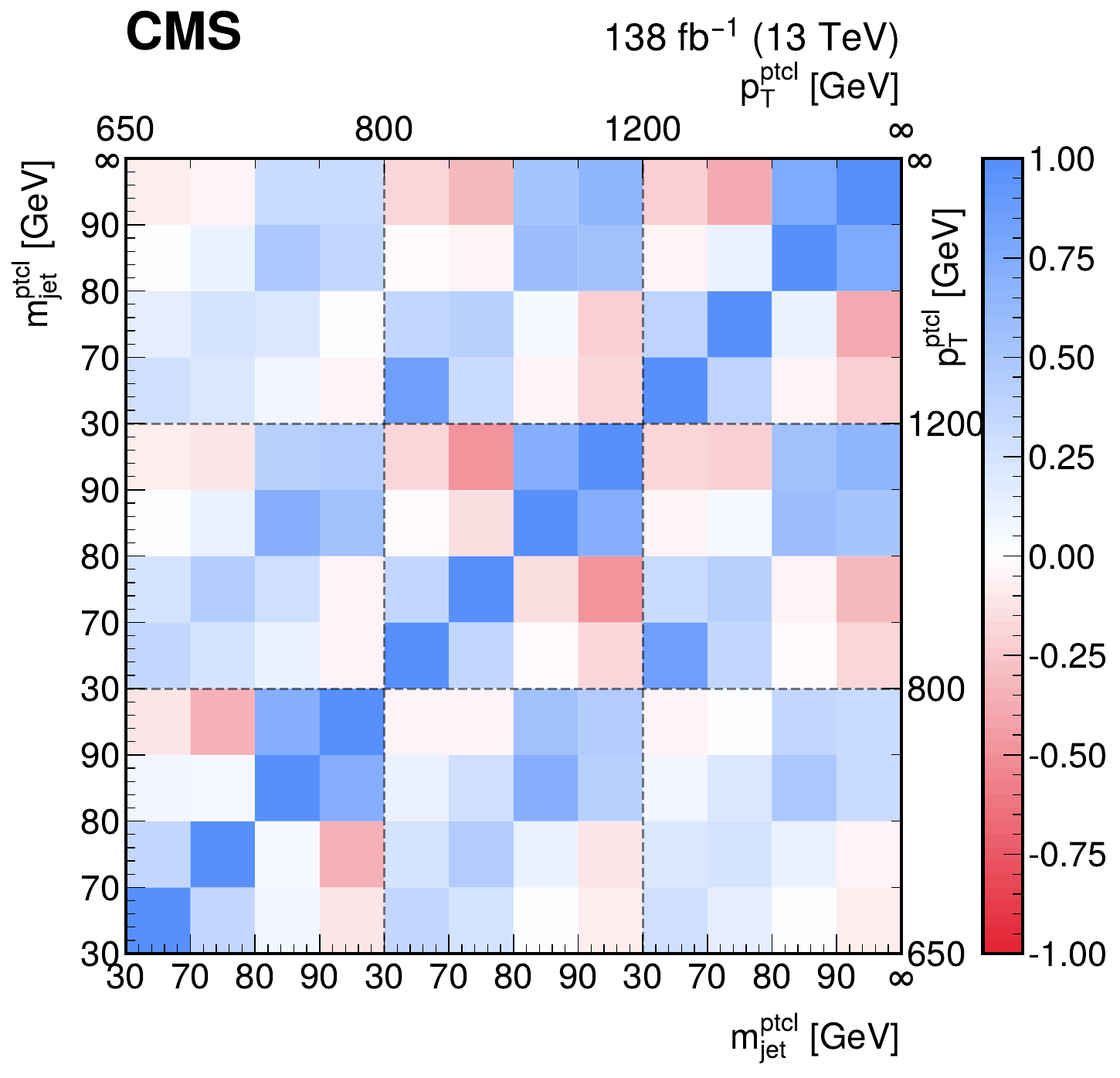}
    \caption{Correlation matrix of the maximum likelihood estimators of the signal strength modifiers \POI{}{} with inclusion of an $\NTwo<0.2$ selection in the particle-level definition. The plot shows the matrix from the fit to the data using \pNetWMDDDT.
    The grey dashed lines indicate the individual \ptgen bins.
    The binning within each \ptgen vs. \ptgen bin corresponds to the \msdgen vs. \msdgen binning.
    \label{img:correlation_matrix}}
\end{figure}
The correlation matrix in Fig.~\ref{img:correlation_matrix} shows the correlation coefficients 
of the individual particle-level bins.
The \msdgen bins in the different \ptgen bins are shown unrolled, where the grey dashed lines indicate the \ptgen 
bins boundaries. 
The highest correlations are found to be among neighboring bins, both along the \msdgen and \ptgen axes.
The latter ones are similar for all \msdgen bins, but smaller on average than the correlations between neighboring 
bins along the \msdgen axis.
The highest anticorrelations are found between the first two and last two \msdgen bins in the third and fourth \ptgen bins.

\begin{figure}[tbp]
    \centering
        \includegraphics[width= 0.47\textwidth]{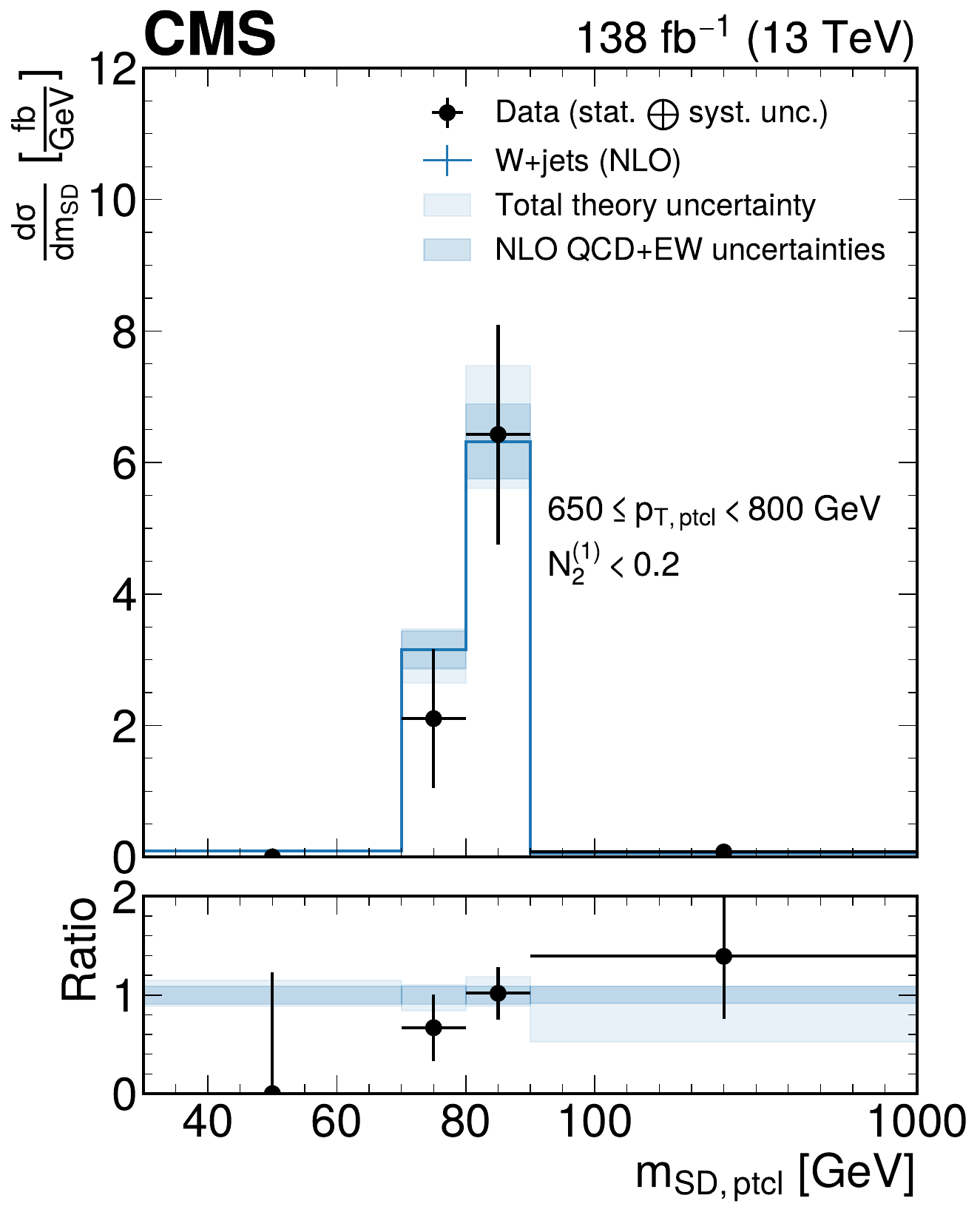}
        \includegraphics[width= 0.47\textwidth]{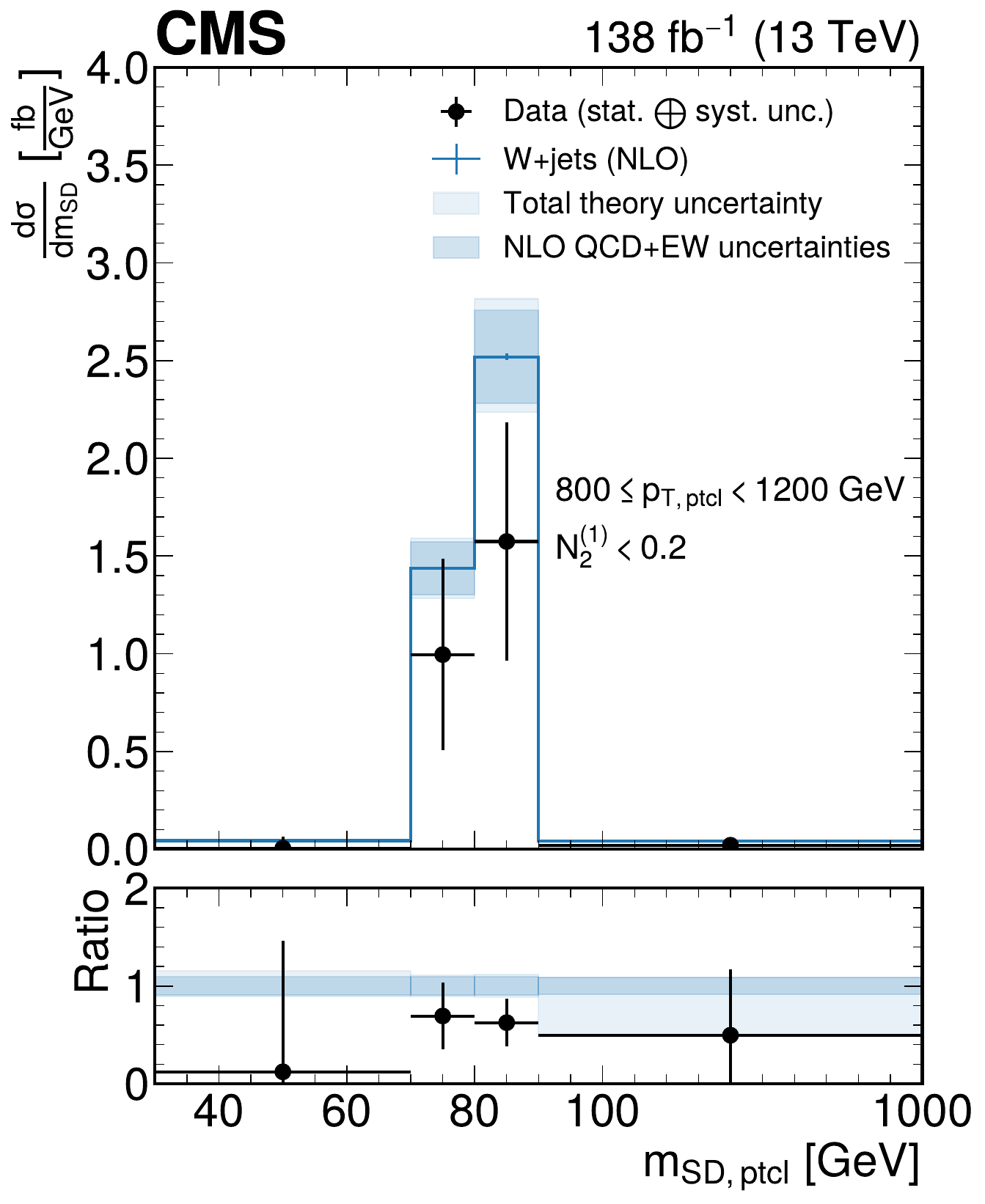}
        \includegraphics[width= 0.47\textwidth]{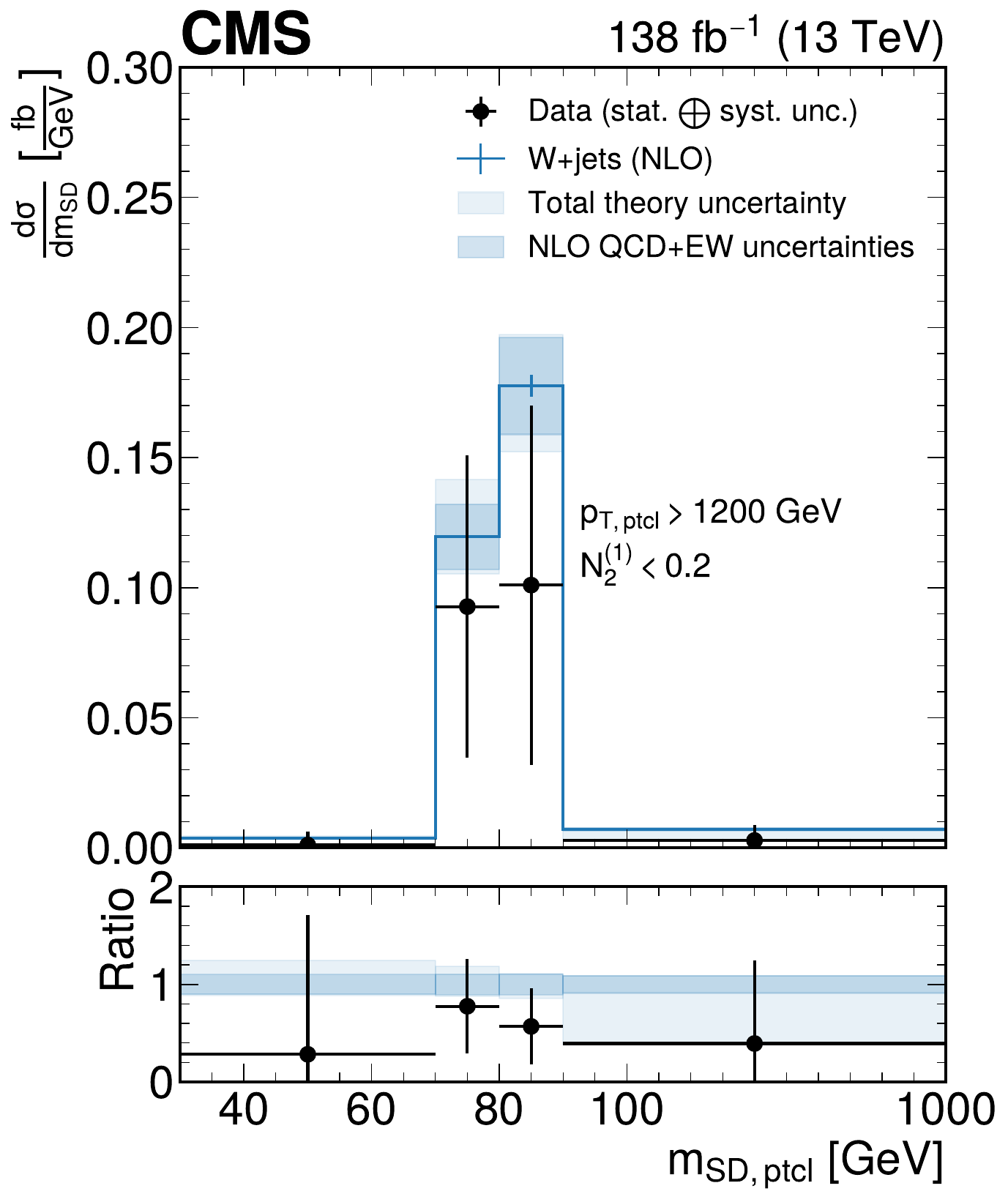}
    \caption{Unfolded and background subtracted jet mass distribution at the particle level for each \ptgen bin obtained with \pNetWMDDDT. 
    The fiducial measurement region at the particle level includes the $\NTwo<0.2$ selection.
    The unfolded data are shown as black markers with the total uncertainty indicated as error bar.
    The blue lines show the predictions 
    from \MGvATNLO{}+\PYTHIA at LO with MLM matching and supplemented by NLO QCD and EW corrections
    with the statistical uncertainty indicated as error bar.
    The theory uncertainty is the sum in quadrature of parton shower variations and the hadronization model uncertainty, 
    as well as the uncertainties in the QCD and EW corrections, and is drawn as a light shaded blue band.
    The purely perturbative uncertainties are overlaid as a dark, shaded blue band.
\label{img:unfolding_ptbinsN2}}
\end{figure}

The extracted jet mass distributions are shown in Fig.~\ref{img:unfolding_ptbinsN2}.
For the purpose of display the differential fiducial cross section $\text{d}\sigma/\text{d}\msd$ is shown in bins of \ptgen.
The unfolded data are shown as black markers with the total uncertainty (the square roots of the diagonal elements of the covariance matrix containing statistical $\oplus$ systematic uncertainty) drawn as black bars.
The \vjets theory prediction from \MGvATNLO{}+\PYTHIA has LO+MLM accuracy with NLO (QCD and EW) 
corrections and is depicted as a blue curve.
The corresponding theory uncertainty is the sum in quadrature of parton shower variations 
and hadronization model uncertainty, as well as the two envelopes of systematic uncertainties 
characterizing missing higher-order QCD and EW corrections~\cite{Lindert:2017olm}, as detailed in the previous section, and is drawn as a light shaded blue band.
The purely perturbative uncertainties are also overlaid as a dark, shaded blue band.

We quantify the agreement of the data with the prediction using a goodness-of-fit test based on the saturated model~\cite{Cousins2013} and observe p-values larger than 0.13 for all four data-taking periods.
The unfolded data distributions agree with the predictions within the total uncertainty in all \ptgen bins.
The data in the first \ptgen bin are generally in better agreement with the 
simulation concerning both the normalization as well as the shape of the distributions.
In the last two \ptgen bins, we observe a shift in the peak position in the unfolded data compared to the simulation,
that is, however, covered by the uncertainties and their correlations.

\begin{figure}[tb]
    \centering
        \includegraphics[width= 0.47\textwidth]{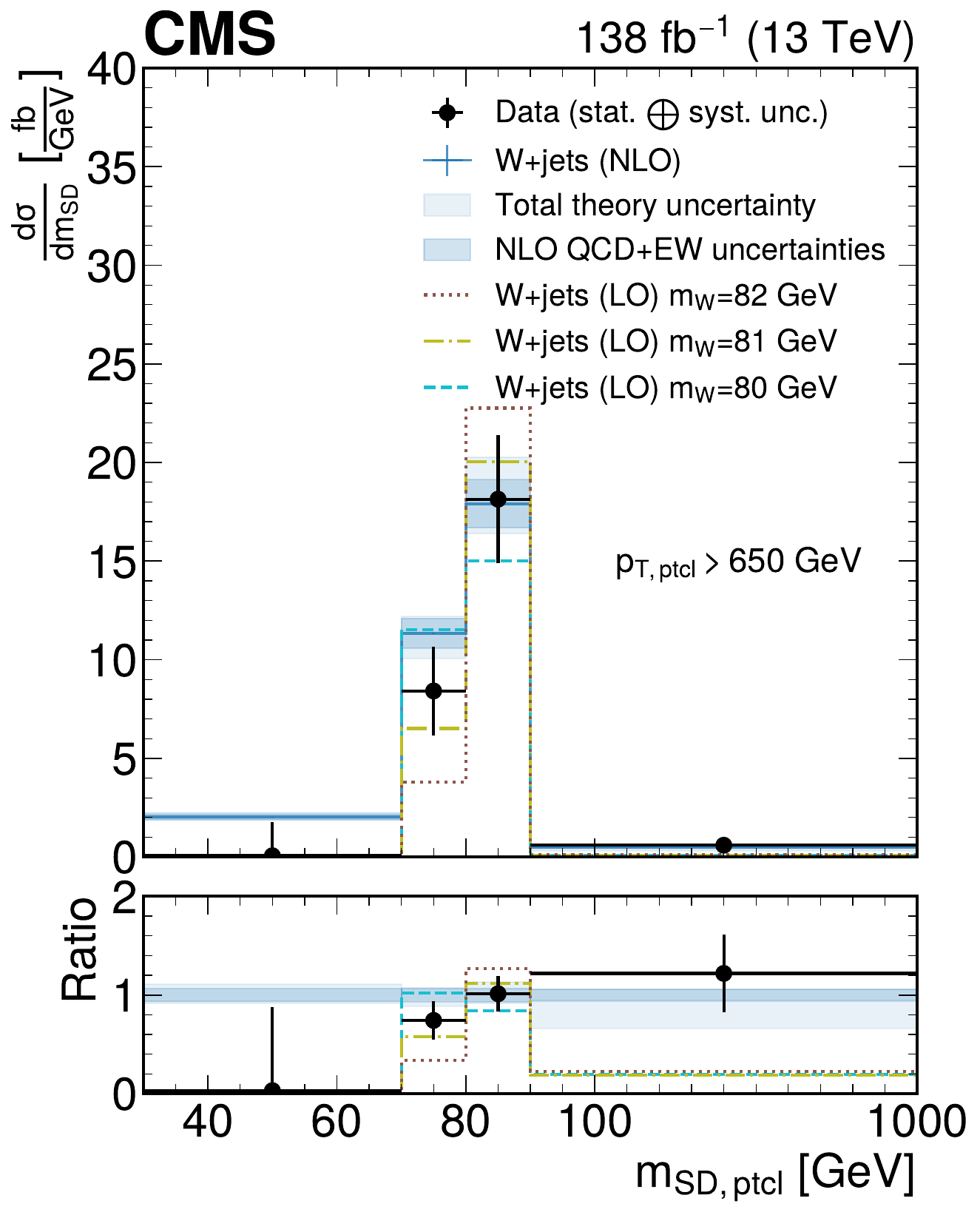}
        \includegraphics[width= 0.47\textwidth]{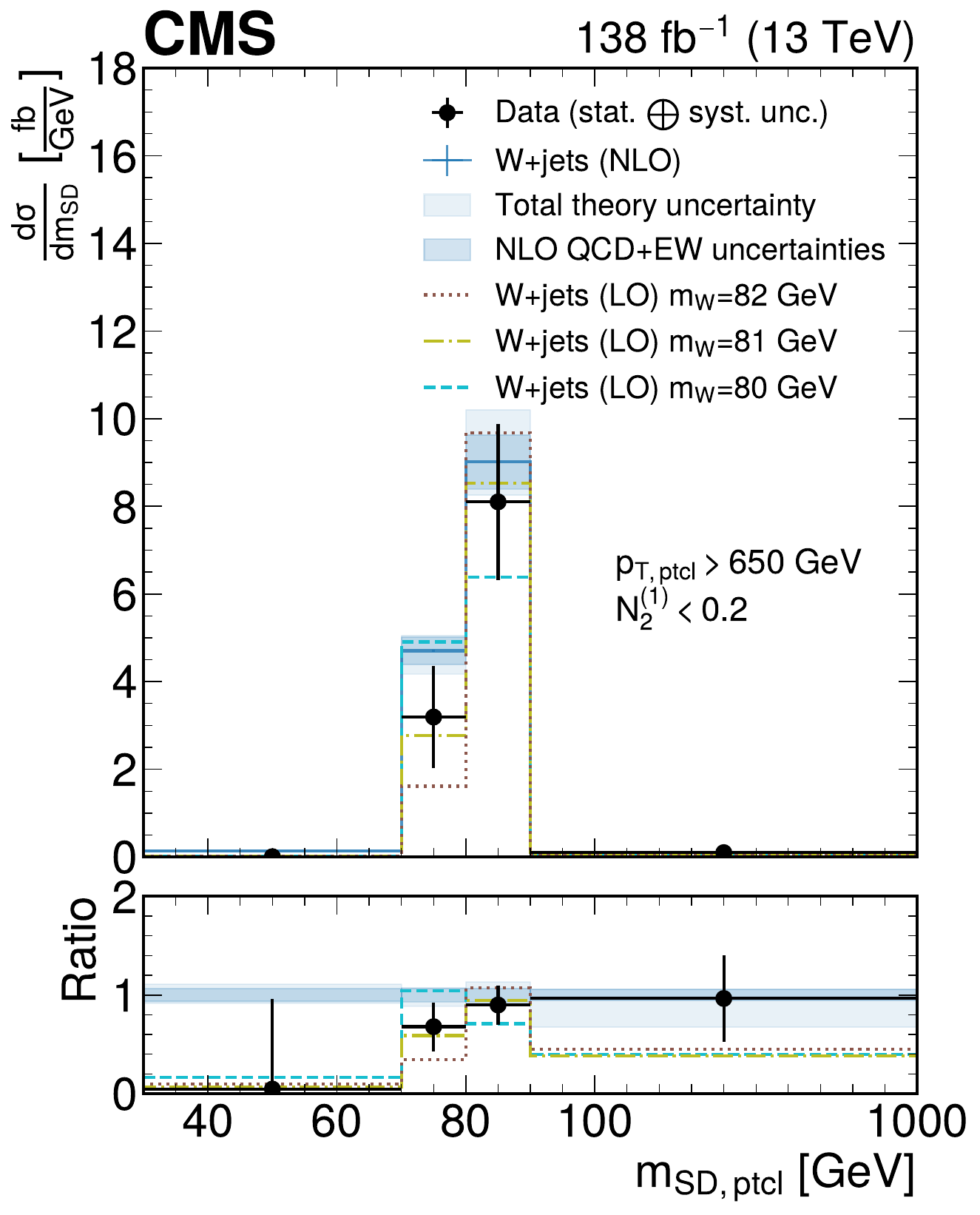}
    \caption{Unfolded and background subtracted jet mass distribution at the particle level for $\ptgen>650\GeV$, obtained by summing all \ptgen bins. 
    The results are obtained using the \pNetWMDDDT tagger without (\cmsLeft) and with (\cmsRight) 
    the $\NTwo<0.2$ selection at the particle level.
    The unfolded data are shown as black markers.
    The blue lines show the predictions from \MGvATNLO{}+\PYTHIA at LO with MLM matching and supplemented by NLO QCD and EW corrections.
    The theory uncertainty is the sum in quadrature of parton shower variations and the hadronization model uncertainty, 
    as well as the uncertainties in the QCD and EW corrections, and is drawn as a light shaded blue band.
    The purely perturbative uncertainties are overlaid as a dark, shaded blue band.
    Predictions with different values of the \PW boson mass generated with \PYTHIA at LO and scaled to match 
    the total number of events in data are overlaid. \label{img:unfolding_sum}}
\end{figure}
The upper and lower ends of the \msdgen spectrum contain events where not the \PW boson jet, 
but the recoiling quark/gluon jet is measured. Thus, in the first \msdgen bin, the predicted yield 
depends on the modeling of nonperturbative effects, which is model dependent.
Figure~\ref{img:unfolding_sum} shows the distributions without and with the $\NTwo<0.2$ selection at the particle level, 
where the individual distributions in each \ptgen bin for $\ptgen>650\GeV$ have been added.
The predictions without the $\NTwo<0.2$ selection give a larger fraction of events in the lowest and highest \msd bins, 
where the signal is contaminated by light-flavor quark and gluon jets.

The measured jet mass distribution is sensitive to the \PW boson mass.
In the following, we quantify the sensitivity of the \msd distribution to the \PW boson mass 
and quote a first measurement of the \PW boson mass in an all-jets final state at a hadron collider.
We compare multiple LO predictions (called templates in the following) generated with \PYTHIA with different 
values of generated \PW boson mass: 79.0, 80.0, 80.385 (\PYTHIA default), 81.0, 82.0\GeV to the unfolded data distributions.
Higher-order corrections to the theory prediction are not accounted for in this estimate, 
as the statistical uncertainty dominates over the theoretical uncertainties. 
In order to minimize the impact of shower and hadronization effects on the result, 
we use cross sections obtained for $\NTwo<0.2$ at the particle level. 
We only use the two bins in \msd that fall into the \PW boson mass peak with $70 < \msd < 90\GeV$, integrating over $\ptgen>650\GeV$, and 
use the normalized cross section, where the normalization is done for the peak region only.
We carry out a fit using the linear template fit package~\cite{Britzger:2021ocj} 
to estimate the \PW boson mass parameter.
We account for the same sources of uncertainties as in the differential cross section measurement.

A closure test is carried out by replacing the data with the \PYTHIA simulation with a certain \PW 
boson mass parameter, and then measuring \mw using the remaining \PYTHIA templates.
We find a linear dependence between the input and measured mass consistent with unity slope.
We also check the consistency of the method by measuring the mass of the \MGvATNLO{}+\PYTHIA 
and \MGvATNLO{}+\HERWIG samples with a generated \PW boson mass of 80.419\GeV using templates from \PYTHIA, 
resulting in $80.30 \pm 0.33\GeV$ and $80.54 \pm 0.37\GeV$, respectively, where the uncertainty 
is the total uncertainty, including the statistical and systematic components.

The result of the template fit to data is $80.81 \pm 0.55\,\text{(total)}\GeV$, 
or when separating statistical and background uncertainty sources, 
$80.81 \pm 0.36\,\text{(stat. + bkg)} \pm 0.41\,\text{(other syst.)}\GeV$.
The impact of individual uncertainties on the \mw measurement is given in Table~\ref{tab:impact_summary_particlenet_mw}.
Uncertainties affecting the position and shape of the \mw peak, such as the statistical uncertainty, the background uncertainty and the jet energy scale are most relevant uncertainties for the \mw parameter estimation.
For consistency with other measurements of the \PW boson mass,
we translate these results to the frequently used mass parameter in a Breit-Wigner distribution with mass-dependent width by adding 27~\MeV~\cite{Dittmaier:2009cr}, resulting in $80.83 \pm 0.55\GeV$.
The result is consistent with previous \mw measurements making use of leptonic final states, though not yet competitive.
It represents a first step towards a potentially competitive measurement at the HL-LHC~\cite{Freytsis:2018cev}.

\begin{table}[htb]
    \centering    
    \topcaption{Summary of systematic uncertainties and their impact on the \mw measurement.
    \label{tab:impact_summary_particlenet_mw}}
    \begin{tabular}{lc}
        Uncertainty & {Impact on $\mw [\GeVns{}]$ } \\
        \hline
        \quad \quad Statistical & 0.31 \\
        \quad \quad QCD multijet estimate & 0.18 \\[\cmsTabSkip]
        Stat. + bkg & \quad \quad 0.36 \\
        \quad \quad Jet energy scale & 0.36  \\
        \quad \quad \PW tagging efficiency & 0.07 \\
        \quad \quad Pileup reweighting & 0.04 \\
        \quad \quad Trigger scale factor & 0.02 \\
        \quad \quad Integrated luminosity & 0.02 \\
        \quad \quad \zjets normalization & 0.17 \\
        \quad \quad Top quark \pt reweight & 0.04 \\
        \quad \quad \vjets QCD NLO & 0.03 \\
        \quad \quad \wjets EW NLO & 0.02 \\
        \quad \quad \zjets EW NLO & 0.01 \\
        \quad \quad FSR & 0.01 \\
        \quad \quad ISR & 0.02 \\
        \quad \quad Hadronization model & 0.02 \\[\cmsTabSkip]
        Other syst. & \quad \quad 0.41 \\
        Total & \quad \quad 0.55 \\
    \end{tabular}
\end{table}

\section{Summary}\label{sec:summary}

The jet mass of \PW bosons decaying to a quark-antiquark pair is measured in \wjets events 
from proton-proton collisions at the LHC at a center-of-mass energy $\sqrt{s} = 13\TeV$.
The data were collected by the CMS experiment in 2016--2018 and correspond to an integrated luminosity of 138\fbinv.
The measurement considers \PW bosons with large momenta, resulting in strongly collimated decay products that 
are reconstructed in a single large-radius jet.
Jets initiated by \PW bosons with a characteristic two-prong substructure are distinguished 
from single light-flavor quark and gluon-initiated background jets using a selection based on the substructure of these jets.
The jet mass is measured using the soft-drop (SD) algorithm, which suppresses soft and wide-angle radiation 
that can obscure the resonance peak of the \PW boson at its rest mass.
We report the first measurement of the double-differential cross section in bins of the jet transverse momentum and SD mass. 
The unfolded data are found to be in agreement with predictions from simulations at 
leading order with additional partons added to the matrix element computation, supplemented by next-to-leading order quantum chromodynamics and electroweak corrections.
A \PW boson mass of $80.83 \pm 0.55\GeV$ in a scheme with mass-dependent width is obtained, achieving the smallest uncertainty available today from an all-jets final state at a hadron collider.

\begin{acknowledgments}
We congratulate our colleagues in the CERN accelerator departments for the excellent performance of the LHC and thank the technical and administrative staffs at CERN and at other CMS institutes for their contributions to the success of the CMS effort. In addition, we gratefully acknowledge the computing centers and personnel of the Worldwide LHC Computing Grid and other centers for delivering so effectively the computing infrastructure essential to our analyses. Finally, we acknowledge the enduring support for the construction and operation of the LHC, the CMS detector, and the supporting computing infrastructure provided by the following funding agencies: SC (Armenia), BMBWF and FWF (Austria); FNRS and FWO (Belgium); CNPq, CAPES, FAPERJ, FAPERGS, and FAPESP (Brazil); MES and BNSF (Bulgaria); CERN; CAS, MoST, and NSFC (China); MINCIENCIAS (Colombia); MSES and CSF (Croatia); RIF (Cyprus); SENESCYT (Ecuador); ERC PRG and PSG, TARISTU24-TK10 and MoER TK202 (Estonia); Academy of Finland, MEC, and HIP (Finland); CEA and CNRS/IN2P3 (France); SRNSF (Georgia); BMFTR, DFG, and HGF (Germany); GSRI (Greece); NKFIH (Hungary); DAE and DST (India); IPM (Iran); SFI (Ireland); INFN (Italy); MSIT and NRF (Republic of Korea); MES (Latvia); LMTLT (Lithuania); MOE and UM (Malaysia); BUAP, CINVESTAV, CONACYT, LNS, SEP, and UASLP-FAI (Mexico); MOS (Montenegro); MBIE (New Zealand); PAEC (Pakistan); MES, NSC, and NAWA (Poland); FCT (Portugal);  MESTD (Serbia); MICIU/AEI and PCTI (Spain); MOSTR (Sri Lanka); Swiss Funding Agencies (Switzerland); MST (Taipei); MHESI (Thailand); TUBITAK and TENMAK (T\"{u}rkiye); NASU (Ukraine); STFC (United Kingdom); DOE and NSF (USA).

\hyphenation{Rachada-pisek} Individuals have received support from the Marie-Curie program and the European Research Council and Horizon 2020 Grant, contract Nos.\ 675440, 724704, 752730, 758316, 765710, 824093, 101115353, 101002207, 101001205, and COST Action CA16108 (European Union); the Leventis Foundation; the Alfred P.\ Sloan Foundation; the Alexander von Humboldt Foundation; the Science Committee, project no. 22rl-037 (Armenia); the Fonds pour la Formation \`a la Recherche dans l'Industrie et dans l'Agriculture (FRIA) and Fonds voor Wetenschappelijk Onderzoek contract No. 1228724N (Belgium); the Beijing Municipal Science \& Technology Commission, No. Z191100007219010, the Fundamental Research Funds for the Central Universities, the Ministry of Science and Technology of China under Grant No. 2023YFA1605804, the Natural Science Foundation of China under Grant No. 12535004, and USTC Research Funds of the Double First-Class Initiative No.\ YD2030002017 (China); the Ministry of Education, Youth and Sports (MEYS) of the Czech Republic; the Shota Rustaveli National Science Foundation, grant FR-22-985 (Georgia); the Deutsche Forschungsgemeinschaft (DFG), among others, under Germany's Excellence Strategy -- EXC 2121 ``Quantum Universe" -- 390833306, and under project number 400140256 - GRK2497; the Hellenic Foundation for Research and Innovation (HFRI), Project Number 2288 (Greece); the Hungarian Academy of Sciences, the New National Excellence Program - \'UNKP, the NKFIH research grants K 131991, K 133046, K 138136, K 143460, K 143477, K 146913, K 146914, K 147048, 2020-2.2.1-ED-2021-00181, TKP2021-NKTA-64, and 2025-1.1.5-NEMZ\_KI-2025-00004 (Hungary); the Council of Science and Industrial Research, India; ICSC -- National Research Center for High Performance Computing, Big Data and Quantum Computing, FAIR -- Future Artificial Intelligence Research, and CUP I53D23001070006 (Mission 4 Component 1), funded by the NextGenerationEU program, the Italian Ministry of University and Research (MUR) under Bando PRIN 2022 -- CUP I53C24002390006, PRIN PRIMULA 2022RBYK7T (Italy); the Latvian Council of Science; the Ministry of Education and Science, project no. 2022/WK/14, and the National Science Center, contracts Opus 2021/41/B/ST2/01369, 2021/43/B/ST2/01552, 2023/49/B/ST2/03273, and the NAWA contract BPN/PPO/2021/1/00011 (Poland); the Funda\c{c}\~ao para a Ci\^encia e a Tecnologia (Portugal); the National Priorities Research Program by Qatar National Research Fund; MICIU/AEI/10.13039/501100011033, ERDF/EU, ``European Union NextGenerationEU/PRTR", projects PID2022-142604OB-C21, PID2022-139519OB-C21, PID2023-147706NB-I00, PID2023-148896NB-I00, PID2023-146983NB-I00, PID2023-147115NB-I00, PID2023-148418NB-C41, PID2023-148418NB-C42, PID2023-148418NB-C43, PID2023-148418NB-C44, PID2024-158190NB-C22, RYC2021-033305-I, RYC2024-048719-I, CNS2023-144781, CNS2024-154769 and Plan de Ciencia, Tecnolog{\'i}a e Innovaci{\'o}n de Asturias, Spain; the Chulalongkorn Academic into Its 2nd Century Project Advancement Project, the National Science, Research and Innovation Fund program IND\_FF\_68\_369\_2300\_097, and the Program Management Unit for Human Resources \& Institutional Development, Research and Innovation, grant B39G680009 (Thailand); the Eric \& Wendy Schmidt Fund for Strategic Innovation through the CERN Next Generation Triggers project under grant agreement number SIF-2023-004; the Kavli Foundation; the Nvidia Corporation; the SuperMicro Corporation; the Welch Foundation, contract C-1845; and the Weston Havens Foundation (USA).
\end{acknowledgments}
\bibliography{auto_generated}

\providecommand{\href}[2]{#2}\begingroup\raggedright\begin{thebibliography}{10}%
\makeatletter
\providecommand{\hrefCMSnoop }[0]{\@secondoftwo}%
\makeatother
\providecommand{\doi}{\texttt{doi:}\begingroup \urlstyle{tt}\Url}

\bibitem{ALEPH:2013dgf}
\hrefCMSnoop {}{{ALEPH, DELPHI, L3, OPAL, SLD, LEP Electroweak Working Group,
  SLD Electroweak Group, SLD Heavy Flavour Group}, ``Electroweak measurements
  in electron-positron collisions at {$\PW$}-boson-pair energies at {LEP}'',}
  \textit{ Phys. Rept.} \textbf{ 532} (2013) 119,
  \href{http://dx.doi.org/10.1016/j.physrep.2013.07.004}{\doi{10.1016/j.physrep.2013.07.004}},
  \href{http://www.arXiv.org/abs/1302.3415}{\texttt{arXiv:1302.3415}}.

\bibitem{D0:2012kms}
\hrefCMSnoop {}{{D0} Collaboration, ``Measurement of the {$\PW$} boson mass
  with the {D0} detector'',} \textit{ Phys. Rev. Lett.} \textbf{ 108} (2012)
  151804,
  \href{http://dx.doi.org/10.1103/PhysRevLett.108.151804}{\doi{10.1103/PhysRevLett.108.151804}},
  \href{http://www.arXiv.org/abs/1203.0293}{\texttt{arXiv:1203.0293}}.

\bibitem{CDF:2022hxs}
\hrefCMSnoop {}{{CDF} Collaboration, ``High-precision measurement of the
  {$\PW$} boson mass with the {CDF II} detector'',} \textit{ Science} \textbf{
  376} (2022) 170,
  \href{http://dx.doi.org/10.1126/science.abk1781}{\doi{10.1126/science.abk1781}}.

\bibitem{ATLAS:2017rzl}
\hrefCMSnoop {}{{ATLAS Collaboration}, ``Measurement of the {$\PW$}-boson mass
  in {$\Pp\Pp$} collisions at {$\sqrt{s} = 7\TeV$} with the {ATLAS}
  detector'',} \textit{ Eur. Phys. J. C} \textbf{ 78} (2018) 110,
  \href{http://dx.doi.org/10.1140/epjc/s10052-017-5475-4}{\doi{10.1140/epjc/s10052-017-5475-4}},
  \href{http://www.arXiv.org/abs/1701.07240}{\texttt{arXiv:1701.07240}}.
  [Erratum: \DOI{10.1140/epjc/s10052-018-6354-3}].

\bibitem{LHCb:2021bjt}
\hrefCMSnoop {}{{LHCb Collaboration}, ``Measurement of the {$\PW$} boson
  mass'',} \textit{ JHEP} \textbf{ 01} (2022) 036,
  \href{http://dx.doi.org/10.1007/JHEP01(2022)036}{\doi{10.1007/JHEP01(2022)036}},
  \href{http://www.arXiv.org/abs/2109.01113}{\texttt{arXiv:2109.01113}}.

\bibitem{ATLAS:2024erm}
\hrefCMSnoop {}{{ATLAS Collaboration}, ``Measurement of the {$\PW$}-boson mass
  and width with the {ATLAS} detector using proton-proton collisions at
  {$\sqrt{s} = 7\TeV$}'',} \textit{ Eur. Phys. J. C} \textbf{ 84} (2024) 1309,
  \href{http://dx.doi.org/10.1140/epjc/s10052-024-13190-x}{\doi{10.1140/epjc/s10052-024-13190-x}},
  \href{http://www.arXiv.org/abs/2403.15085}{\texttt{arXiv:2403.15085}}.

\bibitem{CMS:2024lrd}
\hrefCMSnoop {}{{CMS Collaboration}, ``High-precision measurement of the {\PW}
  boson mass with the {CMS} experiment at the {LHC}'',} 2024.
  \href{http://www.arXiv.org/abs/2412.13872}{\texttt{arXiv:2412.13872}}.
  Submitted to \textit{Nature}.

\bibitem{ParticleDataGroup:2024cfk}
\hrefCMSnoop {}{{Particle Data Group}, ``Review of particle physics'',}
  \textit{ Phys. Rev. D} \textbf{ 110} (2024) 030001,
  \href{http://dx.doi.org/10.1103/PhysRevD.110.030001}{\doi{10.1103/PhysRevD.110.030001}}.

\bibitem{Marzani:2019hun}
S.~Marzani, G.~Soyez, and M.~Spannowsky, ``Looking inside jets: an introduction
  to jet substructure and boosted-object phenomenology'', volume 958.
\newblock Springer, 2019.
\newblock ISBN 978-3-030-15708-1.
  \href{http://dx.doi.org/10.1007/978-3-030-15709-8}{\doi{10.1007/978-3-030-15709-8}}.

\bibitem{Larkoski:2017jix}
\hrefCMSnoop {}{A.~J. Larkoski, I.~Moult, and B.~Nachman, ``Jet substructure at
  the {Large Hadron Collider}: {A} review of recent advances in theory and
  machine learning'',} \textit{ Phys. Rept.} \textbf{ 841} (2020) 1,
  \href{http://dx.doi.org/10.1016/j.physrep.2019.11.001}{\doi{10.1016/j.physrep.2019.11.001}},
  \href{http://www.arXiv.org/abs/1709.04464}{\texttt{arXiv:1709.04464}}.

\bibitem{Kogler:2018hem}
\hrefCMSnoop {}{R.~Kogler { et~al.}, ``Jet substructure at the {Large Hadron
  Collider}'',} \textit{ Rev. Mod. Phys.} \textbf{ 91} (2019) 045003,
  \href{http://dx.doi.org/10.1103/RevModPhys.91.045003}{\doi{10.1103/RevModPhys.91.045003}},
  \href{http://www.arXiv.org/abs/1803.06991}{\texttt{arXiv:1803.06991}}.

\bibitem{Kogler:2021kkw}
R.~Kogler, ``Advances in jet substructure at the {LHC}: Algorithms,
  measurements and searches for new physical phenomena''.
\newblock Springer Cham, 2021.
\newblock ISBN 978-3-030-72858-8.
  \href{http://dx.doi.org/10.1007/978-3-030-72858-8}{\doi{10.1007/978-3-030-72858-8}}.

\bibitem{UA2:1987vmn}
\hrefCMSnoop {}{{UA2} Collaboration, ``Search for decays of the ${\PW^\pm}$ and
  ${\PZ}$ bosons into quark-anti-quark pairs'',} \textit{ Phys. Lett. B}
  \textbf{ 186} (1987) 452,
  \href{http://dx.doi.org/10.1016/0370-2693(87)90325-X}{\doi{10.1016/0370-2693(87)90325-X}}.

\bibitem{CDF:1997iiz}
\hrefCMSnoop {}{{CDF} Collaboration, ``Observation of hadronic {$\PW$} decays
  in $t\bar{t}$ events with the {Collider Detector} at {Fermilab}'',} \textit{
  Phys. Rev. Lett.} \textbf{ 80} (1998) 5720,
  \href{http://dx.doi.org/10.1103/PhysRevLett.80.5720}{\doi{10.1103/PhysRevLett.80.5720}},
  \href{http://www.arXiv.org/abs/hep-ex/9711004}{\texttt{arXiv:hep-ex/9711004}}.

\bibitem{Freytsis:2018cev}
M.~Freytsis\hrefCMSnoop {}{ { et~al.}, ``Prospects for a measurement of the
  {$\PW$} boson mass in the all-jets final state at hadron colliders'',}
  \textit{ JHEP} \textbf{ 02} (2019) 003,
  \href{http://dx.doi.org/10.1007/JHEP02(2019)003}{\doi{10.1007/JHEP02(2019)003}},
  \href{http://www.arXiv.org/abs/1807.07454}{\texttt{arXiv:1807.07454}}.

\bibitem{Dasgupta:2013ihk}
\hrefCMSnoop {}{M.~Dasgupta, A.~Fregoso, S.~Marzani, and G.~P. Salam, ``Towards
  an understanding of jet substructure'',} \textit{ JHEP} \textbf{ 09} (2013)
  029,
  \href{http://dx.doi.org/10.1007/JHEP09(2013)029}{\doi{10.1007/JHEP09(2013)029}},
  \href{http://www.arXiv.org/abs/1307.0007}{\texttt{arXiv:1307.0007}}.

\bibitem{Butterworth:2008iy}
\hrefCMSnoop {}{J.~M. Butterworth, A.~R. Davison, M.~Rubin, and G.~P. Salam,
  ``Jet substructure as a new {Higgs} search channel at the {LHC}'',} \textit{
  Phys. Rev. Lett.} \textbf{ 100} (2008) 242001,
  \href{http://dx.doi.org/10.1103/PhysRevLett.100.242001}{\doi{10.1103/PhysRevLett.100.242001}},
  \href{http://www.arXiv.org/abs/0802.2470}{\texttt{arXiv:0802.2470}}.

\bibitem{Larkoski:2014wba}
\hrefCMSnoop {}{A.~J. Larkoski, S.~Marzani, G.~Soyez, and J.~Thaler, ``Soft
  drop'',} \textit{ JHEP} \textbf{ 05} (2014) 146,
  \href{http://dx.doi.org/10.1007/JHEP05(2014)146}{\doi{10.1007/JHEP05(2014)146}},
  \href{http://www.arXiv.org/abs/1402.2657}{\texttt{arXiv:1402.2657}}.

\bibitem{CMS:2008xjf}
\hrefCMSnoop {}{{CMS Collaboration}, ``The {CMS} experiment at the {CERN}
  {LHC}'',} \textit{ JINST} \textbf{ 3} (2008) S08004,
  \href{http://dx.doi.org/10.1088/1748-0221/3/08/S08004}{\doi{10.1088/1748-0221/3/08/S08004}}.

\bibitem{ATLAS:2012am}
\hrefCMSnoop {}{{ATLAS Collaboration}, ``Jet mass and substructure of inclusive
  jets in {$\sqrt{s}=7\TeV$} {$\Pp\Pp$} collisions with the {ATLAS}
  experiment'',} \textit{ JHEP} \textbf{ 05} (2012) 128,
  \href{http://dx.doi.org/10.1007/JHEP05(2012)128}{\doi{10.1007/JHEP05(2012)128}},
  \href{http://www.arXiv.org/abs/1203.4606}{\texttt{arXiv:1203.4606}}.

\bibitem{ATLAS:2017zda}
\hrefCMSnoop {}{{ATLAS Collaboration}, ``Measurement of the soft-drop jet mass
  in {$\Pp\Pp$} collisions at {$\sqrt{s} = 13\TeV$} with the {ATLAS}
  detector'',} \textit{ Phys. Rev. Lett.} \textbf{ 121} (2018) 092001,
  \href{http://dx.doi.org/10.1103/PhysRevLett.121.092001}{\doi{10.1103/PhysRevLett.121.092001}},
  \href{http://www.arXiv.org/abs/1711.08341}{\texttt{arXiv:1711.08341}}.

\bibitem{CMS:2018vzn}
\hrefCMSnoop {}{{CMS Collaboration}, ``Measurements of the differential jet
  cross section as a function of the jet mass in dijet events from
  proton-proton collisions at $\sqrt{s}={13\TeV}$'',} \textit{ JHEP} \textbf{
  11} (2018) 113,
  \href{http://dx.doi.org/10.1007/JHEP11(2018)113}{\doi{10.1007/JHEP11(2018)113}},
  \href{http://www.arXiv.org/abs/1807.05974}{\texttt{arXiv:1807.05974}}.

\bibitem{CMS:2017pcy}
\hrefCMSnoop {}{{CMS Collaboration}, ``Measurement of the jet mass in highly
  boosted {\ttbar} events from {$\Pp\Pp$} collisions at {$\sqrt{s}=8\TeV$}'',}
  \textit{ Eur. Phys. J. C} \textbf{ 77} (2017) 467,
  \href{http://dx.doi.org/10.1140/epjc/s10052-017-5030-3}{\doi{10.1140/epjc/s10052-017-5030-3}},
  \href{http://www.arXiv.org/abs/1703.06330}{\texttt{arXiv:1703.06330}}.

\bibitem{CMS:2019fak}
\hrefCMSnoop {}{{CMS Collaboration}, ``Measurement of the jet mass distribution
  and top quark mass in hadronic decays of boosted top quarks in {$\Pp\Pp$}
  collisions at {$\sqrt{s} =13\TeV$}'',} \textit{ Phys. Rev. Lett.} \textbf{
  124} (2020) 202001,
  \href{http://dx.doi.org/10.1103/PhysRevLett.124.202001}{\doi{10.1103/PhysRevLett.124.202001}},
  \href{http://www.arXiv.org/abs/1911.03800}{\texttt{arXiv:1911.03800}}.

\bibitem{CMS:2022kqg}
\hrefCMSnoop {}{{CMS Collaboration}, ``Measurement of the differential \ttbar
  production cross section as a function of the jet mass and extraction of the
  top quark mass in hadronic decays of boosted top quarks'',} \textit{ Eur.
  Phys. J. C} \textbf{ 83} (2023) 560,
  \href{http://dx.doi.org/10.1140/epjc/s10052-023-11587-8}{\doi{10.1140/epjc/s10052-023-11587-8}},
  \href{http://www.arXiv.org/abs/2211.01456}{\texttt{arXiv:2211.01456}}.

\bibitem{ATLAS:2019dty}
\hrefCMSnoop {}{{ATLAS Collaboration}, ``Measurement of the jet mass in high
  transverse momentum {$\PZ(\to \PQb\PAQb)\Pgg$} production at {$\sqrt{s}=
  13\TeV$} using the {ATLAS} detector'',} \textit{ Phys. Lett. B} \textbf{ 812}
  (2021) 135991,
  \href{http://dx.doi.org/10.1016/j.physletb.2020.135991}{\doi{10.1016/j.physletb.2020.135991}},
  \href{http://www.arXiv.org/abs/1907.07093}{\texttt{arXiv:1907.07093}}.

\bibitem{CMS:2023gfb}
\hrefCMSnoop {}{{CMS Collaboration}, ``Development of the {CMS} detector for
  the {CERN LHC Run 3}'',} \textit{ JINST} \textbf{ 19} (2024) P05064,
  \href{http://dx.doi.org/10.1088/1748-0221/19/05/P05064}{\doi{10.1088/1748-0221/19/05/P05064}},
  \href{http://www.arXiv.org/abs/2309.05466}{\texttt{arXiv:2309.05466}}.

\bibitem{CMS:2014pgm}
\hrefCMSnoop {}{{CMS Collaboration}, ``Description and performance of track and
  primary-vertex reconstruction with the {CMS} tracker'',} \textit{ JINST}
  \textbf{ 9} (2014) P10009,
  \href{http://dx.doi.org/10.1088/1748-0221/9/10/P10009}{\doi{10.1088/1748-0221/9/10/P10009}},
  \href{http://www.arXiv.org/abs/1405.6569}{\texttt{arXiv:1405.6569}}.

\bibitem{Phase1Pixel}
\hrefCMSnoop {}{{CMS Tracker Group}, ``The {CMS} {Phase-1} pixel detector
  upgrade'',} \textit{ JINST} \textbf{ 16} (2021) P02027,
  \href{http://dx.doi.org/10.1088/1748-0221/16/02/P02027}{\doi{10.1088/1748-0221/16/02/P02027}},
  \href{http://www.arXiv.org/abs/2012.14304}{\texttt{arXiv:2012.14304}}.

\bibitem{DP-2020-049}
\href {https://cds.cern.ch/record/2743740}{{CMS Collaboration}, ``Track impact
  parameter resolution for the full pseudo rapidity coverage in the 2017
  dataset with the {CMS} {Phase-1} pixel detector'',} CMS Detector Performance
  Note CMS-DP-2020-049, 2020.

\bibitem{DP-2017-015}
\href {https://cds.cern.ch/record/2290524}{{CMS Collaboration}, ``2017 tracking
  performance plots'',} CMS Detector Performance Note CMS-DP-2017-015, 2017.

\bibitem{CMS:2020cmk}
\hrefCMSnoop {}{{CMS Collaboration}, ``Performance of the {CMS} {\Lone} trigger
  in proton-proton collisions at $\sqrt{s}={13\TeV}$'',} \textit{ JINST}
  \textbf{ 15} (2020) P10017,
  \href{http://dx.doi.org/10.1088/1748-0221/15/10/P10017}{\doi{10.1088/1748-0221/15/10/P10017}},
  \href{http://www.arXiv.org/abs/2006.10165}{\texttt{arXiv:2006.10165}}.

\bibitem{CMS:2016ngn}
\hrefCMSnoop {}{{CMS Collaboration}, ``The {CMS} trigger system'',} \textit{
  JINST} \textbf{ 12} (2017) P01020,
  \href{http://dx.doi.org/10.1088/1748-0221/12/01/P01020}{\doi{10.1088/1748-0221/12/01/P01020}},
  \href{http://www.arXiv.org/abs/1609.02366}{\texttt{arXiv:1609.02366}}.

\bibitem{CMS:2024aqx}
\hrefCMSnoop {}{{CMS Collaboration}, ``Performance of the {CMS} high-level
  trigger during {LHC} {Run 2}'',} \textit{ JINST} \textbf{ 19} (2024) P11021,
  \href{http://dx.doi.org/10.1088/1748-0221/19/11/P11021}{\doi{10.1088/1748-0221/19/11/P11021}},
  \href{http://www.arXiv.org/abs/2410.17038}{\texttt{arXiv:2410.17038}}.

\bibitem{CMS:2017yfk}
\hrefCMSnoop {}{{CMS Collaboration}, ``Particle-flow reconstruction and global
  event description with the {CMS} detector'',} \textit{ JINST} \textbf{ 12}
  (2017) P10003,
  \href{http://dx.doi.org/10.1088/1748-0221/12/10/P10003}{\doi{10.1088/1748-0221/12/10/P10003}},
  \href{http://www.arXiv.org/abs/1706.04965}{\texttt{arXiv:1706.04965}}.

\bibitem{CMS-TDR-15-02}
\href {http://cds.cern.ch/record/2020886}{{CMS Collaboration}, ``Technical
  proposal for the {Phase-II} upgrade of the {Compact Muon Solenoid}'',} CMS
  Technical Proposal CERN-LHCC-2015-010, CMS-TDR-15-02, 2015.

\bibitem{CMS:2020uim}
\hrefCMSnoop {}{{CMS Collaboration}, ``Electron and photon reconstruction and
  identification with the {CMS} experiment at the {CERN} {LHC}'',} \textit{
  JINST} \textbf{ 16} (2021) P05014,
  \href{http://dx.doi.org/10.1088/1748-0221/16/05/P05014}{\doi{10.1088/1748-0221/16/05/P05014}},
  \href{http://www.arXiv.org/abs/2012.06888}{\texttt{arXiv:2012.06888}}.

\bibitem{Sirunyan:2020foa}
\hrefCMSnoop {}{{CMS Collaboration}, ``Pileup mitigation at {CMS} in {13\TeV}
  data'',} \textit{ JINST} \textbf{ 15} (2020) P09018,
  \href{http://dx.doi.org/10.1088/1748-0221/15/09/P09018}{\doi{10.1088/1748-0221/15/09/P09018}},
  \href{http://www.arXiv.org/abs/2003.00503}{\texttt{arXiv:2003.00503}}.

\bibitem{Bertolini:2014bba}
\hrefCMSnoop {}{D.~Bertolini, P.~Harris, M.~Low, and N.~Tran, ``Pileup per
  particle identification'',} \textit{ JHEP} \textbf{ 10} (2014) 059,
  \href{http://dx.doi.org/10.1007/JHEP10(2014)059}{\doi{10.1007/JHEP10(2014)059}},
  \href{http://www.arXiv.org/abs/1407.6013}{\texttt{arXiv:1407.6013}}.

\bibitem{Cacciari:2008gp}
\hrefCMSnoop {}{M.~Cacciari, G.~P. Salam, and G.~Soyez, ``The anti-\kt jet
  clustering algorithm'',} \textit{ JHEP} \textbf{ 04} (2008) 063,
  \href{http://dx.doi.org/10.1088/1126-6708/2008/04/063}{\doi{10.1088/1126-6708/2008/04/063}},
  \href{http://www.arXiv.org/abs/0802.1189}{\texttt{arXiv:0802.1189}}.

\bibitem{Cacciari:2011ma}
\hrefCMSnoop {}{M.~Cacciari, G.~P. Salam, and G.~Soyez, ``{\FASTJET} user
  manual'',} \textit{ Eur. Phys. J. C} \textbf{ 72} (2012) 1896,
  \href{http://dx.doi.org/10.1140/epjc/s10052-012-1896-2}{\doi{10.1140/epjc/s10052-012-1896-2}},
  \href{http://www.arXiv.org/abs/1111.6097}{\texttt{arXiv:1111.6097}}.

\bibitem{CMS:2016lmd}
\hrefCMSnoop {}{{CMS Collaboration}, ``Jet energy scale and resolution in the
  {CMS} experiment in ${\Pp\Pp}$ collisions at {8\TeV}'',} \textit{ JINST}
  \textbf{ 12} (2017) P02014,
  \href{http://dx.doi.org/10.1088/1748-0221/12/02/P02014}{\doi{10.1088/1748-0221/12/02/P02014}},
  \href{http://www.arXiv.org/abs/1607.03663}{\texttt{arXiv:1607.03663}}.

\bibitem{MadGraph_aMCatNLO}
J.~Alwall\hrefCMSnoop {}{ { et~al.}, ``The automated computation of tree-level
  and next-to-leading order differential cross sections, and their matching to
  parton shower simulations'',} \textit{ JHEP} \textbf{ 07} (2014) 079,
  \href{http://dx.doi.org/10.1007/JHEP07(2014)079}{\doi{10.1007/JHEP07(2014)079}},
  \href{http://www.arXiv.org/abs/1405.0301}{\texttt{arXiv:1405.0301}}.

\bibitem{Bierlich:2022pfr}
\hrefCMSnoop {}{C.~Bierlich { et~al.}, ``A comprehensive guide to the physics
  and usage of {\PYTHIA8.3}'',} \textit{ SciPost Phys. Codeb.} \textbf{ 2022}
  (2022) 8,
  \href{http://dx.doi.org/10.21468/SciPostPhysCodeb.8}{\doi{10.21468/SciPostPhysCodeb.8}},
  \href{http://www.arXiv.org/abs/2203.11601}{\texttt{arXiv:2203.11601}}.

\bibitem{Sirunyan:2019dfx}
\hrefCMSnoop {}{{CMS Collaboration}, ``Extraction and validation of a new set
  of {CMS} {\PYTHIA{}8} tunes from underlying-event measurements'',} \textit{
  Eur. Phys. J. C} \textbf{ 80} (2020) 4,
  \href{http://dx.doi.org/10.1140/epjc/s10052-019-7499-4}{\doi{10.1140/epjc/s10052-019-7499-4}},
  \href{http://www.arXiv.org/abs/1903.12179}{\texttt{arXiv:1903.12179}}.

\bibitem{Andersson:1983ia}
\hrefCMSnoop {}{B.~Andersson, G.~Gustafson, G.~Ingelman, and T.~Sj{\"o}strand,
  ``Parton fragmentation and string dynamics'',} \textit{ Phys. Rept.} \textbf{
  97} (1983) 31,
\href{http://dx.doi.org/10.1016/0370-1573(83)90080-7}{\doi{10.1016/0370-1573(83)90080-7}}.

\bibitem{Sjostrand:1984iu}
\hrefCMSnoop {}{T.~Sj{\"o}strand, ``The merging of jets'',} \textit{ Phys.
  Lett. B} \textbf{ 142} (1984) 420,
\href{http://dx.doi.org/10.1016/0370-2693(84)91354-6}{\doi{10.1016/0370-2693(84)91354-6}}.

\bibitem{MadGraph5_MLM}
J.~Alwall\hrefCMSnoop {}{ { et~al.}, ``Comparative study of various algorithms
  for the merging of parton showers and matrix elements in hadronic
  collisions'',} \textit{ Eur. Phys. J. C} \textbf{ 53} (2008) 473,
  \href{http://dx.doi.org/10.1140/epjc/s10052-007-0490-5}{\doi{10.1140/epjc/s10052-007-0490-5}},
  \href{http://www.arXiv.org/abs/0706.2569}{\texttt{arXiv:0706.2569}}.

\bibitem{Bellm:2015jjp}
\hrefCMSnoop {}{J.~Bellm { et~al.}, ``{\HERWIG7.0}/{\HERWIGpp3.0} release
  note'',} \textit{ Eur. Phys. J. C} \textbf{ 76} (2016) 196,
  \href{http://dx.doi.org/10.1140/epjc/s10052-016-4018-8}{\doi{10.1140/epjc/s10052-016-4018-8}},
  \href{http://www.arXiv.org/abs/1512.01178}{\texttt{arXiv:1512.01178}}.

\bibitem{Sirunyan:2020pqv}
\hrefCMSnoop {}{{CMS Collaboration}, ``Development and validation of {\HERWIG7}
  tunes from {CMS} underlying-event measurements'',} \textit{ Eur. Phys. J. C}
  \textbf{ 81} (2021) 312,
  \href{http://dx.doi.org/10.1140/epjc/s10052-021-08949-5}{\doi{10.1140/epjc/s10052-021-08949-5}},
  \href{http://www.arXiv.org/abs/2011.03422}{\texttt{arXiv:2011.03422}}.

\bibitem{Frederix:2012ps}
\hrefCMSnoop {}{R.~Frederix and S.~Frixione, ``Merging meets matching in
  {\MCATNLO}'',} \textit{ JHEP} \textbf{ 12} (2012) 061,
  \href{http://dx.doi.org/10.1007/JHEP12(2012)061}{\doi{10.1007/JHEP12(2012)061}},
  \href{http://www.arXiv.org/abs/1209.6215}{\texttt{arXiv:1209.6215}}.

\bibitem{POWHEG_main1}
\hrefCMSnoop {}{P.~Nason, ``A new method for combining {NLO} {QCD} with shower
  {Monte Carlo} algorithms'',} \textit{ JHEP} \textbf{ 11} (2004) 040,
  \href{http://dx.doi.org/10.1088/1126-6708/2004/11/040}{\doi{10.1088/1126-6708/2004/11/040}},
  \href{http://www.arXiv.org/abs/hep-ph/0409146}{\texttt{arXiv:hep-ph/0409146}}.

\bibitem{POWHEG_main2}
\hrefCMSnoop {}{S.~Frixione, P.~Nason, and C.~Oleari, ``Matching {NLO} {QCD}
  computations with parton shower simulations: the {\POWHEG} method'',}
  \textit{ JHEP} \textbf{ 11} (2007) 070,
  \href{http://dx.doi.org/10.1088/1126-6708/2007/11/070}{\doi{10.1088/1126-6708/2007/11/070}},
  \href{http://www.arXiv.org/abs/0709.2092}{\texttt{arXiv:0709.2092}}.

\bibitem{POWHEG_main3}
\hrefCMSnoop {}{S.~Alioli, P.~Nason, C.~Oleari, and E.~Re, ``A general
  framework for implementing {NLO} calculations in shower {Monte Carlo}
  programs: the {\POWHEG} \textsc{box}'',} \textit{ JHEP} \textbf{ 06} (2010)
  043,
  \href{http://dx.doi.org/10.1007/JHEP06(2010)043}{\doi{10.1007/JHEP06(2010)043}},
  \href{http://www.arXiv.org/abs/1002.2581}{\texttt{arXiv:1002.2581}}.

\bibitem{POWHGE_BOX:hvq}
\hrefCMSnoop {}{S.~Frixione, G.~Ridolfi, and P.~Nason, ``A positive-weight
  next-to-leading-order {Monte Carlo} for heavy flavour hadroproduction'',}
  \textit{ JHEP} \textbf{ 09} (2007) 126,
  \href{http://dx.doi.org/10.1088/1126-6708/2007/09/126}{\doi{10.1088/1126-6708/2007/09/126}},
  \href{http://www.arXiv.org/abs/0707.3088}{\texttt{arXiv:0707.3088}}.

\bibitem{NNPDF:2017mvq}
\hrefCMSnoop {}{{NNPDF} Collaboration, ``Parton distributions from
  high-precision collider data'',} \textit{ Eur. Phys. J. C} \textbf{ 77}
  (2017) 663,
  \href{http://dx.doi.org/10.1140/epjc/s10052-017-5199-5}{\doi{10.1140/epjc/s10052-017-5199-5}},
  \href{http://www.arXiv.org/abs/1706.00428}{\texttt{arXiv:1706.00428}}.

\bibitem{GEANT4:2002zbu}
\hrefCMSnoop {}{{GEANT4} Collaboration, ``{\GEANTfour}---a simulation
  toolkit'',} \textit{ Nucl. Instrum. Meth. A} \textbf{ 506} (2003) 250,
  \href{http://dx.doi.org/10.1016/S0168-9002(03)01368-8}{\doi{10.1016/S0168-9002(03)01368-8}}.

\bibitem{CMS:2022fxs}
\hrefCMSnoop {}{{CMS Collaboration}, ``Search for boosted {Higgs} boson decay
  to a charm quark-antiquark pair in proton-proton collisions at
  {$\sqrt{s}=13\TeV$}'',} \textit{ Phys. Rev. Lett.} \textbf{ 131} (2023)
  041801,
  \href{http://dx.doi.org/10.1103/PhysRevLett.131.041801}{\doi{10.1103/PhysRevLett.131.041801}},
  \href{http://www.arXiv.org/abs/2211.14181}{\texttt{arXiv:2211.14181}}.

\bibitem{Lindert:2017olm}
\hrefCMSnoop {}{J.~M. Lindert { et~al.}, ``Precise predictions for \vjets dark
  matter backgrounds'',} \textit{ Eur. Phys. J. C} \textbf{ 77} (2017) 829,
  \href{http://dx.doi.org/10.1140/epjc/s10052-017-5389-1}{\doi{10.1140/epjc/s10052-017-5389-1}},
  \href{http://www.arXiv.org/abs/1705.04664}{\texttt{arXiv:1705.04664}}.

\bibitem{TopPTdilepton}
\hrefCMSnoop {}{{CMS Collaboration}, ``Measurement of normalized differential
  \ttbar cross sections in the dilepton channel from ${\Pp\Pp}$ collisions at
  {$\sqrt{s}=13\TeV$}'',} \textit{ JHEP} \textbf{ 04} (2018) 060,
  \href{http://dx.doi.org/10.1007/JHEP04(2018)060}{\doi{10.1007/JHEP04(2018)060}},
  \href{http://www.arXiv.org/abs/1708.07638}{\texttt{arXiv:1708.07638}}.

\bibitem{TopPTlepjets}
\hrefCMSnoop {}{{CMS Collaboration}, ``Measurement of differential cross
  sections for top quark pair production using the lepton+jets final state in
  proton-proton collisions at {13\TeV}'',} \textit{ Phys. Rev. D} \textbf{ 95}
  (2017) 092001,
  \href{http://dx.doi.org/10.1103/PhysRevD.95.092001}{\doi{10.1103/PhysRevD.95.092001}},
  \href{http://www.arXiv.org/abs/1610.04191}{\texttt{arXiv:1610.04191}}.

\bibitem{Dokshitzer:1997in}
\hrefCMSnoop {}{Y.~L. Dokshitzer, G.~D. Leder, S.~Moretti, and B.~R. Webber,
  ``Better jet clustering algorithms'',} \textit{ JHEP} \textbf{ 08} (1997)
  001,
  \href{http://dx.doi.org/10.1088/1126-6708/1997/08/001}{\doi{10.1088/1126-6708/1997/08/001}},
  \href{http://www.arXiv.org/abs/hep-ph/9707323}{\texttt{arXiv:hep-ph/9707323}}.

\bibitem{CMS:2020poo}
\hrefCMSnoop {}{{CMS Collaboration}, ``Identification of heavy, energetic,
  hadronically decaying particles using machine-learning techniques'',}
  \textit{ JINST} \textbf{ 15} (2020) P06005,
  \href{http://dx.doi.org/10.1088/1748-0221/15/06/P06005}{\doi{10.1088/1748-0221/15/06/P06005}},
  \href{http://www.arXiv.org/abs/2004.08262}{\texttt{arXiv:2004.08262}}.

\bibitem{Moult:2016cvt}
\hrefCMSnoop {}{I.~Moult, L.~Necib, and J.~Thaler, ``New angles on energy
  correlation functions'',} \textit{ JHEP} \textbf{ 12} (2016) 153,
  \href{http://dx.doi.org/10.1007/JHEP12(2016)153}{\doi{10.1007/JHEP12(2016)153}},
  \href{http://www.arXiv.org/abs/1609.07483}{\texttt{arXiv:1609.07483}}.

\bibitem{Larkoski:2013eya}
\hrefCMSnoop {}{A.~J. Larkoski, G.~P. Salam, and J.~Thaler, ``Energy
  correlation functions for jet substructure'',} \textit{ JHEP} \textbf{ 06}
  (2013) 108,
  \href{http://dx.doi.org/10.1007/JHEP06(2013)108}{\doi{10.1007/JHEP06(2013)108}},
  \href{http://www.arXiv.org/abs/1305.0007}{\texttt{arXiv:1305.0007}}.

\bibitem{Qu:2019gqs}
\hrefCMSnoop {}{H.~Qu and L.~Gouskos, ``Jet tagging via particle clouds'',}
  \textit{ Phys. Rev. D} \textbf{ 101} (2020) 056019,
  \href{http://dx.doi.org/10.1103/PhysRevD.101.056019}{\doi{10.1103/PhysRevD.101.056019}},
  \href{http://www.arXiv.org/abs/1902.08570}{\texttt{arXiv:1902.08570}}.

\bibitem{10.1145/3326362}
Y.~Wang\hrefCMSnoop {}{ { et~al.}, ``Dynamic graph {CNN} for learning on point
  clouds'',} \textit{ ACM Trans. Graph.} \textbf{ 38} (2019) 12,
  \href{http://dx.doi.org/10.1145/3326362}{\doi{10.1145/3326362}},
  \href{http://www.arXiv.org/abs/1801.07829}{\texttt{arXiv:1801.07829}}.

\bibitem{CMS-DP-2020-002}
\href {https://cds.cern.ch/record/2707946}{{CMS Collaboration},
  ``Identification of highly {L}orentz-boosted heavy particles using graph
  neural networks and new mass decorrelation techniques'',} CMS Detector
  Performance Note CMS-DP-2020-002, 2020.

\bibitem{Dolen:2016kst}
J.~Dolen\hrefCMSnoop {}{ { et~al.}, ``Thinking outside the {ROCs}: Designing
  decorrelated taggers ({DDT}) for jet substructure'',} \textit{ JHEP} \textbf{
  05} (2016) 156,
  \href{http://dx.doi.org/10.1007/JHEP05(2016)156}{\doi{10.1007/JHEP05(2016)156}},
  \href{http://www.arXiv.org/abs/1603.00027}{\texttt{arXiv:1603.00027}}.

\bibitem{CMS:2019emo}
\hrefCMSnoop {}{{CMS Collaboration}, ``Search for low mass vector resonances
  decaying into quark-antiquark pairs in proton-proton collisions at
  $\sqrt{s}={13\TeV}$'',} \textit{ Phys. Rev. D} \textbf{ 100} (2019) 112007,
  \href{http://dx.doi.org/10.1103/PhysRevD.100.112007}{\doi{10.1103/PhysRevD.100.112007}},
  \href{http://www.arXiv.org/abs/1909.04114}{\texttt{arXiv:1909.04114}}.

\bibitem{CMS:2018rym}
\hrefCMSnoop {}{{CMS Collaboration}, ``Performance of the {CMS} muon detector
  and muon reconstruction with proton-proton collisions at
  $\sqrt{s}={13\TeV}$'',} \textit{ JINST} \textbf{ 13} (2018) P06015,
  \href{http://dx.doi.org/10.1088/1748-0221/13/06/P06015}{\doi{10.1088/1748-0221/13/06/P06015}},
  \href{http://www.arXiv.org/abs/1804.04528}{\texttt{arXiv:1804.04528}}.

\bibitem{ATLAS:2019mgf}
\hrefCMSnoop {}{{ATLAS Collaboration}, ``Measurement of soft-drop jet
  observables in $pp$ collisions with the atlas detector at {$\sqrt {s}$ =13
  TeV}'',} \textit{ Phys. Rev. D} \textbf{ 101} (2020) 052007,
  \href{http://dx.doi.org/10.1103/PhysRevD.101.052007}{\doi{10.1103/PhysRevD.101.052007}},
  \href{http://www.arXiv.org/abs/1912.09837}{\texttt{arXiv:1912.09837}}.

\bibitem{CAT-23-001}
\hrefCMSnoop {}{{CMS Collaboration}, ``The {CMS} statistical analysis and
  combination tool: \textsc{combine}'',} \textit{ Comput. Softw. Big Sci.}
  \textbf{ 8} (2024) 19,
  \href{http://dx.doi.org/10.1007/s41781-024-00121-4}{\doi{10.1007/s41781-024-00121-4}},
  \href{http://www.arXiv.org/abs/2404.06614}{\texttt{arXiv:2404.06614}}.

\bibitem{CMS-LUM-17-003}
\hrefCMSnoop {}{{CMS Collaboration}, ``Precision luminosity measurement in
  proton-proton collisions at {$\sqrt{s}=13\TeV$} in 2015 and 2016 at {CMS}'',}
  \textit{ Eur. Phys. J. C} \textbf{ 81} (2021) 800,
  \href{http://dx.doi.org/10.1140/epjc/s10052-021-09538-2}{\doi{10.1140/epjc/s10052-021-09538-2}},
  \href{http://www.arXiv.org/abs/2104.01927}{\texttt{arXiv:2104.01927}}.

\bibitem{CMS-PAS-LUM-17-004}
\href {https://cds.cern.ch/record/2621960/}{{CMS Collaboration}, ``{CMS}
  luminosity measurement for the 2017 data-taking period at
  {$\sqrt{s}=13\TeV$}'',} CMS Physics Analysis Summary CMS-PAS-LUM-17-004,
  2018.

\bibitem{CMS-PAS-LUM-18-002}
\href {https://cds.cern.ch/record/2676164/}{{CMS Collaboration}, ``{CMS}
  luminosity measurement for the 2018 data-taking period at
  {$\sqrt{s}=13\TeV$}'',} CMS Physics Analysis Summary CMS-PAS-LUM-18-002,
  2019.

\bibitem{CMS:2018mlc}
\hrefCMSnoop {}{{CMS Collaboration}, ``Measurement of the inelastic
  proton-proton cross section at {$\sqrt{s}=13\TeV$}'',} \textit{ JHEP}
  \textbf{ 07} (2018) 161,
  \href{http://dx.doi.org/10.1007/JHEP07(2018)161}{\doi{10.1007/JHEP07(2018)161}},
  \href{http://www.arXiv.org/abs/1802.02613}{\texttt{arXiv:1802.02613}}.

\bibitem{Rubin:2010xp}
\hrefCMSnoop {}{M.~Rubin, G.~P. Salam, and S.~Sapeta, ``Giant {QCD} {K}-factors
  beyond {NLO}'',} \textit{ JHEP} \textbf{ 09} (2010) 084,
  \href{http://dx.doi.org/10.1007/JHEP09(2010)084}{\doi{10.1007/JHEP09(2010)084}},
  \href{http://www.arXiv.org/abs/1006.2144}{\texttt{arXiv:1006.2144}}.

\bibitem{S_Schmitt_2012}
\hrefCMSnoop {}{S.~Schmitt, ``\textsc{TUnfold}, an algorithm for correcting
  migration effects in high energy physics'',} \textit{ JINST} \textbf{ 7}
  (2012) T10003,
  \href{http://dx.doi.org/10.1088/1748-0221/7/10/T10003}{\doi{10.1088/1748-0221/7/10/T10003}},
  \href{http://www.arXiv.org/abs/1205.6201}{\texttt{arXiv:1205.6201}}.

\bibitem{Tikhonov1963}
\hrefCMSnoop {}{A.~N. Tikhonov, ``Solution of incorrectly formulated problems
  and the regularization method'',} \textit{ Soviet Math. Dokl.} \textbf{ 4}
  (1963) 1035.

\bibitem{hepdata}
\hrefCMSnoop {}{}{HEPD}ata record for this analysis, 2026.
\newblock
  \href{http://dx.doi.org/10.17182/hepdata.169182}{\doi{10.17182/hepdata.169182}}.

\bibitem{Cousins2013}
\hrefCMSnoop {}{R.~D. Cousins, ``Generalization of Chisquare Goodness-of-Fit
  Test for Binned Data Using Saturated Models, with Application to
  Histograms''.}
  \url{https://www.physics.ucla.edu/%7Ecousins/stats/cousins_saturated.pdf},
  2013.

\bibitem{Britzger:2021ocj}
\hrefCMSnoop {}{D.~Britzger, ``The linear template fit'',} \textit{ Eur. Phys.
  J. C} \textbf{ 82} (2022) 731,
  \href{http://dx.doi.org/10.1140/epjc/s10052-022-10581-w}{\doi{10.1140/epjc/s10052-022-10581-w}},
  \href{http://www.arXiv.org/abs/2112.01548}{\texttt{arXiv:2112.01548}}.

\bibitem{Dittmaier:2009cr}
\hrefCMSnoop {}{S.~Dittmaier and M.~Huber, ``Radiative corrections to the
  neutral-current {Drell-Yan} process in the standard model and its minimal
  supersymmetric extension'',} \textit{ JHEP} \textbf{ 01} (2010) 060,
  \href{http://dx.doi.org/10.1007/JHEP01(2010)060}{\doi{10.1007/JHEP01(2010)060}},
  \href{http://www.arXiv.org/abs/0911.2329}{\texttt{arXiv:0911.2329}}.

\end{thebibliography}\endgroup

\numberwithin{figure}{section}
\appendix
\section{Full set of reconstructed \texorpdfstring{\msd}{m[SD]} distributions}\label{sec:appendix}

\begin{figure}[htbp]
    \centering
        \includegraphics[width=0.40\textwidth]{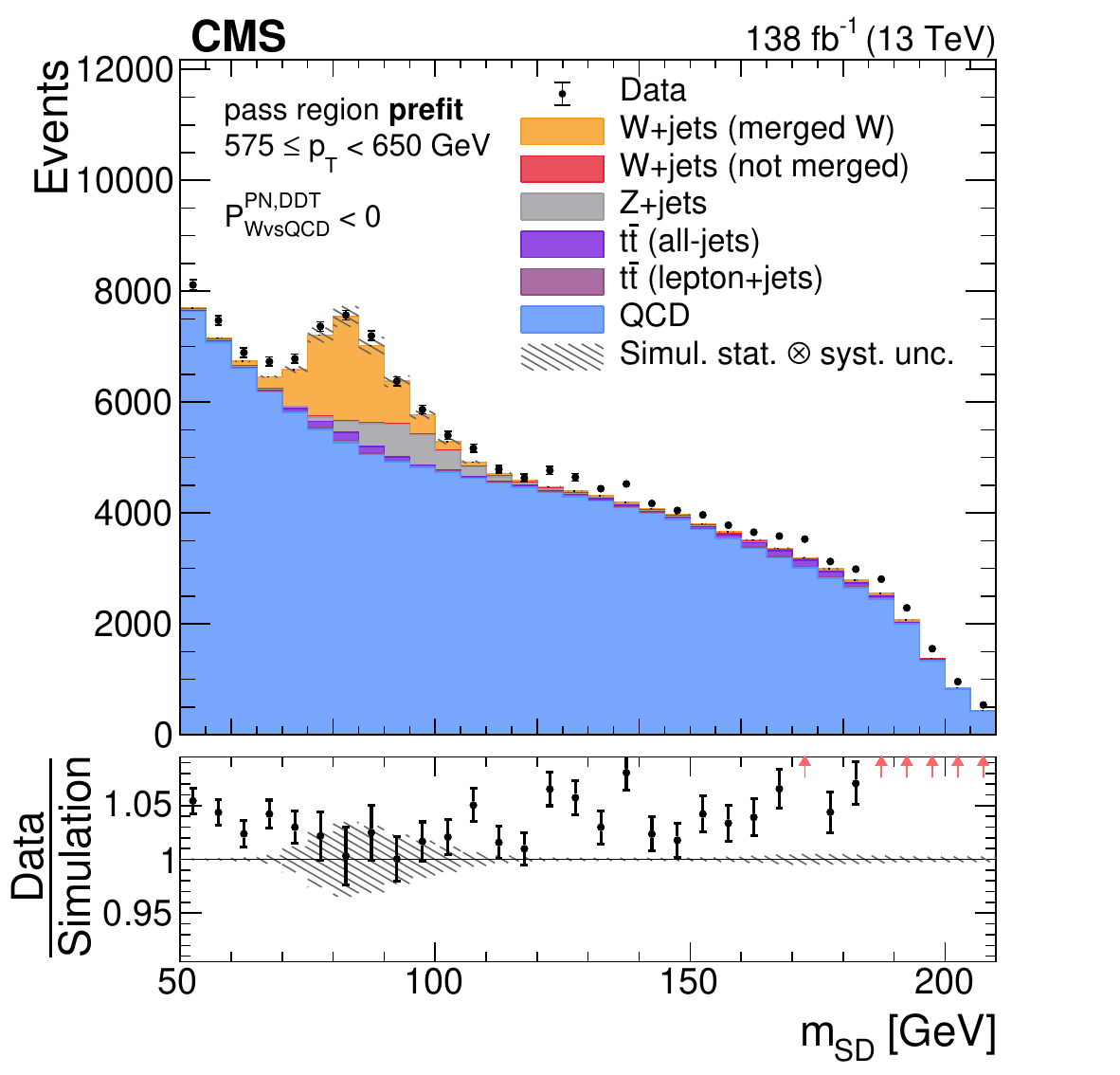}
        \includegraphics[width=0.40\textwidth]{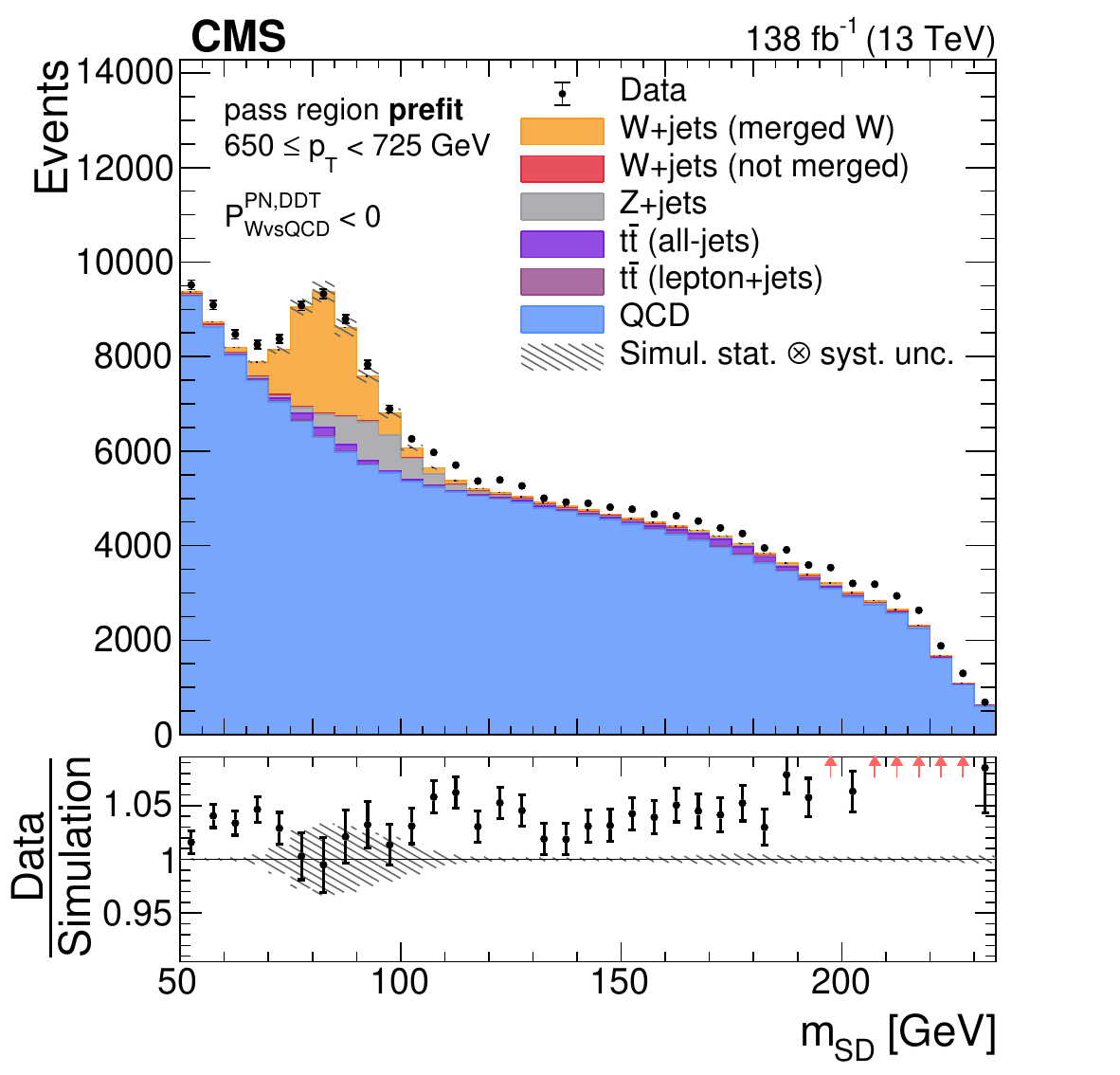} \\
        \includegraphics[width=0.40\textwidth]{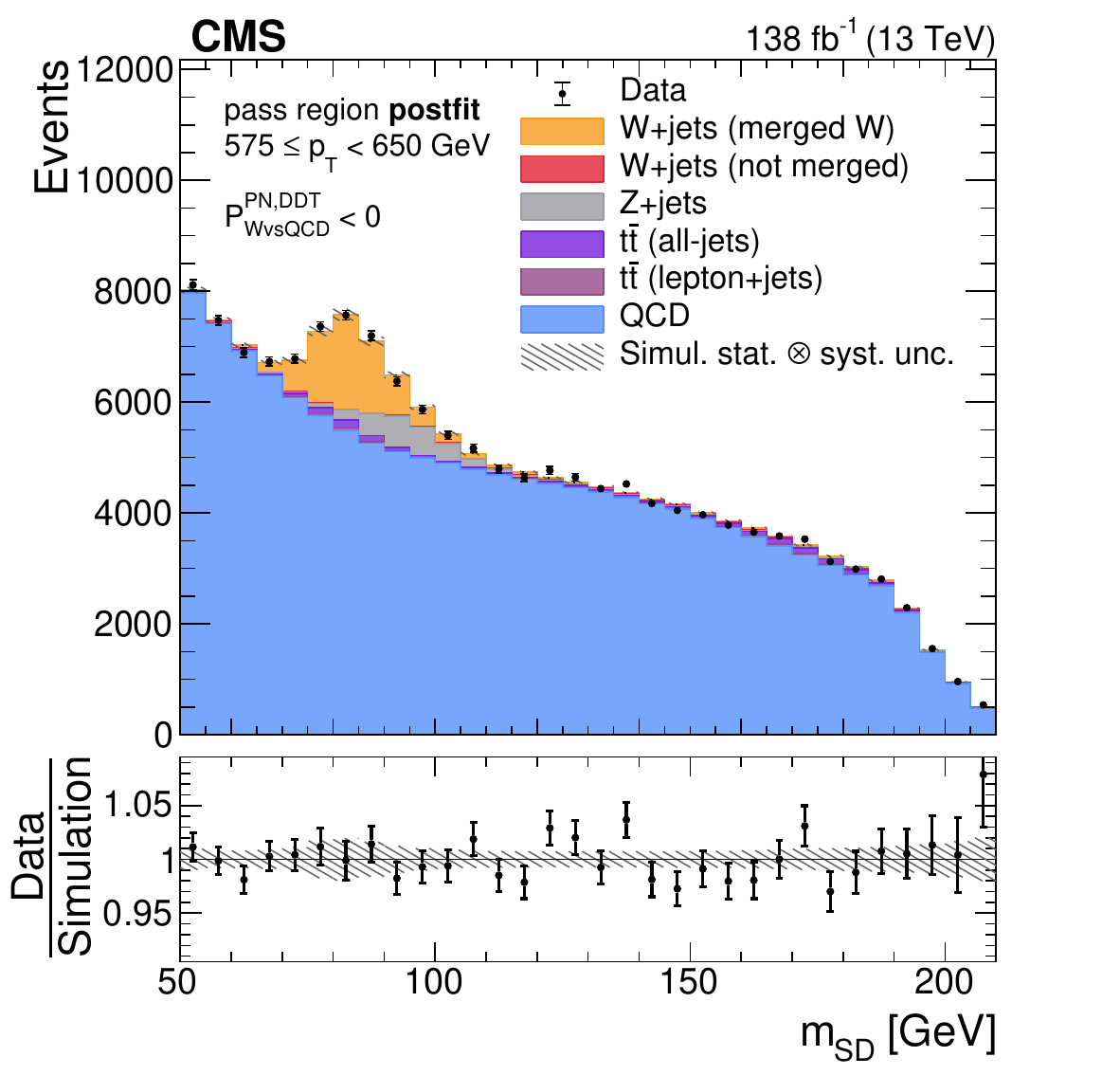}
        \includegraphics[width=0.40\textwidth]{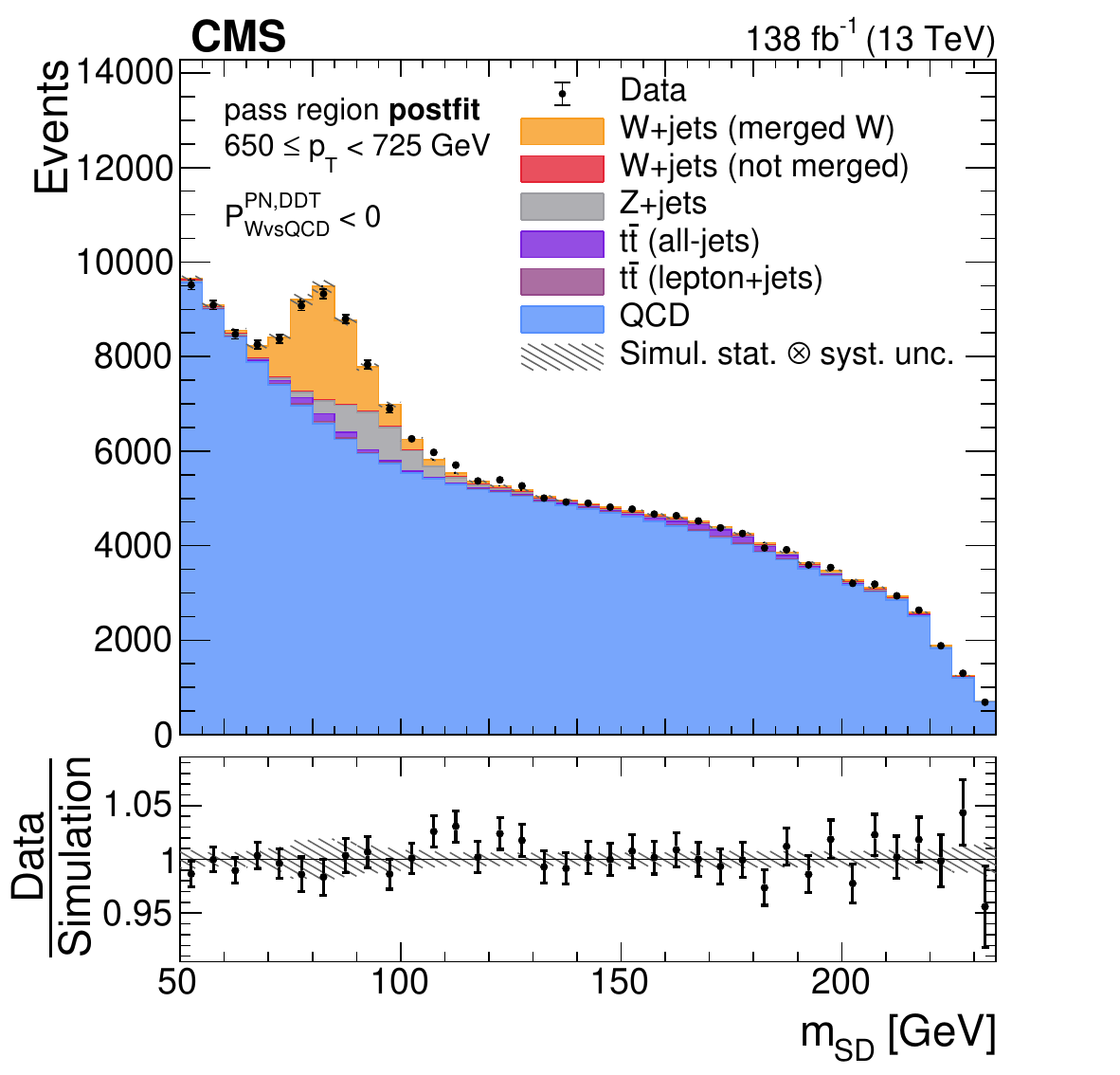} \\
        \includegraphics[width=0.40\textwidth]{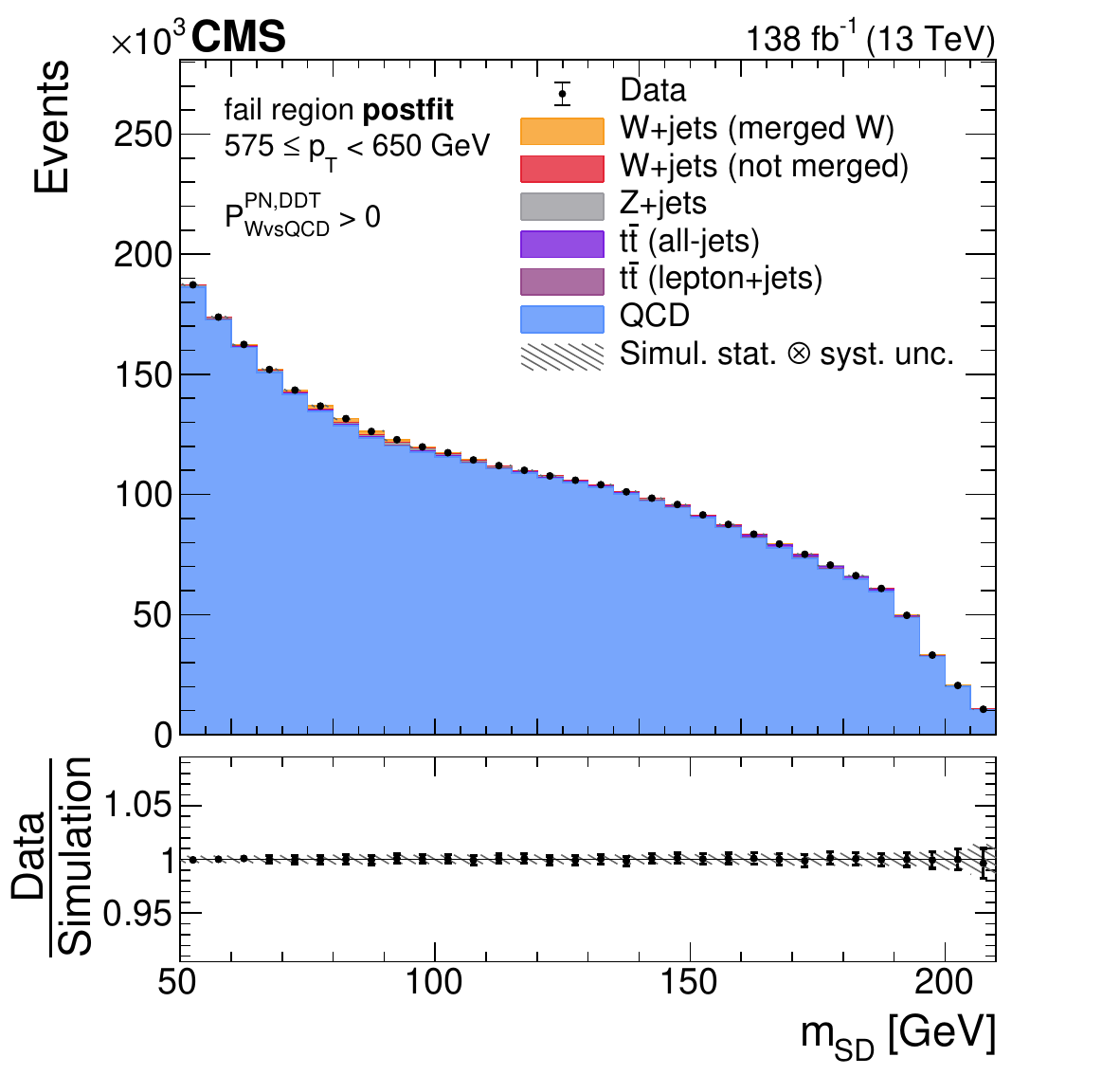} 
        \includegraphics[width=0.40\textwidth]{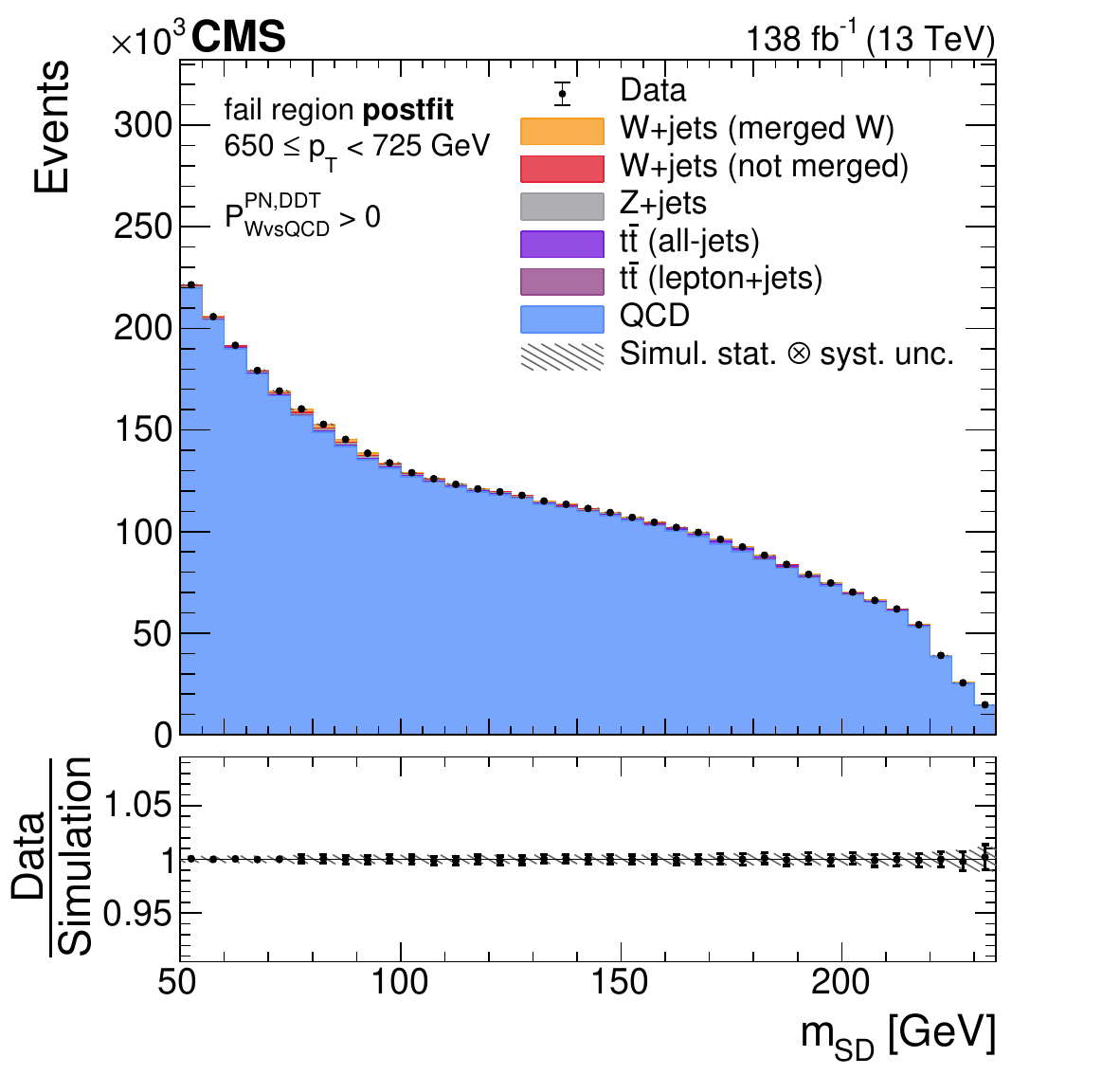}
    \caption{Reconstructed \msd distributions in the \pt bin with $575 < \pt < 650\GeV$ (\cmsLeft) and $650 < \pt < 725\GeV$ (\cmsRight) in the control region (lower row) and signal region before (upper row) and after (middle row) the background estimation and a fit to the data.
    All four data-taking periods are combined, resulting in a total integrated luminosity of 138\fbinv.
    The lower panels show the data-to-simulation ratio. The error bars correspond to the statistical uncertainty in the data.
    In the upper row, the dashed band corresponds to the total uncertainty excluding the uncertainty on the QCD background estimate, while in the middle and lower row it corresponds to the total uncertainty on the background after a fit to the data distribution. \label{img:reco_msd_1}}
\end{figure}

\begin{figure}[htbp]
    \centering
        \includegraphics[width=0.40\textwidth]{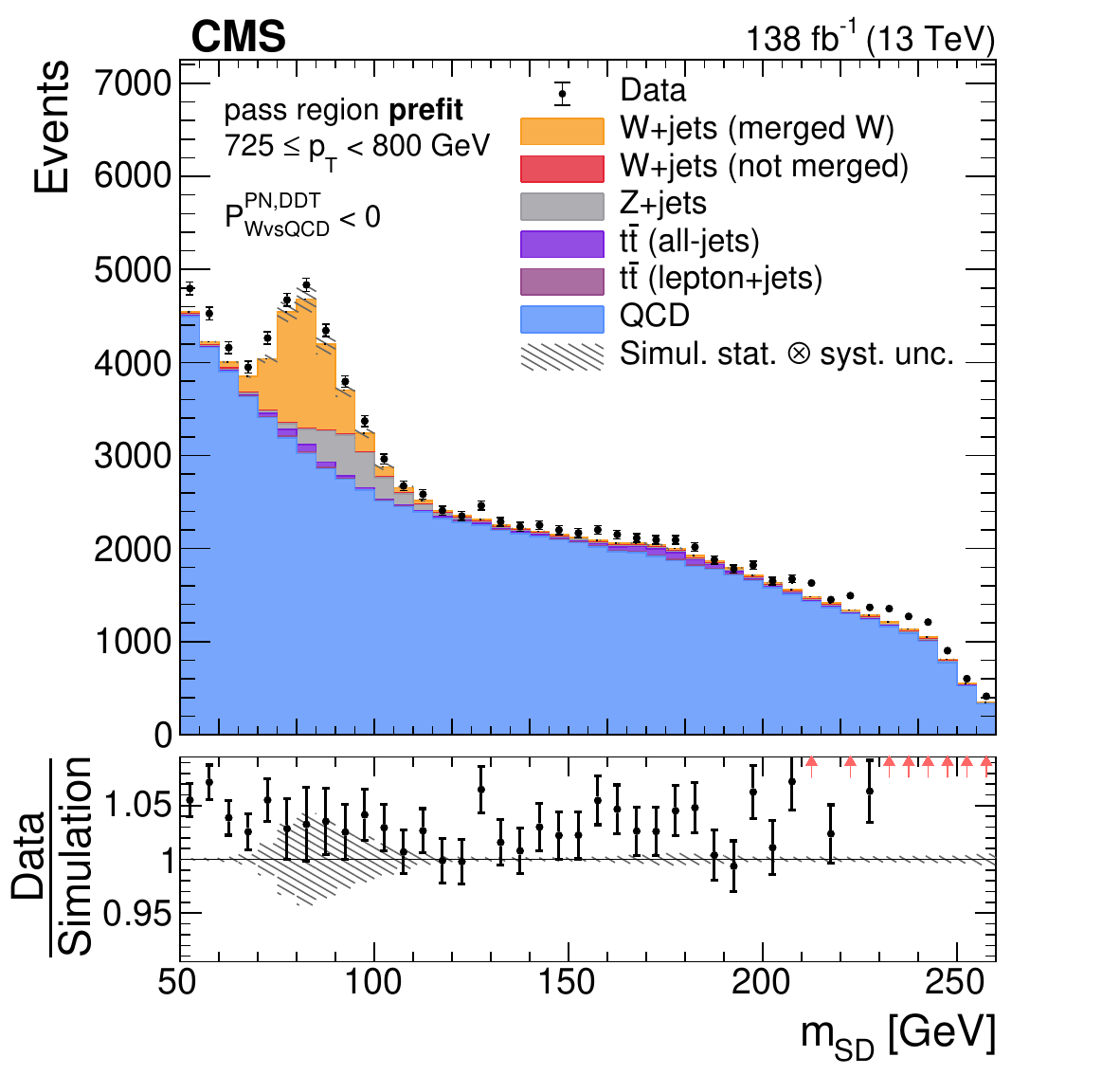}
        \includegraphics[width=0.40\textwidth]{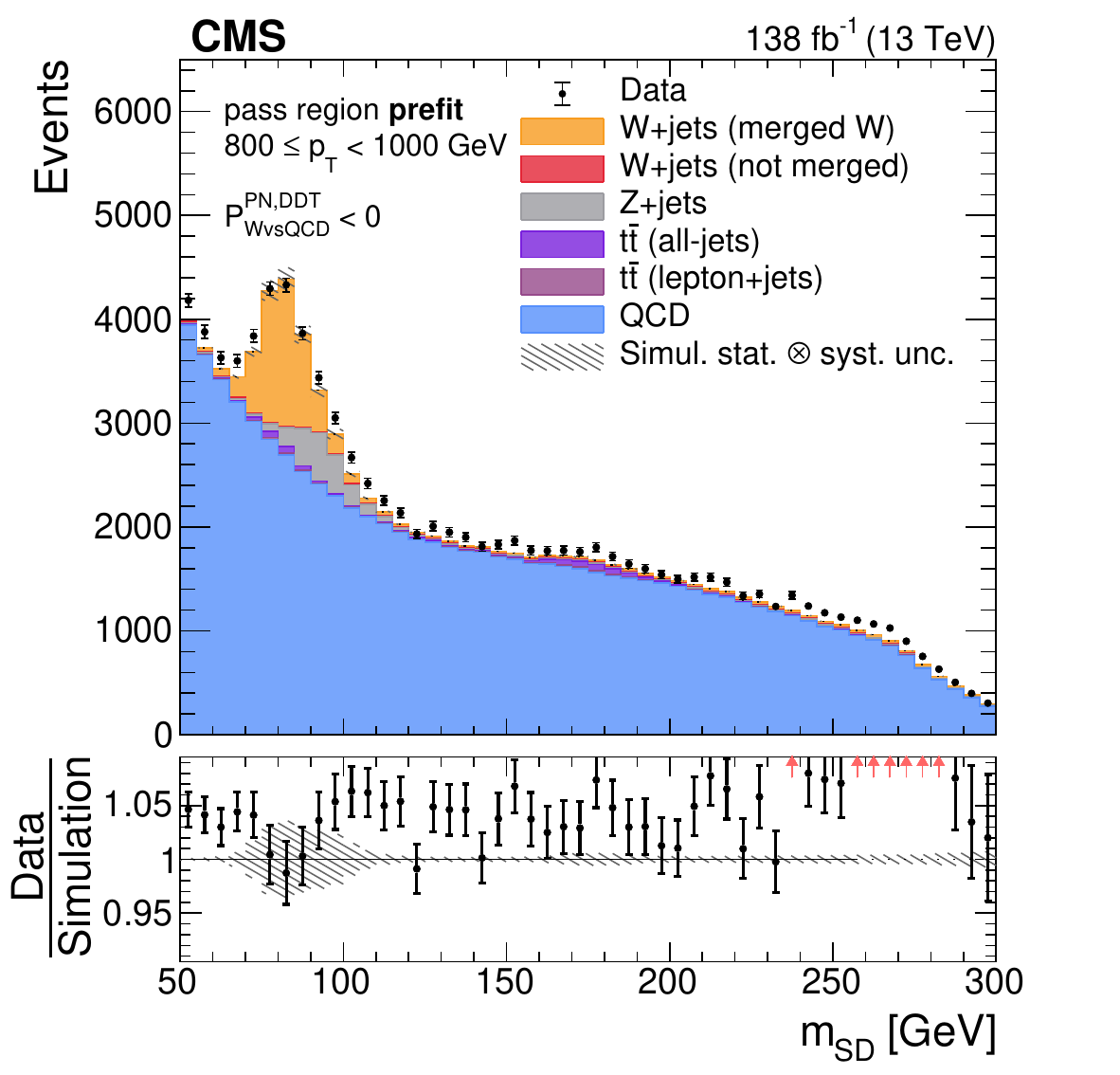} \\
        \includegraphics[width=0.40\textwidth]{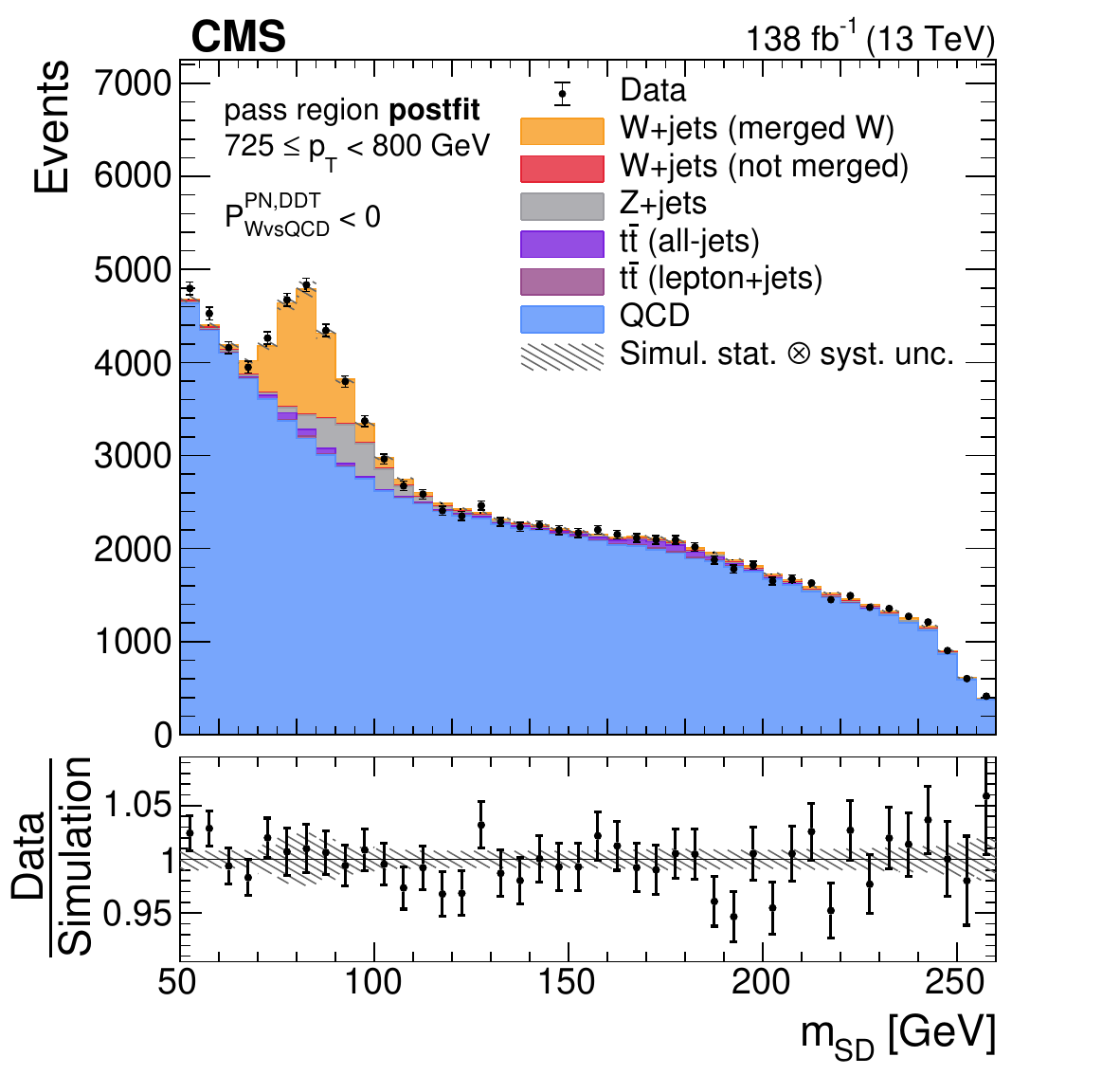}
        \includegraphics[width=0.40\textwidth]{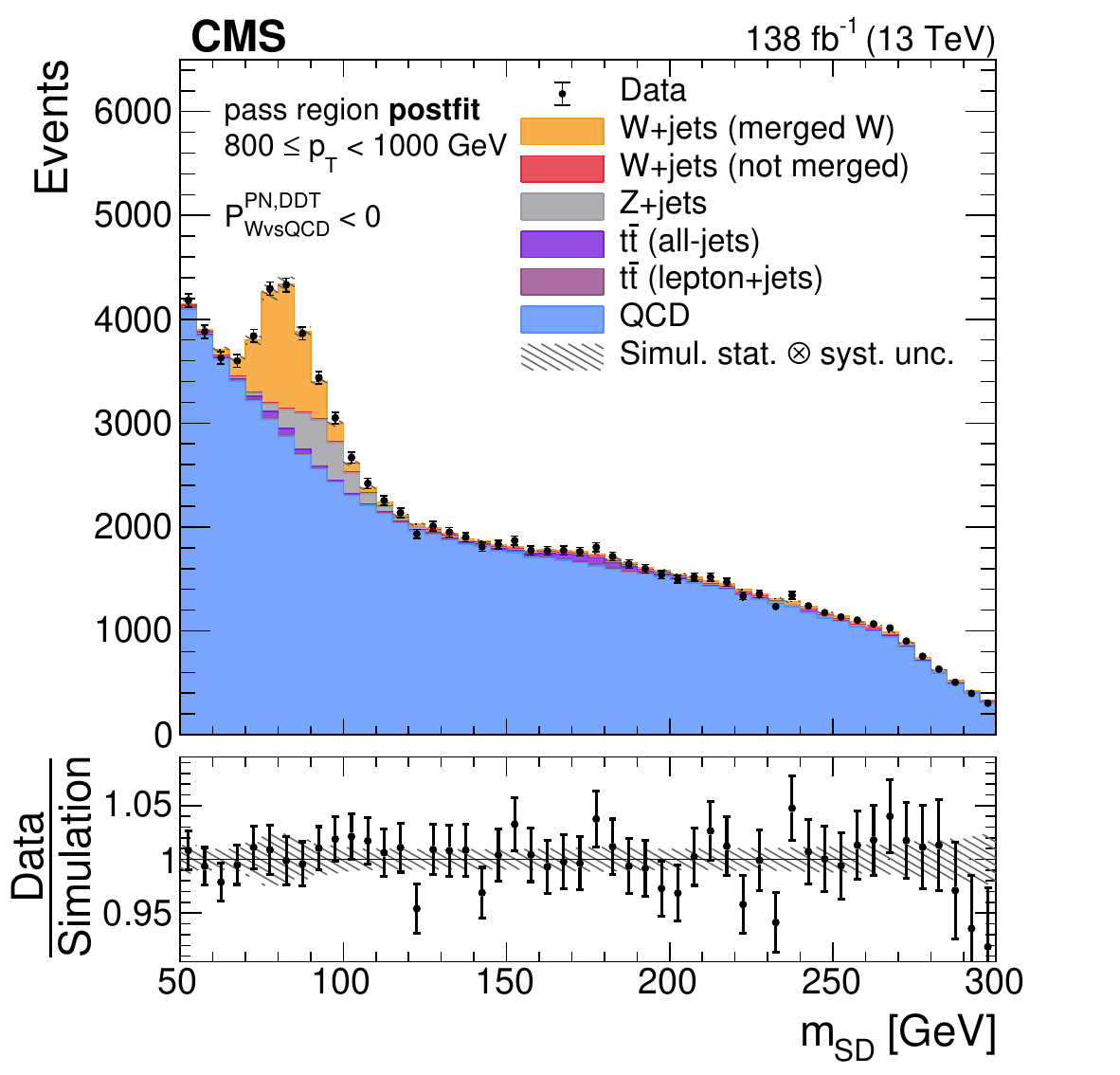} \\
        \includegraphics[width=0.40\textwidth]{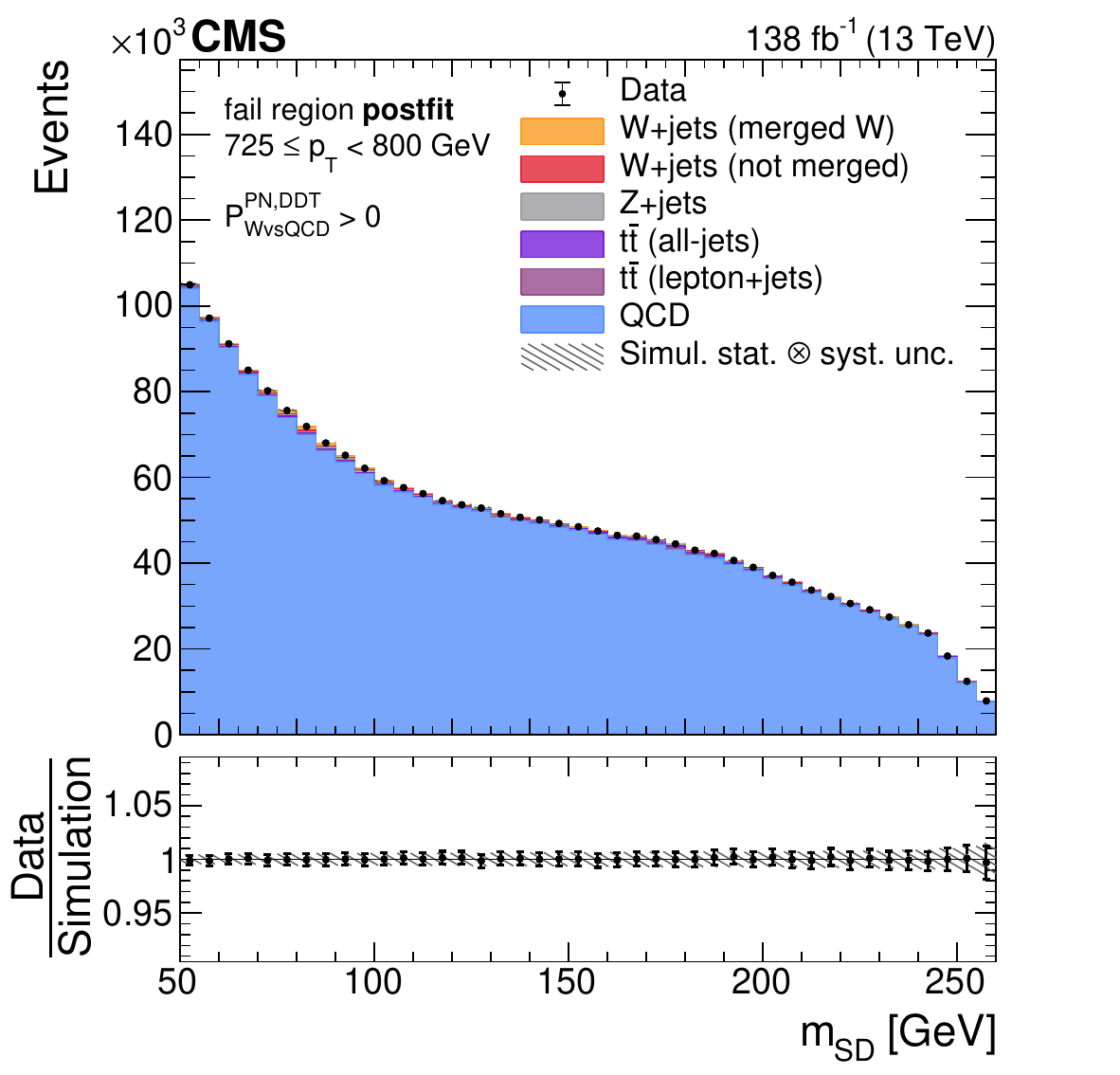} 
        \includegraphics[width=0.40\textwidth]{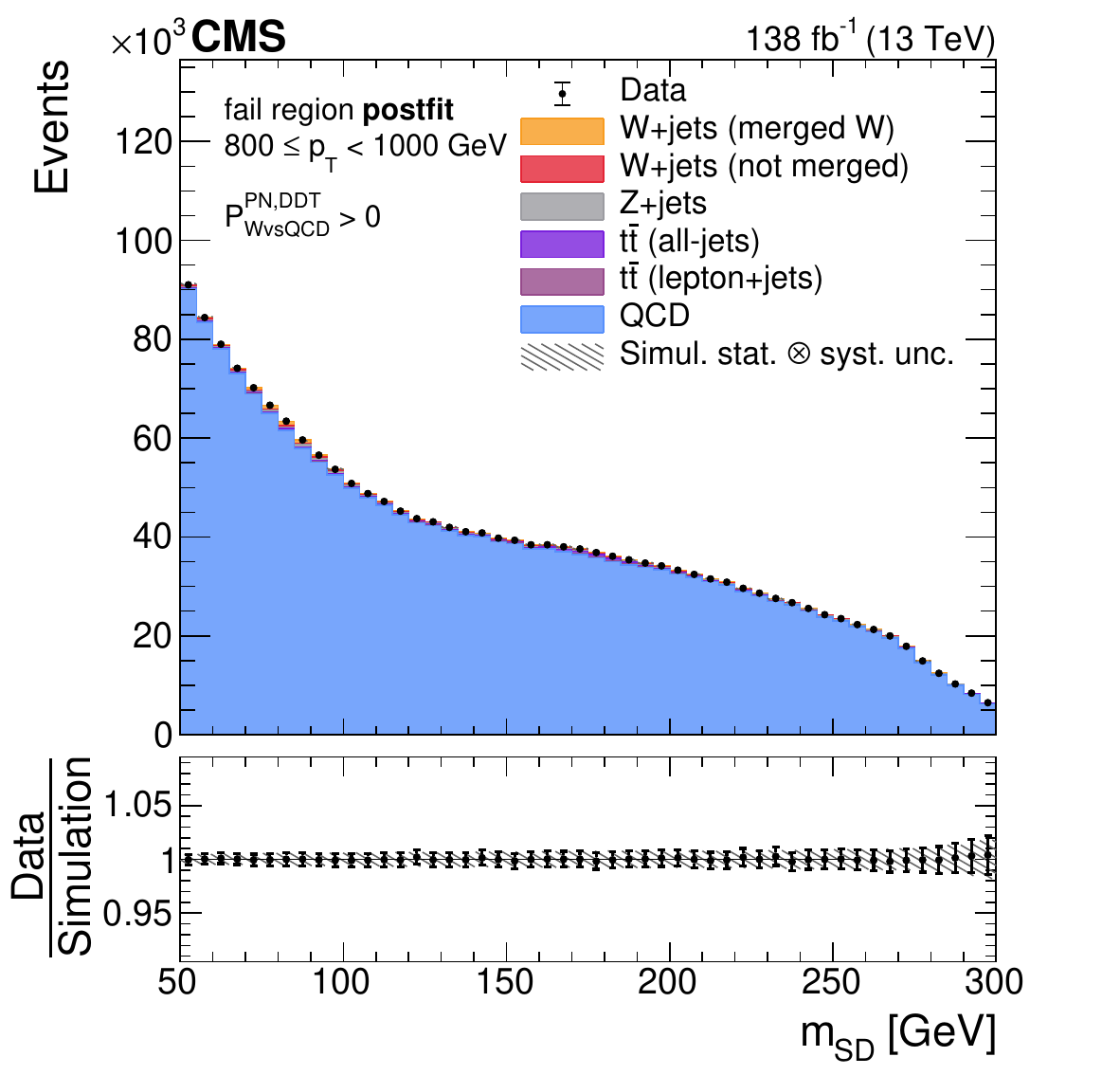}
    \caption{Reconstructed \msd distributions in the \pt bin with $725 < \pt < 800\GeV$ (\cmsLeft) and $800 < \pt < 100\GeV$ (\cmsRight) in the control region (lower row) and signal region before (upper row) and after (middle row) the background estimation and a fit to the data.
    All four data-taking periods are combined, resulting in a total integrated luminosity of 138\fbinv.
    The lower panels show the data-to-simulation ratio. The error bars correspond to the statistical uncertainty in the data.
    In the upper row, the dashed band corresponds to the total uncertainty excluding the uncertainty on the QCD background estimate, while in the middle and lower row it corresponds to the total uncertainty on the background after a fit to the data distribution. \label{img:reco_msd_2}}
\end{figure}

\begin{figure}[htbp]
    \centering
        \includegraphics[width=0.40\textwidth]{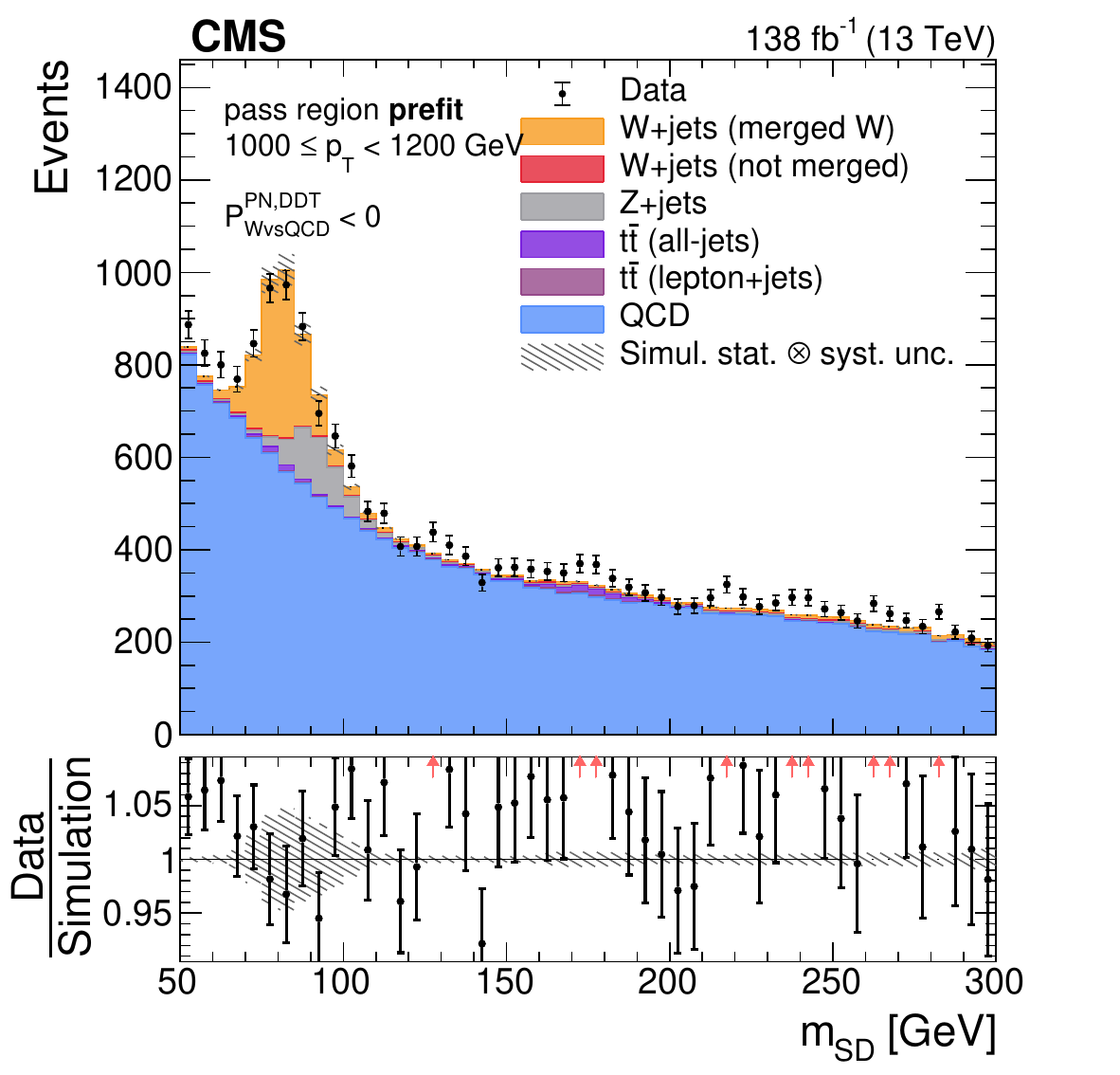}
        \includegraphics[width=0.40\textwidth]{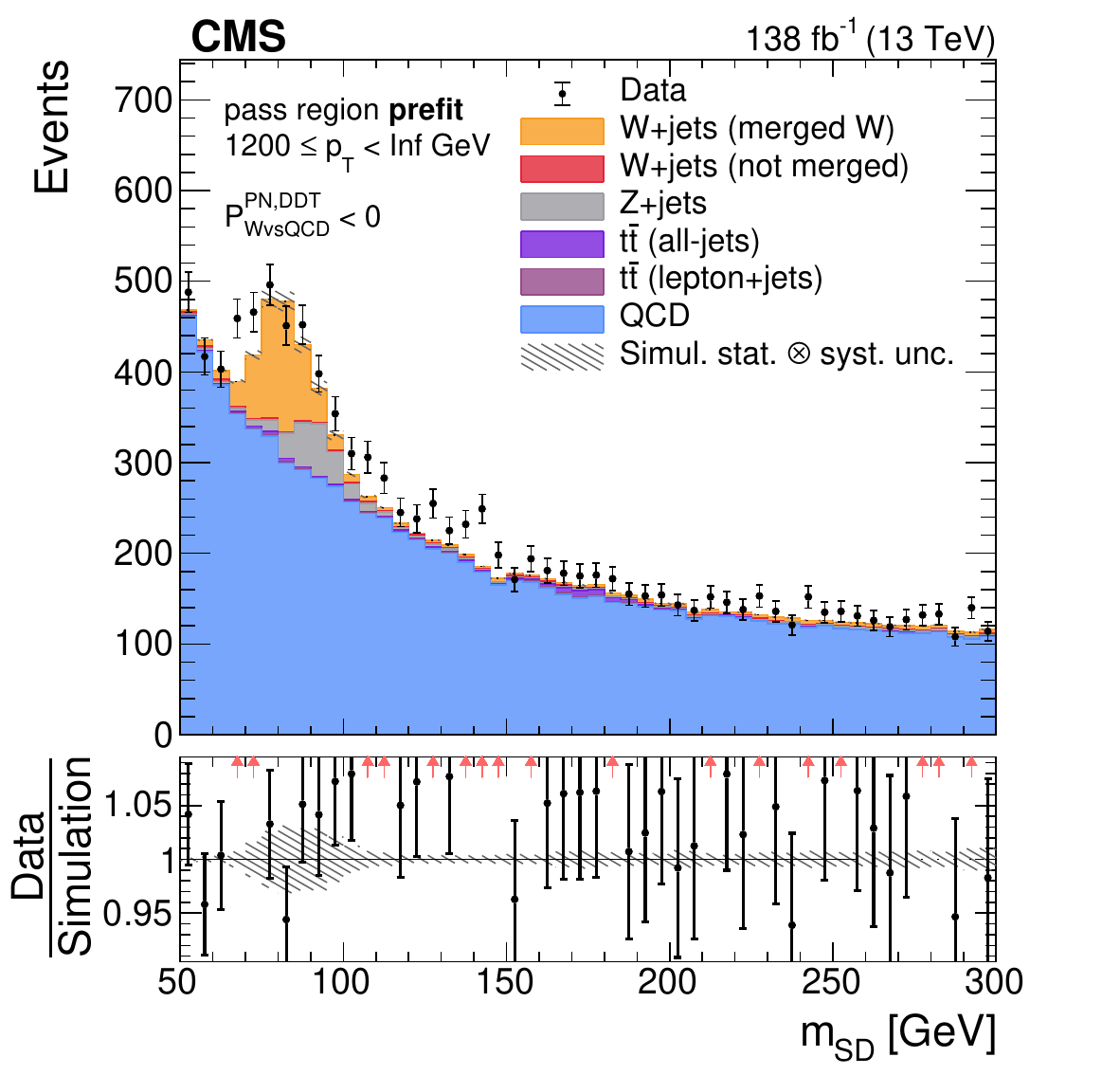} \\
        \includegraphics[width=0.40\textwidth]{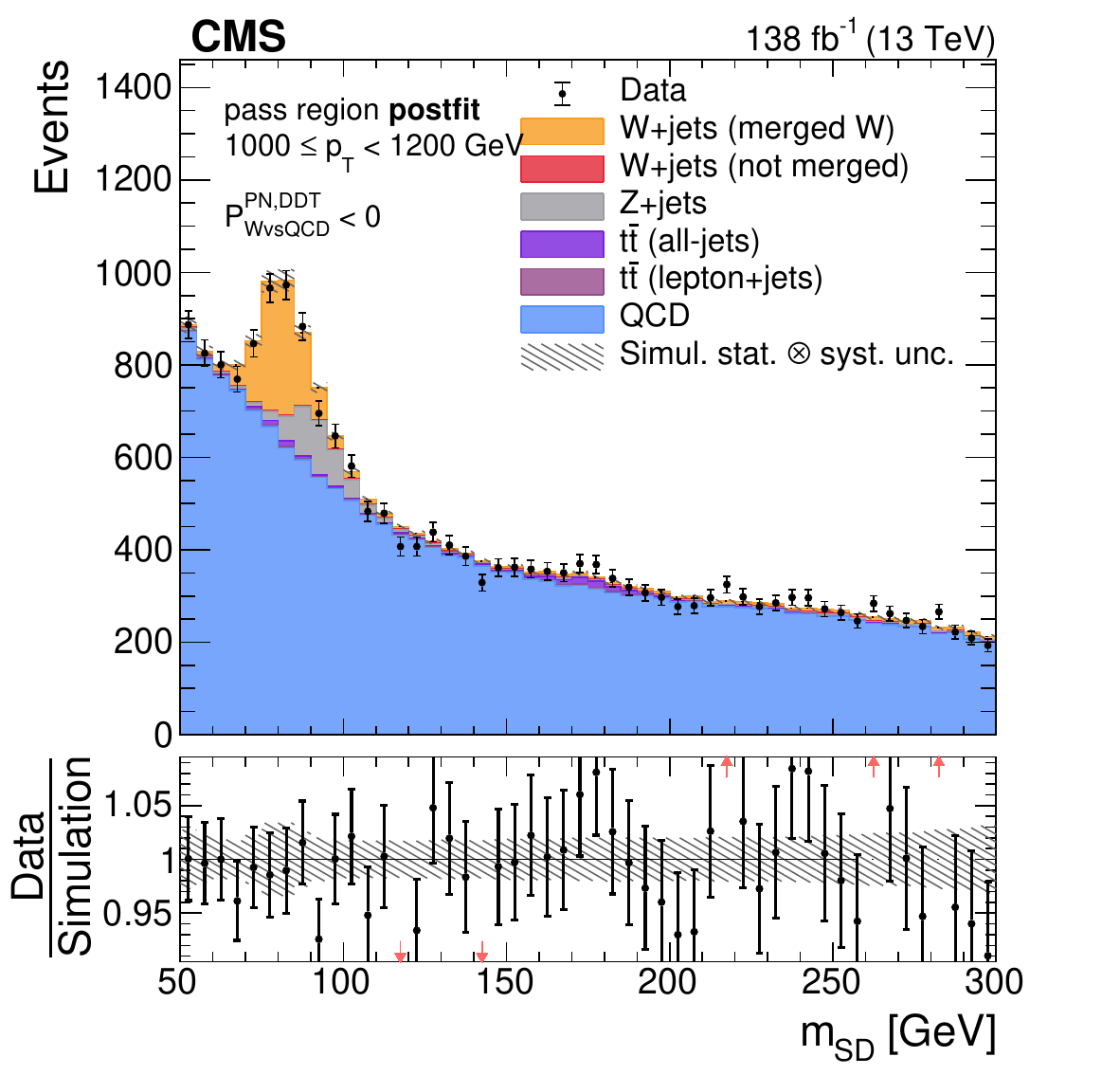}
        \includegraphics[width=0.40\textwidth]{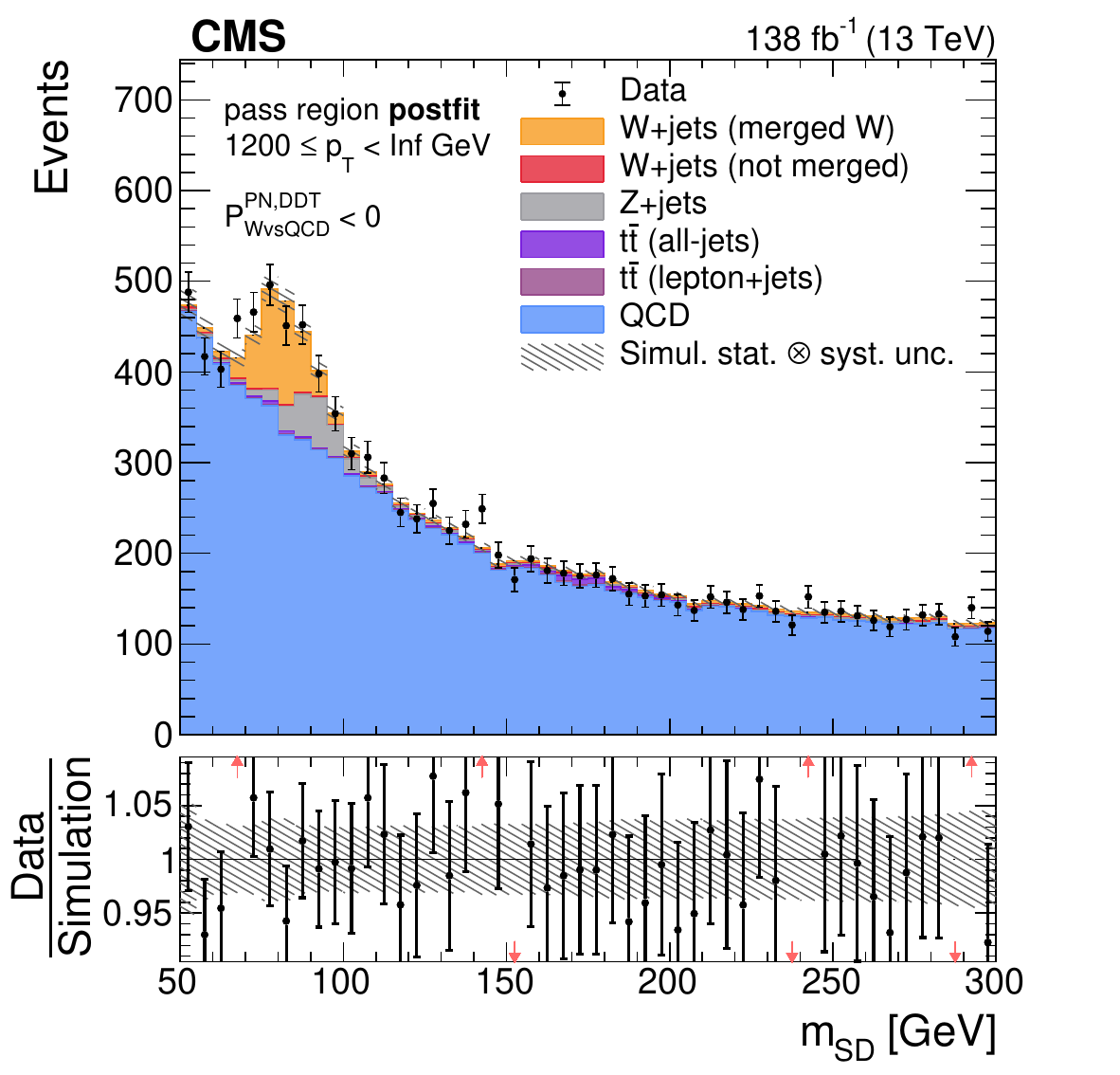} \\
        \includegraphics[width=0.40\textwidth]{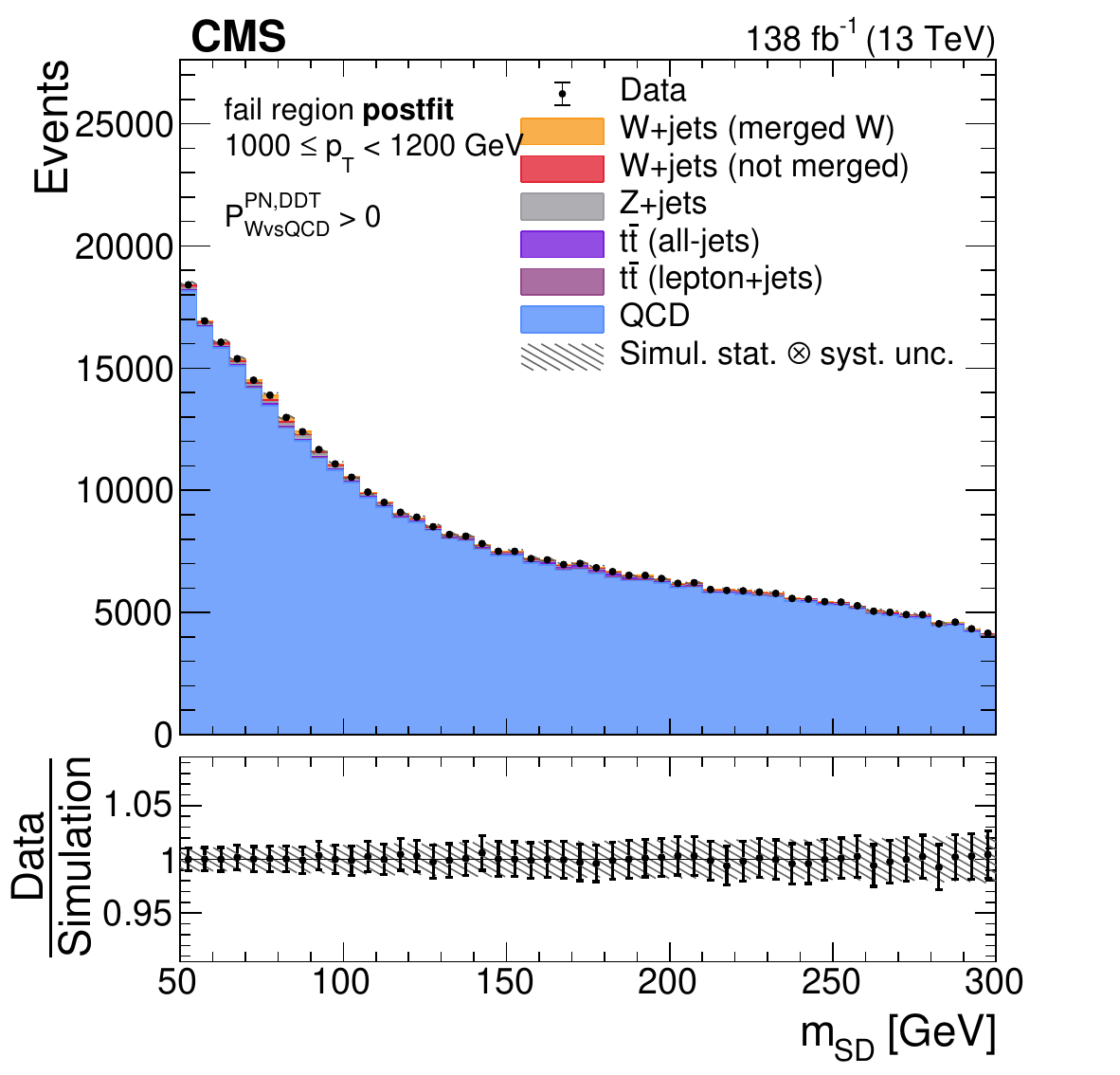} 
        \includegraphics[width=0.40\textwidth]{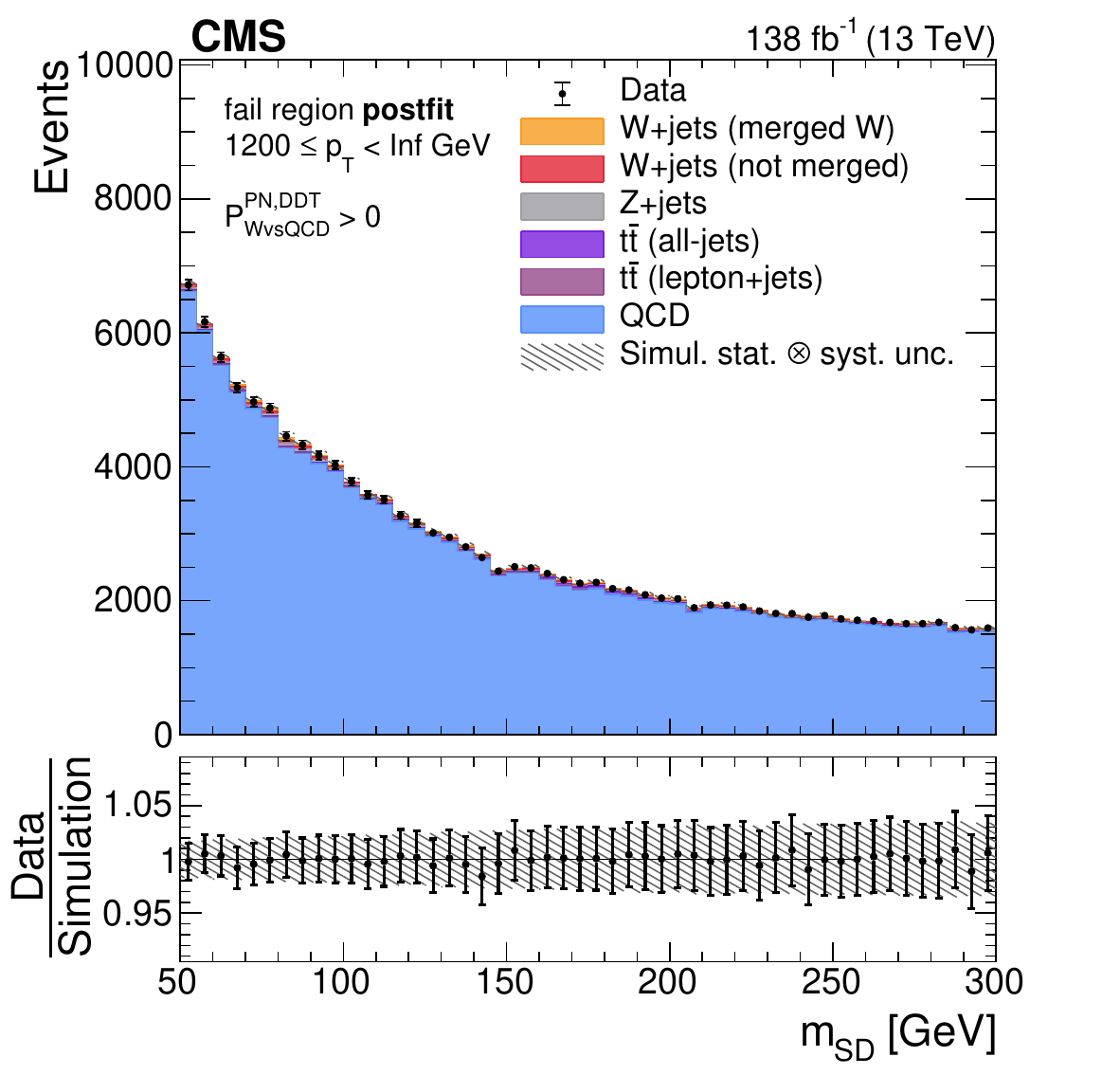}
    \caption{Reconstructed \msd distributions in the \pt bin with $1000 < \pt < 1200\GeV$ (\cmsLeft) and $\pt > 1200\GeV$ (\cmsRight) in the control region (lower row) and signal region before (upper row) and after (middle row) the background estimation and a fit to the data.
    All four data-taking periods are combined, resulting in a total integrated luminosity of 138\fbinv.
    The lower panels show the data-to-simulation ratio. The error bars correspond to the statistical uncertainty in the data.
    In the upper row, the dashed band corresponds to the total uncertainty excluding the uncertainty on the QCD background estimate, while in the middle and lower row it corresponds to the total uncertainty on the background after a fit to the data distribution. \label{img:reco_msd_3}}
\end{figure}

\cleardoublepage \section{The CMS Collaboration \label{app:collab}}\begin{sloppypar}\hyphenpenalty=5000\widowpenalty=500\clubpenalty=5000\cmsinstitute{Yerevan Physics Institute, Yerevan, Armenia}
{\tolerance=6000
A.~Hayrapetyan, V.~Makarenko\cmsorcid{0000-0002-8406-8605}, A.~Tumasyan\cmsAuthorMark{1}\cmsorcid{0009-0000-0684-6742}
\par}
\cmsinstitute{Institut f\"{u}r Hochenergiephysik, Vienna, Austria}
{\tolerance=6000
W.~Adam\cmsorcid{0000-0001-9099-4341}, J.W.~Andrejkovic, L.~Benato\cmsorcid{0000-0001-5135-7489}, T.~Bergauer\cmsorcid{0000-0002-5786-0293}, M.~Dragicevic\cmsorcid{0000-0003-1967-6783}, C.~Giordano\cmsorcid{0000-0001-6317-2481}, P.S.~Hussain\cmsorcid{0000-0002-4825-5278}, M.~Jeitler\cmsAuthorMark{2}\cmsorcid{0000-0002-5141-9560}, N.~Krammer\cmsorcid{0000-0002-0548-0985}, A.~Li\cmsorcid{0000-0002-4547-116X}, D.~Liko\cmsorcid{0000-0002-3380-473X}, M.~Matthewman, I.~Mikulec\cmsorcid{0000-0003-0385-2746}, J.~Schieck\cmsAuthorMark{2}\cmsorcid{0000-0002-1058-8093}, R.~Sch\"{o}fbeck\cmsAuthorMark{2}\cmsorcid{0000-0002-2332-8784}, D.~Schwarz\cmsorcid{0000-0002-3821-7331}, M.~Shooshtari\cmsorcid{0009-0004-8882-4887}, M.~Sonawane\cmsorcid{0000-0003-0510-7010}, W.~Waltenberger\cmsorcid{0000-0002-6215-7228}, C.-E.~Wulz\cmsAuthorMark{2}\cmsorcid{0000-0001-9226-5812}
% ...
\par}
\vskip\cmsinstskip
\dag:~Deceased\\
$^{1}$Also at Yerevan State University, Yerevan, Armenia\\
$^{2}$Also at TU Wien, Vienna, Austria\\
% ...
\end{sloppypar}
\end{document}